\documentclass[aps,prd,reprint,longbibliography,nofootinbib,floatfix,
author-year]{revtex4-2}

\usepackage[T1]{fontenc}
\usepackage{newtxtext,newtxmath}
\usepackage{microtype}

\usepackage{graphicx}
\usepackage{float}            

\usepackage{mathtools}
\usepackage{bm}
\usepackage{booktabs}
\usepackage{siunitx}
\usepackage{natbib}
\usepackage[colorlinks=true,linkcolor=blue,citecolor=magenta,urlcolor=blue]{hyperref}

\def\doi#1{\href{https://doi.org/#1}{\color{blue}#1}}

\usepackage{xstring}
\usepackage{tikz}
\usetikzlibrary{shapes.geometric,shapes.symbols}
\def\urlprefix{}
\makeatletter
\let\ORIGurl\url
\renewcommand{\url}[1]{%
	\begingroup
	\def\UrlArg{#1}%
	\IfBeginWith{\UrlArg}{https://doi.org/}{\href{\UrlArg}{\StrBehind{\UrlArg}{https://doi.org/}}}{%
		\IfBeginWith{\UrlArg}{http://doi.org/}{\href{\UrlArg}{\StrBehind{\UrlArg}{http://doi.org/}}}{%
			\IfBeginWith{\UrlArg}{https://arxiv.org/abs/}{\href{\UrlArg}{arXiv:\StrBehind{\UrlArg}{https://arxiv.org/abs/}}}{%
				\IfBeginWith{\UrlArg}{http://arxiv.org/abs/}{\href{\UrlArg}{arXiv:\StrBehind{\UrlArg}{http://arxiv.org/abs/}}}{%
					\IfBeginWith{\UrlArg}{https://ui.adsabs.harvard.edu/abs/}{\href{\UrlArg}{\StrBehind{\UrlArg}{https://ui.adsabs.harvard.edu/abs/}}}{%
						\IfBeginWith{\UrlArg}{http://ui.adsabs.harvard.edu/abs/}{\href{\UrlArg}{\StrBehind{\UrlArg}{http://ui.adsabs.harvard.edu/abs/}}}{%
							\ORIGurl{\UrlArg}%
	}}}}}}%
	\endgroup
}
\makeatother

\definecolor{mkOrange}{HTML}{E69F00}
\definecolor{mkBlue}{HTML}{0173B2}
\definecolor{mkRed}{HTML}{D55E00}
\definecolor{mkGreen}{HTML}{029E73}
\definecolor{mkPurple}{HTML}{CC78BC}
\newcommand{\mkTriU}[1]{\tikz[baseline=-0.6ex]{\node[draw=black,line width=0.4pt,fill=#1,regular polygon,regular polygon sides=3,inner sep=2.0pt] {};}}
\newcommand{\mkCirc}[1]{\tikz[baseline=-0.6ex]{\node[draw=black,line width=0.4pt,fill=#1,circle,inner sep=2.0pt] {};}}
\newcommand{\mkSqr}[1]{\tikz[baseline=-0.6ex]{\node[draw=black,line width=0.4pt,fill=#1,rectangle,inner sep=2.2pt] {};}}
\newcommand{\mkDia}[1]{\tikz[baseline=-0.6ex]{\node[draw=black,line width=0.4pt,fill=#1,diamond,inner sep=1.6pt] {};}}
\newcommand{\mkPent}[1]{\tikz[baseline=-0.6ex]{\node[draw=black,line width=0.4pt,fill=#1,regular polygon,regular polygon sides=5,inner sep=1.6pt] {};}}
\newcommand{\mkStar}[1]{\tikz[baseline=-0.6ex]{\node[draw=black,line width=0.4pt,fill=#1,star,star points=5,star point ratio=0.5,inner sep=1.5pt] {};}}
\newcommand{\mkHex}[1]{\tikz[baseline=-0.6ex]{\node[draw=black,line width=0.4pt,fill=#1,regular polygon,regular polygon sides=6,inner sep=1.6pt] {};}}
\newcommand{\mkTriD}[1]{\tikz[baseline=-0.6ex]{\node[draw=black,line width=0.4pt,fill=#1,regular polygon,regular polygon sides=3,shape border rotate=180,inner sep=2.0pt] {};}}

\def\araa{ARA\&A}
\def\apj{ApJ}
\def\apjl{ApJ}

\def\aap{A\&A}
\def\mnras{MNRAS}

\def\prb{Phys.~Rev.~B}

\def\prd{Phys.~Rev.~D}
\def\pre{Phys.~Rev.~E}
\def\prl{Phys.~Rev.~Lett.}
\def\physrep{Phys.~Rep.}

\def\lrr{Living Rev.~Relativ.}
\def\jcap{JCAP}
\def\nphysb{Nucl.~Phys.~B}
\def\zphy{Z.~Phys.}

\def\copp{Contrib.~Plasma Phys.}

\def\pop{Phys.~Plasmas}

\newcommand{\mpsi}{m_\psi}
\newcommand{\mchi}{m_\chi}
\newcommand{\mphi}{m_\phi}
\newcommand{\gphi}{g_\phi}
\newcommand{\lamB}{\lambda_{\mathrm{B}}}
\newcommand{\eV}{\mathrm{eV}}
\newcommand{\Msun}{M_\odot}
\newcommand{\Rsun}{R_\odot}
\newcommand{\kappaphi}{\kappa_\phi}
\newcommand{\RT}{R_T}
\newcommand{\xT}{x_T}

\DeclareRobustCommand{\orcidicon}{%
	\begin{tikzpicture}
		\definecolor{orcidgreen}{HTML}{166B3A}
		\draw[orcidgreen, fill=orcidgreen] (0,0)
		circle [radius=0.16]
		node[white] {{\fontfamily{qag}\selectfont \tiny ID}};
	\end{tikzpicture}
	\hspace{-2mm}
}
\foreach \x in {A, ..., Z}{%
	\expandafter\xdef\csname orcid\x\endcsname{\noexpand\href{https://orcid.org/\csname orcidauthor\x\endcsname}{\noexpand\orcidicon}}
}

\begin{document}
	\raggedbottom

	\title{Self-gravitating quantum stars with a globally relevant Bohm potential}

	\author{Il\'{i}dio Lopes\orcidA{}}
	\affiliation{CENTRA, Instituto Superior T\'ecnico, Universidade de Lisboa, 1049-001 Lisbon, Portugal}

	\date{February 2026}

\begin{abstract}
The microphysics underlying non-baryonic dark matter remains unknown.
I derive the two-species Schr\"odinger--Poisson--Yukawa system for spin-$\tfrac{1}{2}$ dark-sector fermion fields, $\psi$ (mass $m_1$) and $\chi$ (mass $m_2$), coupled through a scalar mediator of mass $\mphi$ via a universal Yukawa coupling, within an orbital-free density-functional framework with the Kirzhnits gradient coefficient $\lamB=1/9$.
A central result is that the Bohm potential, far from being negligible in the Thomas--Fermi regime, contributes a species-dependent surface-energy correction analogous to the nuclear liquid-drop model: the heavier fermion species generates an outward quantum-pressure wall whilst the lighter species provides an inward surface tension, with degeneracy pressure furnishing the bulk confinement.
In the single-species Schr\"odinger--Poisson limit the ground state recovers the benchmarked invariants $M_{\mathrm{dim}}\simeq 3.883$ and $\xT\simeq 2.562$, yielding $M \RT\simeq 9.95\,\lamB\hbar^2/(G m_1^2)$.
For polytropic index $\gamma=5/3$ the mass--radius relation satisfies $R\propto M^{-1/3}$; for $\gamma=4/3$ a limiting mass emerges above which no stable equilibrium exists.
Illustrative configurations span $M=10^{-8}$--$5\,\Msun$, $m_1\sim 10^{-14}$--$10^{-6}\,\eV$, and radii from a few~km to $\sim 10^3\,\Rsun$, with gravitational-wave contact frequencies in the Einstein Telescope and LISA bands and microlensing signatures accessible to current surveys.
The predictive rigidity of the resulting mass--radius relation, in which the single microphysical parameter $m_1$ determines the equilibrium radius once the total mass is specified, furnishes a reproducible, first-principles reference for constraining the dark-fermion mass in multi-component dark sectors.
\end{abstract}

	\maketitle

	\section{Introduction}
	\label{sec:intro}

	The case for non-baryonic dark matter rests on independent and mutually reinforcing observational probes: the power spectrum of the cosmic microwave background \citep{2020A&A...641A...6P}, galaxy rotation curves \citep{2017ARA&A..55..343B}, the universal density profiles of dark-matter haloes \citep{1996ApJ...462..563N}, and strong-lensing observations, amongst others.
	Taken together, these measurements establish that roughly $27\%$ of the energy density of the Universe resides in a gravitationally active component that neither emits nor absorbs electromagnetic radiation, yet the particle-physics identity of this component remains stubbornly unknown \citep{2024PhRvD.110c0001N}.
	Were a fraction of that dark component locked in stable, self-gravitating compact objects rather than distributed as a diffuse particle bath, the dark matter problem would be partly recast from particle physics to compact-object astrophysics, with observational signatures accessible to gravitational-wave and microlensing surveys.
	One of the most instructive proving grounds for specific dark-matter candidates is the small-scale structure of galactic haloes, where collisionless cold dark matter simulations predict steeply cusped density profiles that conflict with the flatter cores inferred from dwarf-galaxy rotation curves \citep{2017ARA&A..55..343B,2000PhRvL..84.3760S}.

	Amongst the candidates that naturally suppress small-scale power, ultralight dark matter, whether bosonic (fuzzy/wave dark matter) or fermionic, has attracted considerable attention \citep{2000PhRvL..85.1158H,2016PhR...643....1M,2021A&ARv..29....7F}.
	In the bosonic scenario a particle mass $m\sim 10^{-22}\,\eV$ yields a de~Broglie wavelength of order a kiloparsec, so that quantum-pressure effects smooth the central cusps collisionless models produce; in the fermionic case degeneracy pressure furnishes analogous support.
	The cosmological behaviour of ultralight bosonic dark matter is well described by the Schr\"odinger--Poisson (SP) system \citep{2017PhRvD..95d3541H}, and numerical simulations reveal that virialised haloes harbour a dense solitonic core whose profile is the nodeless ground state of the SP eigenvalue problem \citep{2014NatPh..10..496S,2014PhRvL.113z1302S}.

	The theoretical pedigree of self-gravitating quantum equilibria is considerably older: \citet{1968PhRv..172.1331K} studied the Klein--Gordon geon, \citet{1969PhRv..187.1767R} established the general-relativistic framework for self-gravitating boson systems (see \citet{2023LRR....26....1L} for a comprehensive review), and the Newtonian SP limit was placed on a rigorous footing by \citet{1977StAM...57...93L} and \citet{1998CQGra..15.2733M}.
	Self-interacting extensions were analysed by \citet{2011PhRvD..84d3531C} and \citet{2011PhRvD..84d3532C}, who obtained mass--radius relations for Newtonian condensates.
	The same ground-state soliton appears in the theory of self-bound dark-sector objects, where a Yukawa-type scalar interaction generates compact configurations held together by dark forces rather than gravity \citep{1987PhRvD..35.3678L,2014PhRvD..90e5030W,2023PhRvD.108d4024D}; related non-topological solitons such as $Q$-balls \citep{1985NuPhB.262..263C} further illustrate the observational richness of dark-matter microphysics \citep{2018PhR...730....1T,2010JCAP...05..021K}.

	In a parallel line of investigation, dark-matter imprints on ordinary stars have been explored through gravitational capture and energy transport \citep{1985ApJ...296..679P,1987ApJ...321..560G}, with observable consequences for helioseismology \citep{2010PhRvL.105a1301F,2010PhRvD..82h3509T,2002PhRvL..88o1303L,2011ApJ...733L..51L,2012ApJ...757..130L,2014ApJ...786...25L,2014ApJ...780L..15L,2014ApJ...795..162L,2019ApJ...880L..25L}, asteroseismology \citep{2011MNRAS.410..535C,2013ApJ...765L..21C,2021MNRAS.507.3434R} and neutron-star structure \citep{2010PhRvD..81l3521D,2011PhRvD..83h3512K,2024PhR..1052....1B}.
	Recent work has begun to connect these stellar-scale diagnostics with the broader programme of characterising exotic compact objects in the dark sector \citep{2024PhLB..85638901P,2026PhRvD.113d3049B}.
	The present work extends this programme to the Newtonian regime, deriving the equilibrium structure of self-gravitating dark-fermion configurations from first principles within the quantum-hydrodynamic framework.

	In the quantum-hydrodynamic (QHD) description, the support provided by wave-mechanical effects is encoded by the Bohm potential, a gradient correction to the classical Euler equation first derived in hydrodynamic form by \citet{1927ZPhy...40..322M} and later recast within the hidden-variable interpretation by \citet{1952PhRv...85..166B}.
	The same gradient term arises naturally in extended Thomas--Fermi theory \citep{1935ZPhy...96..431W,2025PhyU...68..691P} and in quantum-hydrodynamic models of charged fluids \citep{2001PhRvB..64g5316M,2015CoPP...55..437M}.
	For a degenerate fermion fluid the Bohm correction, with the Kirzhnits prefactor $\lamB=1/9$, supplements the Thomas--Fermi degeneracy pressure; in a multi-component dark sector each species carries its own Bohm potential and degeneracy pressure whilst sharing the gravitational field, so that the equilibrium structure acquires a rich dependence on the particle masses $m_i$, the mass ratio $q=m_1/m_2$, and the central composition $f$.

	Before proceeding, it is worth remarking on the principal novelties of this study.
	Whilst previous works have elegantly characterised single-species or purely bosonic configurations, the equilibrium structure of a two-component degenerate dark-fermion fluid has received comparatively little attention, despite being physically well motivated by multi-species dark-sector models.
	In this article I restrict attention to the Thomas--Fermi regime, wherein degeneracy pressure furnishes the dominant bulk support, and demonstrate that the Bohm potential, far from vanishing entirely, contributes a species-dependent surface-energy correction whose sign differs between the two fermion species, in close analogy with the nuclear liquid-drop model.
	The original contributions of the present work are threefold: (i)~the derivation of the two-species Schr\"odinger--Poisson--Yukawa system from a Yukawa Lagrangian with universal coupling, (ii)~the identification and quantification of the species-dependent Bohm surface-energy correction in the Thomas--Fermi regime, and (iii)~the computation of benchmarked dimensionless invariants together with the gravitational-wave and microlensing signatures of these exotic configurations.

	The treatment is deliberately Newtonian, restricted to equilibrium configurations within the orbital-free density-functional framework where the Bohm term contributes to the bulk force balance rather than merely regularising a sharp surface \citep{1987PhRvD..35.3678L,2023PhRvD.108d4024D}.
	Setting $f=\eta_2(0)=0$ recovers the single-species limit, in which an exact scaling symmetry renders the family tightly constrained: once $m_1$ and the total mass $M$ are specified, the radius follows uniquely.
	It follows that any detection of a compact object whose mass and radius are inconsistent with neutron stars or black holes, yet compatible with the sequences derived here, would directly constrain the dark-fermion mass $m_1$, thereby furnishing a bridge between compact-object astrophysics and the dark-sector Lagrangian.

	The remainder of the paper is organised as follows.
	Sections~\ref{sec:model} and~\ref{sec:QHD} derive the Newtonian quantum-hydrodynamic model from a two-species Yukawa Lagrangian, identify the SP--Yukawa reduction at finite $\mphi$, and set out the hydrostatic equilibrium incorporating the Bohm potential.
	Section~\ref{sec:dimensionless} introduces dimensionless variables and the scaling symmetry governing the mass--radius relation.
	The numerical implementation is described in Section~\ref{sec:numerics}, whilst Section~\ref{sec:results} presents the SP baseline, the Thomas--Fermi reference model, and parameter-sensitivity tests.
	Section~\ref{sec:conclusions} discusses stability, observational prospects, and the bearing of these results on the dark matter problem.

	I adopt natural units ($\hbar=c=1$) throughout, so that all dimensionful quantities are expressed as powers of the electron-volt; the Yukawa coupling $\gphi$ is dimensionless.
	Explicit factors of $\hbar$, $c$ and $G$ are retained where clarity demands it.
	Standard conversion factors \citep{2024PhRvD.110c0001N} and the IAU~2015 nominal solar values $\Msun\simeq 1.9885\times 10^{30}\,\mathrm{kg}$, $\Rsun=6.957\times 10^{8}\,\mathrm{m}$ \citep{2016AJ....152...41P} are used for SI and astrophysical scales.

	\section{Model: from the Yukawa Lagrangian to the Schr\"odinger--Poisson limit}
	\label{sec:model}

	Before proceeding to the numerical study, it is instructive to situate the SP system within the broader framework of dark-sector Lagrangians, both to clarify the approximations involved and to establish contact with the fermionic soliton models of \citet{1987PhRvD..35.3678L} and \citet{2023PhRvD.108d4024D}.
	I follow the field-theoretic conventions of \citet{1997IJMPE...6..515S} throughout.

	\subsection{Yukawa Lagrangian and field equations}

	Consider two dark-sector spin-$\tfrac{1}{2}$ fields, $\psi$ of bare mass $\mpsi$ and $\chi$ of bare mass $\mchi$, both coupled to a common real scalar mediator $\phi$ of mass $\mphi$ through a universal Yukawa coupling of strength $\gphi$.
	In this notation, $\bar\psi=\psi^\dagger\gamma^0$ is the Dirac adjoint, and the bilinears $\bar\psi\psi$, $\bar\chi\chi$ reduce, in the non-relativistic limit, to the respective particle number densities $n_1$, $n_2$.
	The baseline renormalisable Lagrangian density, standard in quantum field theory and nuclear mean-field models of the Walecka type \citep{1997IJMPE...6..515S}, reads
	\begin{equation}
		\begin{aligned}
			\mathcal{L}={}&\frac{1}{2}\,\partial_\mu\phi\,\partial^\mu\phi - V(\phi) +\bar\psi\,(i\gamma^\mu\partial_\mu-\mpsi)\psi\\
			&+\bar\chi\,(i\gamma^\mu\partial_\mu-\mchi)\chi -\gphi\,(\bar\psi\psi + \bar\chi\chi)\,\phi,
		\end{aligned}
		\label{eq:Lagrangian}
	\end{equation}
	with the most general renormalisable scalar potential
	\begin{equation}
		V(\phi)=\frac{1}{2}\mphi^2\phi^2
		+\frac{\kappa}{3!}\phi^3
		+\frac{\lambda_\phi}{4!}\phi^4,
		\qquad \lambda_\phi>0.
		\label{eq:Vphi}
	\end{equation}
	The Yukawa coupling explicitly breaks $\phi\to-\phi$, so the cubic interaction is permitted by all symmetries; boundedness from below requires $\lambda_\phi>0$ \citep{1997IJMPE...6..515S}.
	A linear term can be removed by a constant field shift that eliminates the tadpole and redefines the couplings.
	It is worth noting that the baseline Lagrangian~\eqref{eq:Lagrangian} contains no direct four-fermion cross-coupling of the form $(\bar\psi\psi)(\bar\chi\chi)$: such a vertex has mass dimension six, is therefore non-renormalisable, and lies outside the minimal model \citep{1997IJMPE...6..515S}.
	The two species consequently interact only through the shared mediator $\phi$, and in the gravitational limit through the common potential $\Phi$, with no residual contact interaction between $m_1$ and $m_2$.
	With $\kappaphi^2=\mphi^2$ in natural units, Eq.~\eqref{eq:KG_linear} below makes this structure explicit.

	The Euler--Lagrange equations yield the Dirac equations in the scalar background,
	\begin{align}
		(i\gamma^\mu\partial_\mu-\mpsi)\psi &= \gphi\,\phi\,\psi,
		\label{eq:EL_Dirac_psi}\\
		(i\gamma^\mu\partial_\mu-\mchi)\chi &= \gphi\,\phi\,\chi,
		\label{eq:EL_Dirac_chi}
	\end{align}
	showing that the scalar field generates effective mass shifts $\tilde{m}_\psi=\mpsi+\gphi\phi$ and $\tilde{m}_\chi=\mchi+\gphi\phi$.
	For compactness I write $m_1\equiv\mpsi$ and $m_2\equiv\mchi$ henceforth, and define the mass ratio $q\equiv m_1/m_2$.
	The mediator obeys the Klein--Gordon equation
	\begin{equation}
		(\partial^2+\mphi^2)\phi
		+\frac{\kappa}{2}\phi^2+\frac{\lambda_\phi}{6}\phi^3
		=-\gphi\,(\bar\psi\psi + \bar\chi\chi).
		\label{eq:EL_KG}
	\end{equation}
	In the static regime, introducing the mean-field potential $\Phi\equiv\gphi\,\phi$ and replacing $\langle\bar\psi\psi\rangle$, $\langle\bar\chi\chi\rangle$ by the number densities $n_1$, $n_2$ in the non-relativistic limit, Eq.~\eqref{eq:EL_KG} becomes
	\begin{equation}
		\bigl(\nabla^2-\mphi^2\bigr)\Phi
		=\gphi^2\,(n_1 + n_2)
		+\frac{\kappa}{2\gphi}\Phi^2
		+\frac{\lambda_\phi}{6\gphi^2}\Phi^3.
		\label{eq:KG_full}
	\end{equation}
	The last two terms encode the cubic and quartic self-couplings of the mediator.
	Their fractional importance relative to the linear source scales as $J_\kappa\sim\kappa\gphi n_{\mathrm{tot}}/(2\mphi^4)$ and $J_\lambda\sim\lambda_\phi\gphi^2 n_{\mathrm{tot}}^2/(6\mphi^6)$, respectively, where $n_{\mathrm{tot}}=n_1+n_2$.
	In the weak-field regime ($\Phi\ll\mphi/\gphi$) both ratios are negligible, and Eq.~\eqref{eq:KG_full} reduces to the linear Yukawa form \citep{1997IJMPE...6..515S},
	\begin{equation}
		(\nabla^2-\mphi^2)\Phi=\gphi^2\,(n_1 + n_2).
		\label{eq:KG_linear}
	\end{equation}
	The inverse Compton wavenumber of the mediator,
	\begin{equation}
		\kappaphi \equiv \frac{\mphi\,c}{\hbar}\,,
		\label{eq:kappaphi_def}
	\end{equation}
	sets the Yukawa screening length $\ell_\phi=\kappaphi^{-1}$; in natural units ($\hbar=c=1$) one has simply $\kappaphi=\mphi$ and $\kappaphi^2=\mphi^2$, consistent with Eq.~\eqref{eq:KG_linear}.
	For the gravitational interpretation, the universal Yukawa coupling is identified with the Newtonian strength,
	\begin{equation}
		\gphi^2=4\pi G m_1^2\,,
		\label{eq:gphi_identification}
	\end{equation}
	so that the source in Eq.~\eqref{eq:KG_linear} becomes $4\pi G m_1^2(n_1+n_2)$ for the mean-field potential $\Phi=\gphi\phi$.
	Newtonian gravity, however, couples to mass density rather than to number density; the equivalence principle demands that each species contribute to the gravitational source proportionally to $m_i\,n_i$ \citep{Weinberg1972gc,Will2018teg}.
	This distinction is immaterial for the single-species case but has physical consequences when $m_1\neq m_2$.
	I therefore impose the equivalence principle at the level of the source whilst retaining the Yukawa screening.
	Since the gravitational potential $\Phi_g\equiv\Phi/m_1$ must satisfy $\nabla^2\Phi_g=4\pi G(m_1\,n_1+m_2\,n_2)$ in the Newtonian limit, the mean-field potential $\Phi=\gphi\phi$ obeys the two-species Yukawa--Poisson equation
	\begin{equation}
		(\nabla^2-\kappaphi^2)\,\Phi = \gphi^2\!\left(n_1 + \frac{n_2}{q}\right),
		\label{eq:Yukawa_Poisson}
	\end{equation}
	where $1/q=m_2/m_1$ weights species~2 by the mass ratio, as the equivalence principle demands.
	For a single species ($n_2=0$) the identification $\gphi^2=4\pi G m_1^2$ recovers Eq.~\eqref{eq:KG_linear} identically.
	In the massless-mediator limit $\kappaphi\to 0$, dividing Eq.~\eqref{eq:Yukawa_Poisson} by $m_1$ gives the standard Poisson equation for $\Phi_g$,
	\begin{equation}
		\nabla^2\Phi_g=4\pi G(m_1\,n_1 + m_2\,n_2)\,,
		\label{eq:Poisson_limit}
	\end{equation}
	which is the Newtonian form used in the numerical calculations of Sec.~\ref{sec:numerics}.

	\section{Hydrostatic equilibrium in QHD form}
	\label{sec:QHD}

	In this section I recast the dynamics of each species as a pair of fluid equations, a continuity equation and an Euler equation, within the quantum-hydrodynamic formulation.
	Within the orbital-free density-functional framework for a degenerate fermion fluid, one introduces for each species $i\in\{1,2\}$ an effective Madelung field $\psi_{\mathrm{eff},i}=\sqrt{n_i}\,\mathrm{e}^{iS_i/\hbar}$, a hydrodynamic auxiliary rather than a true single-particle state \citep{1927ZPhy...40..322M,1952PhRv...85..166B,1935ZPhy...96..431W}.
	The densities $n_i=|\psi_{\mathrm{eff},i}|^2$ and the velocity fields $\mathbf{v}_i=\nabla S_i/m_i$ satisfy, for each species,
	\begin{align}
		\frac{\partial n_i}{\partial t} + \nabla\cdot(n_i\mathbf{v}_i) &= 0,
		\label{eq:continuity}\\
		\frac{\partial\mathbf{v}_i}{\partial t} + (\mathbf{v}_i\cdot\nabla)\mathbf{v}_i &= -\frac{1}{m_i n_i}\nabla P_i -\frac{1}{m_i}\nabla\Phi - \frac{1}{m_i}\nabla Q_i,
		\label{eq:euler}
	\end{align}
	where $\Phi$ is the shared gravitational potential satisfying the two-species Yukawa--Poisson equation~\eqref{eq:Yukawa_Poisson}.

	\subsection{The Bohm potential}

	The dispersive stress in Eq.~\eqref{eq:euler} is encoded by the Bohm potential for each species,
	\begin{equation}
		Q_i = -\frac{\lamB\hbar^2}{2m_i}\frac{\nabla^2\sqrt{n_i}}{\sqrt{n_i}},
		\qquad i\in\{1,2\}.
		\label{eq:bohm_potential}
	\end{equation}
	The dimensionless prefactor $\lamB$ is set to $1/9$ throughout this work, the value mandated by the second-order Kirzhnits gradient expansion of the kinetic-energy density functional for a degenerate Fermi gas \citep{2018PhPl...25c2115M,2025PhyU...68..691P,2015CoPP...55..437M}.

	\subsection{Equation of state and pseudo-enthalpy}

	For each species I adopt a polytropic equation of state \citep{1939isss.book.....C,1983bhwd.book.....S},
	\begin{equation}
		P_i(n_i)=K_{p,i}\,n_i^{\gamma_{p,i}},
		\qquad i\in\{1,2\},
		\label{eq:polytrope}
	\end{equation}
	where $K_{p,i}$ is the polytropic constant and $\gamma_{p,i}$ the polytropic index of species $i$; $\gamma_{p,i}=5/3$ reproduces a non-relativistic degenerate Fermi gas, whilst $\gamma_{p,i}=4/3$ corresponds to the ultra-relativistic limit.
	It is convenient to work with the pseudo-enthalpy per particle,
	\begin{equation}
		U_{P,i}(n_i)=\int_0^{n_i}\frac{\mathrm{d}P_i(n_i')}{n_i'}=\frac{K_{p,i}\gamma_{p,i}}{\gamma_{p,i}-1}\,n_i^{\gamma_{p,i}-1}.
		\label{eq:Up_def}
	\end{equation}

	\subsection{Closure relation from the Euler equation}

	Setting $\partial_t\mathbf{v}_i=0$ and $\mathbf{v}_i=0$ in Eq.~\eqref{eq:euler} for a static equilibrium, the left-hand side vanishes identically.
	Noting that $\nabla U_{P,i}=(1/n_i)\nabla P_i$ from Eq.~\eqref{eq:Up_def}, one obtains upon integration
	\begin{equation}
		\mu_i = U_{P,i} + \Phi + Q_i,
		\qquad i\in\{1,2\},
		\label{eq:chemical_potential}
	\end{equation}
	where $\mu_i$ is the chemical potential (integration constant) for each species.
	The gradient of Eq.~\eqref{eq:chemical_potential} gives the global balance condition for species $i$, expressing the vanishing of the net force per unit mass at every radius.
	For the SP ground state $\psi_i$ attains its maximum at $r=0$, whence $\nabla^2\sqrt{n_i}/\sqrt{n_i}<0$ near the centre and $Q_i>0$ there; $Q_i$ decreases outward and becomes negative in the exponential tail.
	The outward-supporting force is $-m_i^{-1}\nabla Q_i$, which is positive wherever $Q_i$ decreases with radius, so the Bohm potential provides global support for each species independently.
	By contrast, in a fermionic Thomas--Fermi configuration the nearly uniform density core gives $\nabla^2\sqrt{n_i}/\sqrt{n_i}\approx 0$ in the bulk, rendering $Q_i$ negligible in the interior; the Bohm term then contributes only as a surface-energy correction \citep{1987PhRvD..35.3678L,2023PhRvD.108d4024D}, a distinction explored quantitatively in Section~\ref{sec:TF_reference}.

	\section{Dimensionless formulation, scaling symmetry and mass--radius relation}
	\label{sec:dimensionless}

	\subsection{Dimensionless Schr\"odinger--Poisson system}

	For spherical symmetry, combining the closure relation~\eqref{eq:chemical_potential}, the Bohm potential~\eqref{eq:bohm_potential}, and the Yukawa--Poisson equation~\eqref{eq:Yukawa_Poisson} (with $n_i=\psi_{\mathrm{eff},i}^2$) gives, for species~1,
	\begin{equation}
		\frac{1}{r^2}\frac{d}{d r}\!\left(r^2\frac{d\psi_1}{d r}\right)
		=
		\frac{2m_1}{\lamB\hbar^2}\bigl(U_{P,1}+\Phi-\mu_1\bigr)\psi_1\,,
		\label{eq:radial_schro1}
	\end{equation}
	and, for species~2 with $m_2=m_1/q$,
	\begin{equation}
		\frac{1}{r^2}\frac{d}{d r}\!\left(r^2\frac{d\psi_2}{d r}\right)
		=
		\frac{2m_1}{q\,\lamB\hbar^2}\bigl(U_{P,2}+\Phi-\mu_2\bigr)\psi_2\,,
		\label{eq:radial_schro2}
	\end{equation}
	together with the Yukawa--Poisson equation
	\begin{equation}
		\frac{1}{r^2}\frac{d}{d r}\!\left(r^2\frac{d\Phi}{d r}\right)
		=
		\gphi^2\!\left(\psi_1^2 + \frac{\psi_2^2}{q}\right) + \kappaphi^2\,\Phi\,.
		\label{eq:radial_poisson}
	\end{equation}
	To expose the parameter dependence and facilitate numerical integration I introduce a characteristic radial scale $r_0$, initially free, and define the dimensionless coordinate $x=r/r_0$.
	All characteristic scales are referred to $m_1$; the mass ratio $q\equiv m_1/m_2$ then enters the equations for species~2 as a dimensionless parameter.
	The fields are normalised to their central values,
	\begin{equation}
		\eta_i(x)\equiv \frac{\psi_i(r)}{\psi_0}=\sqrt{\frac{n_i(r)}{n_0}}\,,\qquad
		\varphi(x)\equiv\frac{\Phi(r)}{\Phi_c}\,,
		\label{eq:dimless_vars}
	\end{equation}
	where $n_0=\psi_0^2$ is the reference central number density and $\Phi_c\equiv\Phi(0)$ the central gravitational potential.
	The characteristic scales, defined via $m_1$, are
	\begin{equation}
		\Phi_0 = \frac{\lamB\hbar^2}{m_1\,r_0^2} = \mu_0\,,\qquad
		\psi_0^2 = \frac{4\pi\lamB\hbar^2}{\gphi^2\,m_1\,r_0^4}\,,
		\label{eq:dimensionless_scales}
	\end{equation}
	and I adopt the normalisation convention $\eta_1(0)=1$.
	The central value of the second species, $f\equiv\eta_2(0)$, is a free input parameter that controls the relative abundance of the two components.
	With these definitions, the radial system~\eqref{eq:radial_schro1}--\eqref{eq:radial_poisson} becomes the coupled dimensionless system of three second-order ODEs (primes denoting $d/dx$):
	\begin{align}
		\eta_1''+\frac{2}{x}\eta_1'
		&=2 v_c\,\eta_1\,\varphi + 2\sigma_1\,\eta_1^{2\gamma_{p,1}-1} - 2\varepsilon_1\,\eta_1\,,
		\label{eq:eta1_eq}\\
		\eta_2''+\frac{2}{x}\eta_2'
		&=\frac{2}{q}\!\left(v_c\,\eta_2\,\varphi + \sigma_2\,\eta_2^{2\gamma_{p,2}-1} - \varepsilon_2\,\eta_2\right),
		\label{eq:eta2_eq}\\
		\varphi''+\frac{2}{x}\varphi'
		&=\frac{4\pi}{v_c}\!\left(\eta_1^2+\frac{\eta_2^2}{q}\right)+\xi^2\,\varphi\,.
		\label{eq:varphi_eq}
	\end{align}
	The factor $2/q$ in Eq.~\eqref{eq:eta2_eq} is inherited directly from the $1/q$ already visible in the dimensional equation~\eqref{eq:radial_schro2}, where $m_2=m_1/q$ enters the kinetic prefactor.
	The corresponding factor in the Poisson source~\eqref{eq:varphi_eq} follows from the equivalence principle imposed in Eq.~\eqref{eq:Yukawa_Poisson}: because the gravitational source is $4\pi G(m_1\,n_1+m_2\,n_2)$, species~2 enters weighted by $m_2/m_1=1/q$ relative to species~1 once all quantities are normalised to the reference mass.
	The dimensionless central potential $v_c\equiv\Phi_c/\Phi_0$ is determined self-consistently by the boundary-value problem.
	The dimensionless pressure parameters for the two species are
	\begin{equation}
		\sigma_i \equiv \frac{m_1\,r_0^2}{\lamB\hbar^2}\,\frac{K_{p,i}\gamma_{p,i}}{\gamma_{p,i}-1}\,\psi_0^{2(\gamma_{p,i}-1)},
		\qquad i\in\{1,2\},
		\label{eq:sigma_def}
	\end{equation}
	where the pure SP limit is recovered for $\sigma_1=\sigma_2=0$.
	In this regime the sole mechanism opposing gravitational collapse is the quantum kinetic (Bohm) pressure encoded in the gradient term $\nabla^{2}\!\sqrt{n_i}/\sqrt{n_i}$, precisely the framework employed in fuzzy dark matter and ultralight boson models.
	Each species has its own dimensionless eigenvalue,
	\begin{equation}
		\varepsilon_i\equiv\mu_i/\mu_0\,,
		\qquad i\in\{1,2\},
		\label{eq:tilde_epsilon_def}
	\end{equation}
	both measured in units of the same reference scale $\mu_0=\Phi_0$.
	The dimensionless mediator-mass parameter is
	\begin{equation}
		\xi \equiv \kappaphi\,r_0 = \frac{\mphi\,c\,r_0}{\hbar}\,,
		\label{eq:xi_def}
	\end{equation}
	equal to the ratio of the radial scale $r_0$ to the screening length $\ell_\phi=\kappaphi^{-1}$; setting $\xi=0$ recovers the massless-mediator limit.
	Regularity at the origin and the central normalisation impose six boundary conditions,
	\begin{equation}
		\begin{aligned}
			\eta_1(0)&=1,\quad \eta_1'(0)=0,\quad
			\eta_2(0)=f,\quad \eta_2'(0)=0,\\[4pt]
			\varphi(0)&=1,\quad \varphi'(0)=0\,,
		\end{aligned}
		\label{eq:BC}
	\end{equation}
	and bound states satisfy $\eta_1(x)\to 0$, $\eta_2(x)\to 0$ and $\varphi(x)\to 0$ as $x\to\infty$.

	\subsection{Scaling symmetry and its breaking by $\mphi$}

	For $\xi=0$ and $\sigma_1=\sigma_2=0$, the Yukawa screening term vanishes and Eqs.~\eqref{eq:eta1_eq}--\eqref{eq:varphi_eq} admit a scaling symmetry: the normalised profiles $\eta_1(x)$, $\eta_2(x)$ and $\varphi(x)$ are shape-invariant, while the physical central values and the radial scale transform as
	\begin{equation}
		\begin{aligned}
			&\psi_c\to\alpha^2\psi_c,\quad \Phi_c\to\alpha^2\Phi_c,\\
			&\mu_i\to\alpha^2\mu_i,\quad r_0\to r_0/\alpha
		\end{aligned}
		\label{eq:scaling_symmetry}
	\end{equation}
	for any $\alpha>0$.
	In terms of the un-normalised potential $v\equiv v_c\varphi=\Phi/\Phi_0$, the scaling reads $v_\alpha(x)=\alpha^2 v(\alpha x)$ and $\varepsilon_{i,\alpha}=\alpha^2\varepsilon_i$.
	Under this rescaling $\xi\to\xi/\alpha$; for $\xi\neq 0$ the Yukawa term breaks the symmetry, and each value of $\xi$ defines a distinct family.
	Since $q$ is itself invariant, the dimensionless profiles remain shape-invariant at fixed $q$ and $f$.
	The normalisation convention $\eta_1(0)=1$ is not unique (one could equally normalise to $\int \eta_1^2 x^2\,dx=1$), but it yields particularly transparent expressions for the physical scales.

	Because $\xi$ itself depends on $r_0$ via Eq.~\eqref{eq:xi_def}, and $r_0$ depends on $M_{\mathrm{dim}}(\xi)$ via Eq.~\eqref{eq:r0_relation}, the dimensionless mediator parameter satisfies the self-consistency condition
	\begin{equation}
		\xi = \Xi\cdot M_{\mathrm{dim}}(\xi),\qquad
		\Xi \equiv \frac{\kappaphi\,\lamB\hbar^2}{G m_1^2 M},
		\label{eq:xi_selfconsistency}
	\end{equation}
	which must be solved iteratively for given $(\mphi,\,m_1,\,q,\,f,\,M)$.
	In the limit $\Xi\to 0$ ($\mphi\to 0$), $\xi\to 0$ and the SP invariants of Eq.~\eqref{eq:invariants} are recovered.

	\begin{table}
	\centering
	\caption{Comparison of dimensionless invariants for the SP ground state with $\eta(0)=1$.
		The eigenvalue $\varepsilon$ depends on conventions, but $M_{\mathrm{dim}}$ and $M_{\mathrm{dim}}\,\xT$ are convention-independent.}
	\label{tab:comparison}
	\begin{tabular}{@{}lccc@{}}
		\toprule
		Source & $M_{\mathrm{dim}}$ & $\xT$ & $M_{\mathrm{dim}}\xT$ \\
		\midrule
		This work & 3.883 & 2.562 & 9.95 \\
		\citet{1998CQGra..15.2733M} & 3.88$^a$ & $\cdots$ & $\cdots$ \\
		\citet{2011PhRvD..84d3531C} & 3.88 & $\cdots$ & 9.9$^b$ \\
		\citet{2004PhRvD..69l4033G} & 3.88 & $\cdots$ & $\cdots$ \\
		\bottomrule
		\multicolumn{4}{@{}l}{\footnotesize $^a$Converted from their normalisation using $\alpha=1$.}\\
		\multicolumn{4}{@{}l}{\footnotesize $^b$Inferred from their mass--radius formula.}
	\end{tabular}
	\end{table}

	\subsection{Mass--radius relation}

	The total dimensionless mass, accounting for both species with their respective gravitational couplings, is
	\begin{equation}
		M_{\mathrm{dim}} = 4\pi\int_0^\infty \!\left(\eta_1^2+\frac{\eta_2^2}{q}\right) x^2\,dx,
		\label{eq:dimensionless_mass}
	\end{equation}
	where the weight $1/q=m_2/m_1$ mirrors the Poisson source weighting of Eq.~\eqref{eq:varphi_eq}.
	The physical mass is
	\begin{equation}
		M = \frac{\lamB\hbar^2}{G m_1^2 r_0}\,M_{\mathrm{dim}}(\xi,q,f)\,.
		\label{eq:mass_scale}
	\end{equation}
	For a chosen total mass $M$ this fixes the radial scale,
	\begin{equation}
		r_0(\xi) = \frac{\lamB\hbar^2}{G m_1^2}\,\frac{M_{\mathrm{dim}}(\xi,q,f)}{M}\,,
		\label{eq:r0_relation}
	\end{equation}
	where $M_{\mathrm{dim}}$ is obtained by solving the coupled system~\eqref{eq:eta1_eq}--\eqref{eq:varphi_eq} at given $(\xi,q,f)$.
	Because the Yukawa screening term $\xi^2\varphi$ weakens the gravitational well, $M_{\mathrm{dim}}$ increases monotonically with $\xi$ and the radial scale inherits this dependence.
	Substituting numerical values and normalising to the single-species massless-mediator invariant $M_{\mathrm{dim}}(0)\simeq 3.883$ (at $f=0$),
	\begin{equation}
		\begin{aligned}
			r_0(\xi)
			\simeq 4.5\,\Rsun
			&\left(\frac{M_{\mathrm{dim}}(\xi,q,f)}{3.883}\right)\\
			&\times\left(\frac{m_1}{6\times 10^{-14}\,\eV}\right)^{\!-2}
			\!\left(\frac{M}{1\,\Msun}\right)^{\!-1}\!,
		\end{aligned}
		\label{eq:r0_numerical}
	\end{equation}
	The physical radius $\RT(\xi)=\xT(\xi)\,r_0(\xi)$ depends on both the Yukawa screening and the two-species parameters $(q,f)$.

	\begin{figure*}
		\centering
		\includegraphics[width=0.45\linewidth]{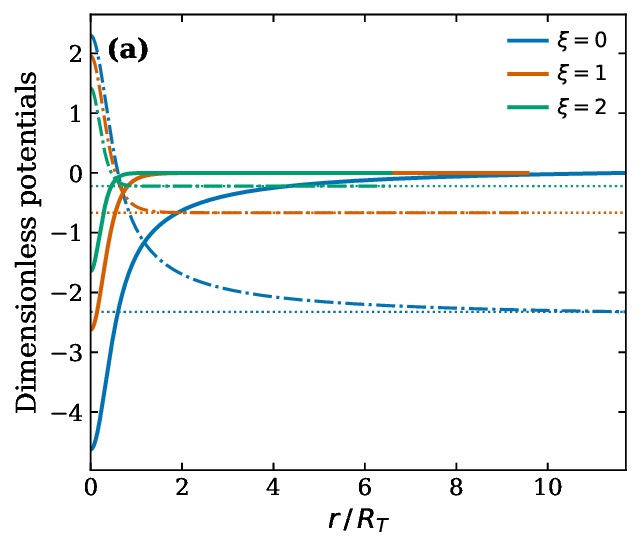}
		\includegraphics[width=0.45\linewidth]{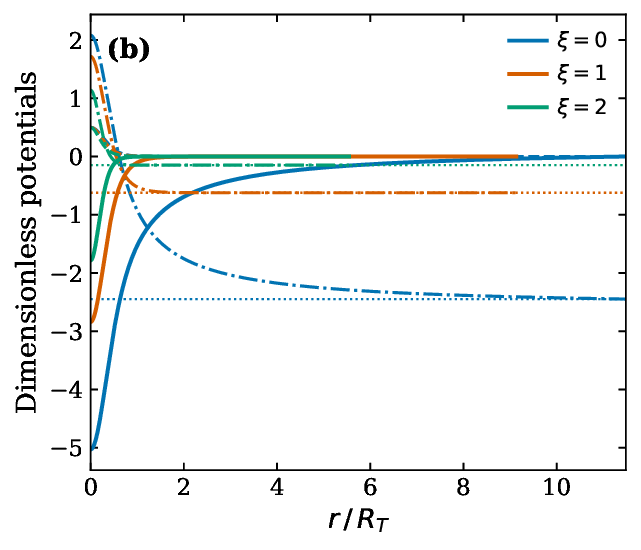}
		\caption{Dimensionless potential balance for the nodeless ground state at $\xi=0$ (blue), $\xi=1$ (vermillion) and $\xi=2$ (green), plotted against the normalised radius $r/\RT$.
			Solid curves show $v_c\varphi$, dot-dashed curves the Bohm diagnostic $Q_{\mathrm{dim}}$ (Eq.~\eqref{eq:Qdim}), and dotted horizontal lines mark $\varepsilon$.
			(a)~Pure SP case ($\sigma=0$).
			(b)~With polytropic pressure ($\sigma=0.5$, $\gamma_p=5/3$); dashed curves show the pseudo-enthalpy contribution $\sigma\,\eta^{2(\gamma_p-1)}$.}
		\label{fig:potential_balance}
	\end{figure*}

	A physically meaningful radius is $\RT$, defined as the sphere enclosing $99\%$ of the total mass.
	Writing $\xT$ for the corresponding dimensionless radius, one obtains the explicit mass--radius relation
	\begin{equation}
		M \RT = \frac{\lamB\hbar^2}{G m_1^2}\Bigl(M_{\mathrm{dim}}(\xi,q,f)\,\xT(\xi,q,f)\Bigr),
		\label{eq:MR_relation}
	\end{equation}
	where $M_{\mathrm{dim}}$ and $\xT$ are determined by the boundary-value problem at the self-consistent $\xi$ satisfying Eq.~\eqref{eq:xi_selfconsistency}.
	For the single-species massless-mediator ground state ($\xi=0$, $f=0$, $\eta_1(0)=1$) I find numerically (Sec.~\ref{sec:numerics})
	\begin{equation}
		\begin{aligned}
			&M_{\mathrm{dim}}(0)\simeq 3.883,\quad \xT(0)\simeq 2.562,\\
			&M_{\mathrm{dim}}(0)\,\xT(0)\simeq 9.95,
		\end{aligned}
		\label{eq:invariants}
	\end{equation}
	so that the bracketed factor in Eq.~\eqref{eq:MR_relation} at $\xi=0$ is fixed to better than $0.1\%$ by a single boundary-value solve.
	For $\xi>0$ the Yukawa screening shifts $M_{\mathrm{dim}}(\xi)$ and $\xT(\xi)$ relative to these reference values, and the $R\propto M^{-1}$ scaling, which holds exactly only at $\xi=0$, acquires an additional $\mphi$-dependence through the self-consistency condition~\eqref{eq:xi_selfconsistency}.
	In the two-species case the invariants additionally depend on $q$ and $f$, so the single mass--radius curve is replaced by a family of curves parameterised by $(q,f)$.

	The $R\propto M^{-1}$ scaling at $\xi=0$, $\sigma_i=0$ coincides with the standard Newtonian boson-star relation \citep{1968PhRv..172.1331K,1969PhRv..187.1767R,1986PhRvL..57.2485C,2011PhRvD..84d3531C,2011PhRvD..84d3532C} and the soliton scaling in wave-dark-matter phenomenology \citep{2000PhRvL..85.1158H,2014NatPh..10..496S,2017PhRvD..95d3541H}; in the fermionic case this limit serves as the purely Bohm-supported reference, whilst the physical regime requires $\sigma_i\neq 0$.

	Table~\ref{tab:comparison} collects the dimensionless mass and the mass--radius product from several independent calculations.
	All three comparison references solve the \emph{bosonic} SP system; however, at $\sigma_i=0$ the coefficient $\lamB$ cancels from the dimensionless equations, rendering the invariants $M_{\mathrm{dim}}$ and $M_{\mathrm{dim}}\,\xT$ independent of particle statistics, so the comparison is entirely appropriate.

	\subsection{Dimensionless energy and escape velocity}
	\label{sec:energy_escape}

	It is useful to define the dimensionless kinetic (quantum-gradient) energy and gravitational potential energy for each species.
	The dimensionless kinetic energy density for species $i$ is
	\begin{equation}
		\mathcal{T}_i(x) = \left(\frac{d\eta_i}{dx}\right)^{\!2},
		\label{eq:kinetic_density}
	\end{equation}
	and the dimensionless gravitational energy density is
	\begin{equation}
		\mathcal{W}_i(x) = v_c\,\eta_i^2(x)\,\varphi(x),
		\label{eq:grav_density}
	\end{equation}
	so that the integrated energies are $T_i = 4\pi\int_0^\infty \mathcal{T}_i\,x^2\,dx$ and $W_i = 4\pi\int_0^\infty \mathcal{W}_i\,x^2\,dx$.
	These satisfy the virial relation $2T + W = 0$ for bound states, up to corrections of order $\sigma_i$ when the polytropic pressure is non-negligible.

	The escape speed from the surface of a configuration of mass $M$ and radius $\RT$ is
	\begin{equation}
		v_{\mathrm{esc}} = \sqrt{\frac{2GM}{\RT}}\,.
		\label{eq:escape_velocity}
	\end{equation}
	The escape speed characterises the binding of the configuration; illustrative values are reported in Section~\ref{sec:physical_scales_results}.

	\section{Numerical method and diagnostic checks}
	\label{sec:numerics}

	The ground state of Eqs.~\eqref{eq:eta1_eq}--\eqref{eq:varphi_eq} is obtained by treating $\varepsilon_1$ and $\varepsilon_2$ as eigenvalues and solving a two-point boundary-value problem on a truncated domain $x\in[0,x_{\max}]$ with $x_{\max}\gg \xT$.
	For each target value of $(\xi,q,f)$, including the single-species reference $f=0$, $\xi=0$, a separate eigenvalue solve is performed; the self-consistent $\xi$ for a given physical configuration is then obtained by iterating Eq.~\eqref{eq:xi_selfconsistency}.
	A broadly similar numerical approach was employed by \citet{2004PhRvD..69l4033G} for the single-species $\xi=0$ system.

\begin{table}
	\centering
	\caption{Dimensionless invariants and physical radial scale for three Yukawa parameters, with $m_1=6\times 10^{-14}\,\eV$ and $M=1\,\Msun$ in the single-species SP reference ($\sigma_i=0$, $f=0$).
		The radial scale $r_0$ follows from Eq.~\eqref{eq:r0_numerical}; the mediator mass is $\mphi=\xi\hbar/(c\,r_0)$.}
	\label{tab:xi_invariants}
	\begin{tabular}{@{}cccccccc@{}}
		\toprule
		$\xi$ & $B_\varphi$ & $\mphi$ [$\eV$] & $M_{\mathrm{dim}}$ & $\xT$ & $M_{\mathrm{dim}}\xT$ & $r_0$ [$\Rsun$] & $\RT$ [$\Rsun$] \\
		\midrule
		0 & 0 & 0                        & 3.883 & 2.562 & 9.95  & 4.5   & 11.6  \\
		1 & 1 & $4.48\times 10^{-17}$    & 5.414 & 3.139 & 17.00 & 6.3   & 19.9  \\
		2 & 4 & $4.69\times 10^{-17}$    & 10.346 & 4.572 & 47.30 & 12.1  & 55.3  \\
		\bottomrule
	\end{tabular}
\end{table}

	\subsection{Implementation}

	For the purposes of numerical integration, the second-order system~\eqref{eq:eta1_eq}--\eqref{eq:varphi_eq} is rewritten in first-order form by introducing the enclosed radial fluxes
	\begin{equation}
		\zeta_i(x)\equiv x^2\frac{{\rm d}\eta_i}{{\rm d}x}\,,\qquad
		\varpi(x)\equiv x^2\frac{{\rm d}\varphi}{{\rm d}x}\,.
		\label{eq:aux_vars}
	\end{equation}
	The six-component first-order system becomes
	\begin{align}
		\frac{{\rm d}\eta_1}{{\rm d}x}&=\frac{\zeta_1}{x^2}\,,&
		\frac{{\rm d}\zeta_1}{{\rm d}x}&=x^2\!\left(2v_c\,\eta_1\varphi+2\sigma_1\,\eta_1^{2\gamma_{p,1}-1}-2\varepsilon_1\,\eta_1\right),
		\label{eq:first_order_eta1}\\
		\frac{{\rm d}\eta_2}{{\rm d}x}&=\frac{\zeta_2}{x^2}\,,&
		\frac{{\rm d}\zeta_2}{{\rm d}x}&=\frac{x^2}{q}\!\left(2v_c\,\eta_2\varphi+2\sigma_2\,\eta_2^{2\gamma_{p,2}-1}-2\varepsilon_2\,\eta_2\right),
		\label{eq:first_order_eta2}\\
		\frac{{\rm d}\varphi}{{\rm d}x}&=\frac{\varpi}{x^2}\,,&
		\frac{{\rm d}\varpi}{{\rm d}x}&=x^2\!\left(\frac{4\pi}{v_c}\bigl(\eta_1^2+\eta_2^2/q\bigr)+\xi^2\,\varphi\right),
		\label{eq:first_order_phi}
	\end{align}
	with coefficients as defined in the preceding section.
	The boundary-value problem is solved using a collocation method with $x_{\max}=30$, $900$ initial mesh points and a relative residual tolerance $\mathrm{tol}=5\times 10^{-7}$.
	Setting $f=0$ and $\xi=0$ reproduces the single-species SP solver.
	Convergence is verified by increasing $x_{\max}$ and the mesh until $M_{\mathrm{dim}}$ and $\xT$ stabilise to the quoted precision.
	The outer boundary condition $\varphi(x_{\max})=0$ approximates $\varphi(\infty)=0$ and fixes the additive constant of the potential; changing $x_{\max}$ shifts $\varepsilon_i$ slightly but leaves $v_c\varphi-\varepsilon_i$ and derived force-balance diagnostics invariant to numerical accuracy.

	The nine boundary conditions that close the system are:
	$\eta_1(0)=1$ (central normalisation),
	$\eta_1'(0)=0$ (regularity),
	$\eta_2(0)=f$ (second-species central value),
	$\eta_2'(0)=0$ (regularity),
	$\varphi(0)=1$ (central normalisation),
	$\varphi'(0)=0$ (regularity),
	$\eta_1(x_{\max})=0$ (bound-state decay),
	$\eta_2(x_{\max})=0$ (bound-state decay),
	$\varphi(x_{\max})=0$ (potential zero at infinity).
	The eigenvalues $\varepsilon_1$, $\varepsilon_2$ and the dimensionless central potential $v_c$ are treated as free parameters adjusted by the solver to satisfy the outer boundary conditions.

	\begin{figure}
		\centering
		\includegraphics[width=\linewidth]{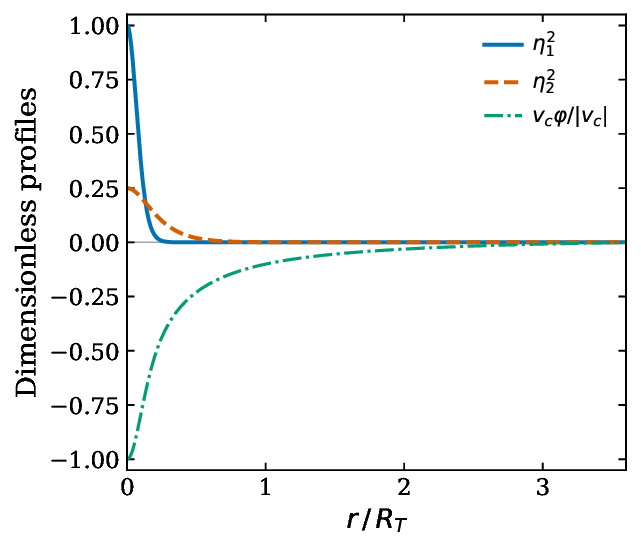}
		\caption{Two-species dimensionless profiles for $q=10$, $f=0.5$ and $\xi=0$ (SP baseline): density profiles $\eta_1^2(x)$ (blue, solid) and $\eta_2^2(x)$ (vermillion, dashed), and the shared gravitational potential $v_c\varphi(x)$ (green, dot-dashed).
			The lighter species ($m_2=m_1/10$) is more spatially extended.}
		\label{fig:two_species_profiles}
	\end{figure}

	\subsection{Internal diagnostics}

	Two internal checks serve as stringent consistency tests for each species.
	First, the ground state is nodeless: $\eta_i(x)>0$ for all $x\in(0,x_{\max})$ and $i\in\{1,2\}$.
	Second, the force balance implied by Eq.~\eqref{eq:chemical_potential} is verified by computing the dimensionless Bohm diagnostic,
	\begin{equation}
		Q_{\mathrm{dim},i}(x)\equiv\varepsilon_i - v_c\,\varphi(x) - \sigma_i\,\eta_i^{2(\gamma_{p,i}-1)}(x)\,,
		\qquad i\in\{1,2\},
		\label{eq:Qdim}
	\end{equation}
	which is the dimensionless form of the closure relation $Q_i=\mu_i-\Phi-U_{P,i}$, Eq.~\eqref{eq:chemical_potential}, divided by $\Phi_0$.
	For the pure SP case ($\sigma_i=0$) this reduces to $Q_{\mathrm{dim},i}=\varepsilon_i-v_c\varphi$.
	The identity $Q_{\mathrm{dim},i}(x)+v_c\,\varphi(x)+\sigma_i\,\eta_i^{2(\gamma_{p,i}-1)}(x)=\varepsilon_i$ holds at all radii up to a numerical error of order $10^{-15}$, confirming that the solver has converged to a self-consistent solution for each species independently.
	A third check, the Bohm potential cross-check, reconstructs $Q_i$ directly from the numerical profile of $\sqrt{n_i}\propto \eta_i$ and compares it with $Q_{\mathrm{dim},i}$; the maximum discrepancy is below $10^{-6}$ on the mesh used.

	\begin{figure}[H]
		\centering
		\includegraphics[width=\linewidth]{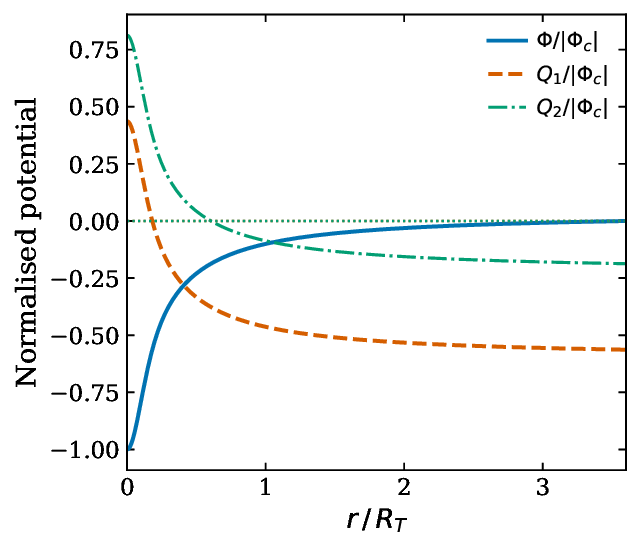}
		\caption{Combined integrated balance for $q=10$, $f=0.5$, $\xi=0$ (SP baseline).
			Blue (solid): shared gravitational potential $\Phi/|\Phi_c|$; vermillion (dashed): Bohm potential $Q_1/|\Phi_c|$ for species~1; green (dash-dotted): Bohm potential $Q_2/|\Phi_c|$ for species~2; dotted curves: closure residuals.
			The Bohm potential of the lighter species exceeds $|\Phi|$ by roughly an order of magnitude.}
		\label{fig:q10_balance}
	\end{figure}

	\section{Results}
	\label{sec:results}

	The results are organised in order of increasing physical realism, each layer of complexity being introduced against a well-understood baseline.
	Section~\ref{sec:SP_baseline} establishes the SP ($\sigma_i=0$) reference; Section~\ref{sec:TF_reference} presents the Thomas--Fermi model, in which degeneracy pressure provides bulk support and the Bohm potential contributes a quantum surface correction; Sections~\ref{sec:physical_scales_results} and~\ref{sec:polytropic_mr} translate to physical units and summarise the polytropic mass--radius relation together with parameter-sensitivity tests.

	\begin{figure*}
		\centering
		\includegraphics[width=0.45\linewidth]{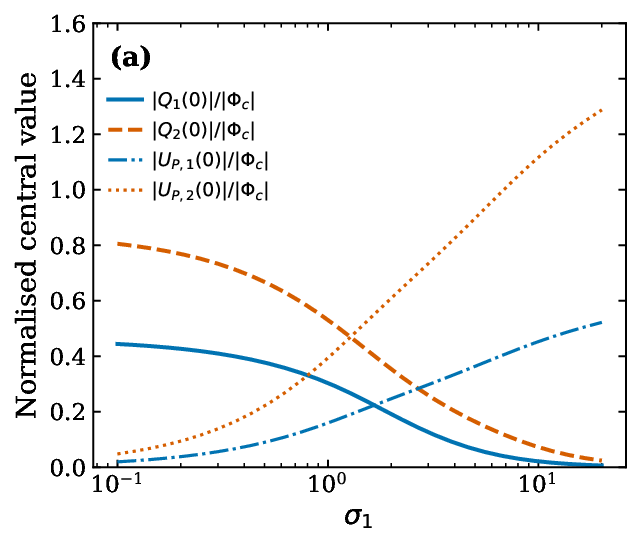}
		\includegraphics[width=0.45\linewidth]{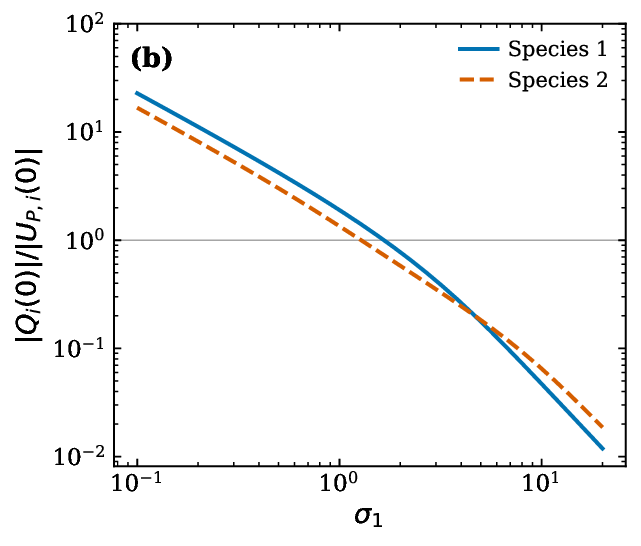}
		\caption{Bohm--Thomas-Fermi crossover for $q=10$, $f=0.35$, $\xi=0$.
			(a)~Normalised central Bohm potential $|Q_i(0)|/|\Phi_c|$ as a function of $\sigma_1$; thick curves: physical scaling $\sigma_2=q\,\sigma_1$ (Eq.~\ref{eq:sigma_ratio}), thin faded curves: equal-$\sigma$ reference.
			(b)~Ratio $|Q_i(0)|/|U_{P,i}(0)|$ for the physical scaling.
			The horizontal grey line marks $|Q_i|=|U_{P,i}|$: above this line, the Bohm potential dominates over degeneracy pressure, below it, degeneracy pressure prevails.
			Species~1 (blue solid line) crosses over near $\sigma_1\simeq 2$, identifying the transition from the Bohm-supported to the Thomas--Fermi regime.}
		\label{fig:q10_crossover}
	\end{figure*}

	\subsection{Schr\"odinger--Poisson baseline ($\sigma_i=0$)}
	\label{sec:SP_baseline}

	In the formal limit $\sigma_i=0$ (vanishing degeneracy pressure), the Bohm term is the sole source of repulsive stress, a regime that is not physically representative of degenerate fermions but that furnishes an indispensable numerical reference.
	The dimensionless potential balance for the single-species ground state (Fig.~\ref{fig:potential_balance}) confirms that the closure relation~\eqref{eq:chemical_potential} holds at every radius; increasing $\xi$ weakens the gravitational confinement, as one would expect.
	At $\xi=0$ the dimensionless invariants are $M_{\mathrm{dim}}(0)\simeq 3.883$, $\xT(0)\simeq 2.562$ and $M_{\mathrm{dim}}\xT\simeq 9.95$, in excellent agreement with independent determinations (Table~\ref{tab:comparison}), a concordance that is rather pleasing and confirms that the numerical procedure is well controlled.
	The enclosed-mass profile is universal at $\xi=0$ owing to the scaling symmetry~\eqref{eq:scaling_symmetry}, with half the mass enclosed within $r\simeq 0.45\,\RT$; for $\xi>0$ the profile broadens as the confining potential is softened.

	In the two-species SP baseline ($q=10$, $f=0.5$, $\sigma_i=0$, $\xi=0$) the heavier species ($\eta_1$) is centrally concentrated whilst the lighter species ($\eta_2$) extends to larger radii, a direct consequence of the $1/q$ factor in Eq.~\eqref{eq:eta2_eq} (Fig.~\ref{fig:two_species_profiles}).
	The solver yields $\varepsilon_1\simeq -3.035$, $\varepsilon_2\simeq -1.008$, $v_c\simeq -5.383$, $M_{\mathrm{dim}}\simeq 6.232$ and $\xT\simeq 8.330$, with species~2 contributing $M_{2,\mathrm{grav}}/M\simeq 40.4\%$ of the gravitational mass (after $1/q$ weighting).
	The Bohm potential provides global support for both species, $|Q_1|/|\Phi|\sim\mathcal{O}(1)$ and $|Q_2|/|\Phi|\sim q$ throughout the interior (Fig.~\ref{fig:q10_balance}).

	\subsection{Reference model: Thomas--Fermi regime with quantum surface correction}
	\label{sec:TF_reference}

	I now turn to the physically more representative Thomas--Fermi regime, where degeneracy pressure furnishes the dominant bulk support.
	The reference model adopts $q=10$, $f=0.35$, $\sigma_1=12$, $\sigma_2=q\,\sigma_1=120$, $\gamma_{p,i}=5/3$ and $\xi=0$.

	\subsubsection{Bohm--Thomas-Fermi crossover}

	It is worth remarking that $\sigma_1$ and $\sigma_2$ are not independent parameters: for a non-relativistic degenerate Fermi gas ($\gamma_{p,i}=5/3$) the polytropic constant scales as $K_{p,i}\propto\hbar^2/m_i$, whence
	\begin{equation}
		\frac{\sigma_2}{\sigma_1}=\frac{K_{p,2}}{K_{p,1}}=\frac{m_1}{m_2}=q\,.
		\label{eq:sigma_ratio}
	\end{equation}
	Figure~\ref{fig:q10_crossover} quantifies this crossover.
	As $\sigma_1$ increases from zero, $|Q_1(0)|/|\Phi_c|$ drops from $\simeq 0.46$ to $\simeq 0.016$ at $\sigma_1=12$, whilst $|U_{P,1}(0)|/|\Phi_c|$ rises from zero to $\simeq 0.47$; species~1 crosses over near $\sigma_1\simeq 2$.
	For species~2 the physical scaling $\sigma_2=q\,\sigma_1$ implies that its degeneracy pressure is $q$ times stronger at the same $\sigma_1$; the competition between the enhanced kinetic coefficient $\hbar^2/(2m_2)=q\,\hbar^2/(2m_1)$ and the stronger degeneracy pressure determines the crossover locus for each species independently.

	\subsubsection{The Bohm force and the liquid-drop analogy}

	\begin{figure}
		\centering
		\includegraphics[width=\linewidth]{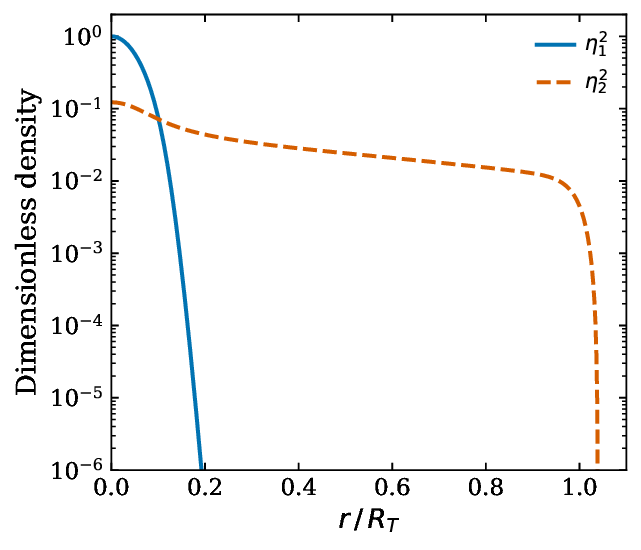}
		\caption{Density profiles $\eta_i^2(x)$ for the reference Thomas--Fermi model ($q=10$, $f=0.35$, $\sigma_1=12$, $\sigma_2=120$, $\xi=0$).
			Species~1 (blue, solid) and species~2 (vermillion, dashed).}
		\label{fig:TF_reference}
	\end{figure}

	The Bohm potential $Q_i$ (Eq.~\ref{eq:bohm_potential}) follows the standard Madelung--Bohm convention \citep{1927ZPhy...40..322M,1952PhRv...85..166B} with the Kirzhnits coefficient $\lamB=1/9$ \citep{2025PhyU...68..691P,2015CoPP...55..437M,2015PhPl...22d4501M,2018CoPP...58..290M}.
	The physically observable quantity is the Bohm force per unit mass,
	\begin{equation}
		\mathbf{F}_{B,i} = -\frac{1}{m_i}\nabla Q_i\,,
		\label{eq:bohm_force}
	\end{equation}
	which encodes the dispersive stress arising from spatial inhomogeneity in the density field \citep{Wyatt:2005uc,2001PhRvB..64g5316M}.
	Within the quantum-hydrodynamic framework the Bohm force plays a role analogous to that of the quantum-pressure tensor in the self-consistent fluid model of \citet{2001PhRvB..64g5316M}, and its regularising properties at steep density gradients have been analysed in the smooth quantum-potential formulation of \citet{1996PhRvE..53..157G}.
	For spherical equilibria one has $F_{B,i}=-m_i^{-1}\,dQ_i/dr$; outward support requires $dQ_i/dr<0$, whilst $dQ_i/dr>0$ produces an inward restoring force.

	\begin{figure}
		\centering
		\includegraphics[width=\linewidth]{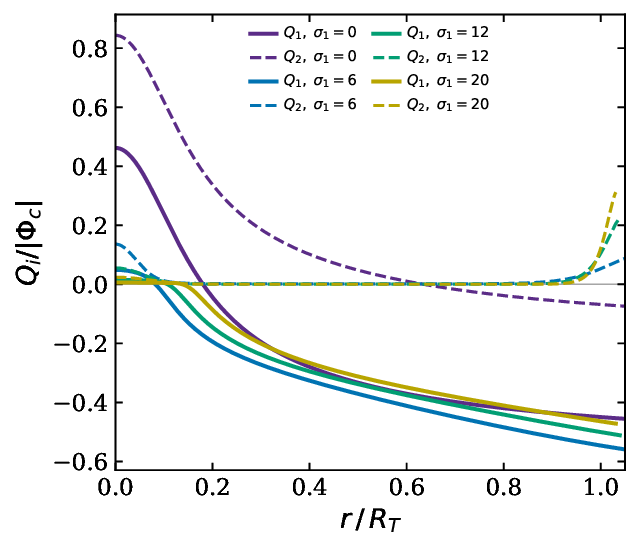}
		\caption{Signed Bohm potential $Q_i(r)/|\Phi_c|$ for the reference Thomas--Fermi model ($q=10$, $f=0.35$, $\sigma_1=12$, $\sigma_2=120$, $\xi=0$), showing the surface localisation characteristic of the Thomas--Fermi regime.}
		\label{fig:TF_Bohm_signed}
	\end{figure}

\begin{figure}
		\centering
		\includegraphics[width=\linewidth]{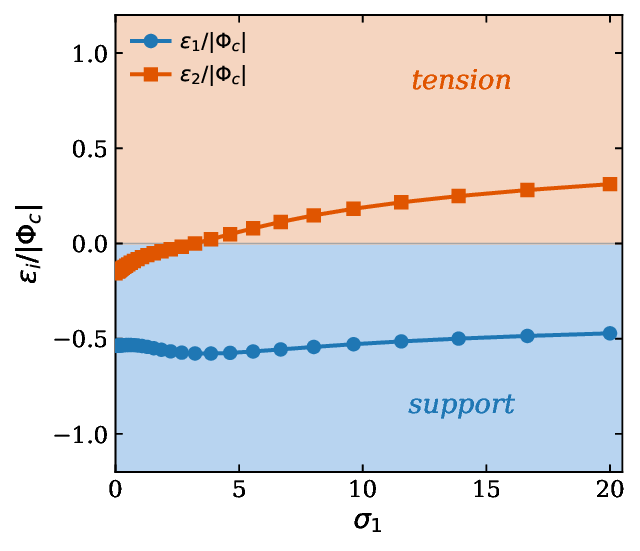}
		\caption{Eigenvalues $\varepsilon_i/|\Phi_c|$ as a function of $\sigma_1$ ($q=10$, $f=0.35$, $\sigma_2=q\,\sigma_1$).
			The eigenvalue $\varepsilon_2$ crosses zero near $\sigma_1\simeq 3$, marking the onset of the surface-tension regime:
			in the blue-shaded region ($\varepsilon<0$), the Bohm force acts outward; in the vermillion-shaded region ($\varepsilon>0$), it acts inward as surface tension.}
		\label{fig:TF_eigenvalues}
	\end{figure}

	In the asymptotic region the closure relation reduces to $Q_i(r\to\infty)\to\varepsilon_i$ (cf.\ Eq.~\ref{eq:chemical_potential}), so the eigenvalue sign determines the direction of the Bohm force near the stellar surface.
	For species~1, whose eigenvalue remains negative at all $\sigma_1$, the Bohm force $F_{B,1}>0$ is directed outward, establishing a quantum-pressure wall that defines the stellar boundary; for species~2, whose eigenvalue crosses zero near $\sigma_1\simeq 3$ (Fig.~\ref{fig:TF_eigenvalues}), the force $F_{B,2}<0$ is directed inward when $\varepsilon_2>0$, acting as quantum surface tension whilst degeneracy pressure $U_{P,2}$ furnishes the bulk confinement.

		\begin{figure}
		\centering
		\includegraphics[width=\linewidth]{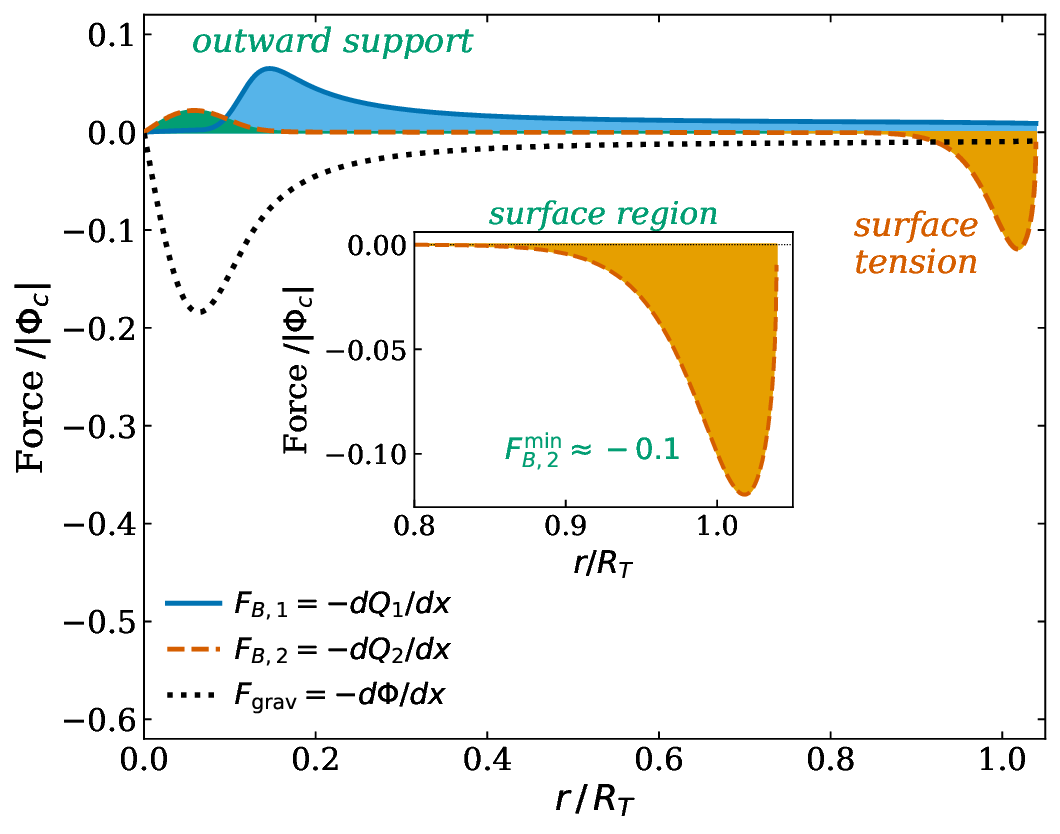}
		\caption{Bohm force $F_{B,i}=-dQ_i/dx$ normalised to $|\Phi_c|$, for $\sigma_1=12$ ($q=10$, $f=0.35$, $\sigma_2=120$).
			Species~1 (solid blue) exerts a purely outward force throughout the interior.
			Species~2 (dashed orange) exhibits two regimes: modest outward support in the bulk (green shading) and a sharp inward surface-tension dip near the stellar surface (orange shading), whose magnitude exceeds the local gravitational force.
			The inset magnifies the surface region to reveal the full depth of this dip, which the main panel clips.
			The gravitational force $F_{\mathrm{grav}}=-d\Phi/dx$ (dotted) is included for comparison.}
		\label{fig:TF_Bohm_force}
	\end{figure}

	This species-dependent directionality of $\mathbf{F}_{B,i}$ is the hallmark of the two-component model, and the physical picture is best illuminated by the nuclear liquid-drop decomposition \citep{1987PhRvD..35.3678L,2020EPJP..135..290C}: degeneracy pressure furnishes the volume binding energy, whilst the Bohm force provides a surface-energy correction whose sign differs between the two species.
	For the heavier species, whose degeneracy pressure is weaker, $\mathbf{F}_{B,1}$ retains its outward-supporting role across the entire profile.
	For the lighter, deeply degenerate species the Bohm force transitions to an inward surface-tension regime, reflecting the sign change of $\varepsilon_2$ as the Thomas--Fermi parameter increases.
	The competition between volume and surface terms governs the equilibrium shape and stability of the configuration \citep{2023PhRvD.108d4024D}, much as it does for nuclear matter.

	\subsubsection{Full diagnostic of the reference model}

\begin{figure*}
		\centering
		\includegraphics[width=0.45\linewidth]{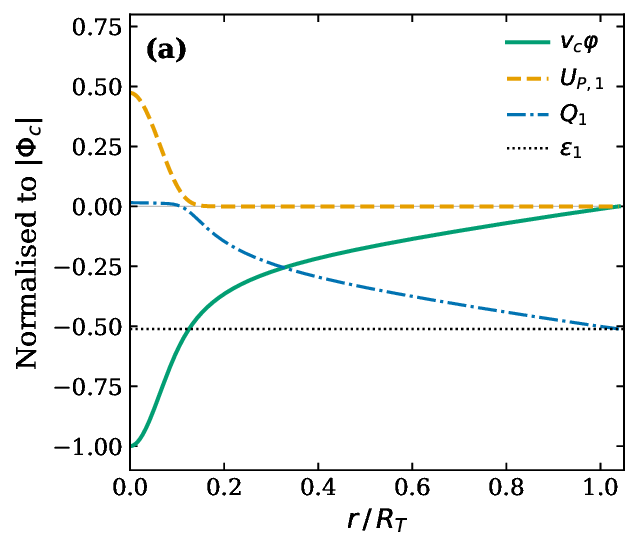}
		\includegraphics[width=0.45\linewidth]{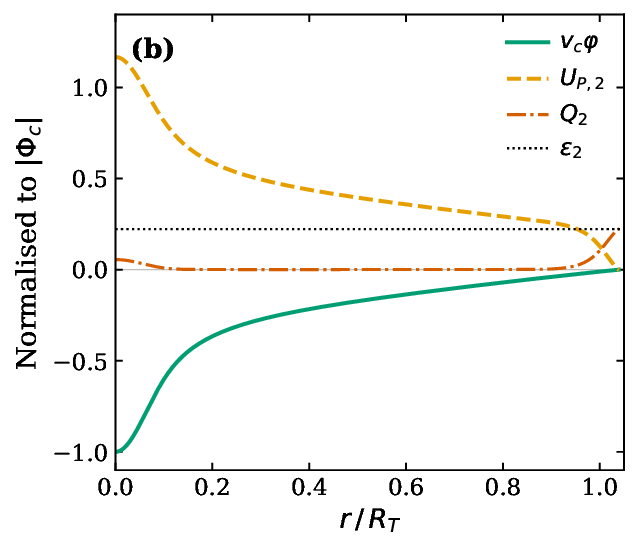}
		\caption{Closure balance $\varepsilon_i = U_{P,i} + v_c\varphi + Q_i$ at $\sigma_1=12$ ($q=10$, $f=0.35$, $\sigma_2=120$):
			gravity (blue, solid), degeneracy pressure (green, dot-dashed) and Bohm potential (vermillion, dashed).
			(a)~Species~1 ($m_1$).
			(b)~Species~2 ($m_2=m_1/q$).}
		\label{fig:TF_closure}
	\end{figure*}

	Figures~\ref{fig:TF_reference}--\ref{fig:TF_closure} present the complete diagnostic for $q=10$, $f=0.35$, $\sigma_2=q\,\sigma_1$.
	At $\sigma_1=12$ ($\sigma_2=120$) the solver yields $\varepsilon_1\simeq -13.0$, $\varepsilon_2\simeq 5.7$, $v_c\simeq -25.4$, $M_{\mathrm{dim}}\simeq 205.0$ and $\xT\simeq 28.9$.
	The central Bohm potentials are small, $|Q_1(0)|/|\Phi_c|\simeq 0.016$ and $|Q_2(0)|/|\Phi_c|\simeq 0.056$, confirming that the configuration resides firmly in the Thomas--Fermi regime.
	Both density profiles (Fig.~\ref{fig:TF_reference}) decrease monotonically from their central values, as befits the nodeless ground state; species~1 is more centrally concentrated whilst species~2, being lighter, extends to larger radii.
	The signed $Q_i(r)/|\Phi_c|$ (Fig.~\ref{fig:TF_Bohm_signed}) exhibits progressive localisation to a thin shell around $r\simeq\RT$, with $Q_1$ becoming negative (outward force) and $Q_2$ becoming positive (inward surface tension).
	The Bohm force $F_{B,i}=-dQ_i/dx$ (Fig.~\ref{fig:TF_Bohm_force}) confirms that $F_{B,1}>0$ throughout the interior whilst $F_{B,2}<0$ near the surface.
	The eigenvalues (Fig.~\ref{fig:TF_eigenvalues}) show $\varepsilon_1$ remaining negative whilst $\varepsilon_2$ crosses zero near $\sigma_1\simeq 3$, marking the onset of the surface-tension regime.
	The closure relation $\varepsilon_i = U_{P,i} + v_c\varphi + Q_i$ (Fig.~\ref{fig:TF_closure}) reveals that the interior balance is between degeneracy pressure and gravity, with $Q_i$ negligible; near the surface, $Q_1\to\varepsilon_1<0$ provides the quantum wall and $Q_2\to\varepsilon_2>0$ acts as surface tension.
	This tripartite balance represents the regime most relevant to astrophysical dark-fermion stars.

	\subsection{Physical scales and illustrative objects}
	\label{sec:physical_scales_results}

	I now translate the dimensionless solutions to physical units using Eqs.~\eqref{eq:dimensionless_scales}--\eqref{eq:r0_numerical}.
	Table~\ref{tab:xi_invariants} collects the dimensionless invariants and radial scales for three Yukawa parameters at $m_1=6\times 10^{-14}\,\eV$ and $M=1\,\Msun$ in the single-species SP reference ($\sigma_i=0$, $f=0$); at $\xi=0$ one obtains $r_0\simeq 4.5\,\Rsun$ via Eq.~\eqref{eq:r0_numerical}.

	To illustrate the breadth of the parameter space, Table~\ref{tab:examples} presents physical configurations at three representative total masses: $M=10^{-8}\,\Msun$ (sub-solar), $M=1\,\Msun$ (solar) and $M=5\,\Msun$ (supra-solar).
	All entries satisfy $\kappa<10^{-3}$, confirming that the Newtonian approximation is well justified.
	The particle mass $m_1$ is chosen so that $\RT$ spans from $\sim 10\,\mathrm{km}$ to $\sim 10^2\,\Rsun$, corresponding to $m_1\sim 10^{-14}$--$10^{-7}\,\eV$, an enormous range that reflects the $\RT\propto m_1^{-2}$ scaling at fixed $M$.
	The lower panel of Table~\ref{tab:examples} additionally presents two-species configurations in which neither $\sigma_i$ nor $\xi$ vanish, demonstrating the richer parameter space accessible when degeneracy pressure and a finite-mass mediator both contribute.

	The central densities span thirteen orders of magnitude and escape speeds range from $\sim 16$ to $\sim 1.2\times 10^{4}\,\mathrm{km\,s^{-1}}$.
	The steep central-density scaling, $\rho_c\propto m_1^6$ at fixed $M$, follows from $r_0\propto m_1^{-2}$ combined with $\rho_c\propto 1/(m_1\,r_0^4)$.

	The two-species entries with $\sigma_1=1$, $\sigma_2=10$ (lower panel) illustrate how degeneracy pressure inflates the equilibrium radius by roughly an order of magnitude relative to the pure Bohm reference at the same $(m_1,M)$, reflecting the larger $M_{\mathrm{dim}}$ of the pressure-supported solution.
	Entries with a finite-mass mediator ($\xi=1$) lie intermediate between the pure Bohm and high-pressure extremes.

	\subsection{Polytropic scaling, parameter sensitivity and limiting cases}
	\label{sec:polytropic_mr}

	In the Thomas--Fermi regime ($\sigma_i\gg 1$), the bulk equilibrium reduces to the Lane--Emden equation: for $\gamma = 5/3$ the radius scales as $R \propto M^{-1/3}$ with no upper mass bound, whereas for $\gamma = 4/3$ a limiting mass emerges above which no stable equilibrium exists.
	These two limiting cases delimit the non-relativistic and ultra-relativistic regimes, respectively, and are well established in the polytropic literature (see, e.g., \citealt{1939isss.book.....C}); the corresponding mass--radius curves are therefore not reproduced here.

	The two-species mass--radius relation recovers the universal single-species scaling $M\RT=\mathrm{const}/m_1^2$ for $f=0$; for $f>0$ the displacement grows with $q$, owing to the broader spatial envelope of the lighter species.
	Yukawa screening ($\xi>0$) shifts the mass--radius curves to larger radii, the shift increasing with $\xi$ through the self-consistency condition~\eqref{eq:xi_selfconsistency}.
	These dependences follow directly from the invariants tabulated in Table~\ref{tab:xi_invariants}; the corresponding mass--radius curves, being single-parameter rescalings of the universal profile, are not reproduced here.


	\subsection{Scope and limitations of the present treatment}

	The Thomas--Fermi reference model ($\sigma_1=12$, $\sigma_2=120$, $q=10$, $f=0.35$) is parameterised in the single-species limit by the triple $(\mphi, m_1, M)$ (with $\lamB$ fixed) and in the two-component case by the quintuple $(\mphi, m_1, m_2, M, f)$.
	Equations~\eqref{eq:MR_relation}--\eqref{eq:xi_selfconsistency} provide a compact mapping from these parameters to a characteristic radius; the massless-mediator limit reproduces the SP invariants of Eq.~\eqref{eq:invariants}.
	The existence of such equilibria does not, of course, establish that Nature produces them in abundance; at least three neglected ingredients appear to matter in realistic settings: self-interactions beyond the polytropic approximation \citep{1986PhRvL..57.2485C,2011PhRvD..84d3531C}, finite temperature or incoherent phase-space structure, and external tidal fields capable of stripping weakly bound configurations \citep{2017PhRvD..95d3541H}.
	The solar-mass configurations have escape speeds of $60$--$604\,\mathrm{km\,s^{-1}}$, confirming that they are only modestly bound; tidal-survival estimates would be required for any astrophysical application.

	\begin{table*}
		\centering
		\caption{Illustrative dark-fermion star configurations spanning three total-mass regimes.
			The gravitational compactness is $\kappa\equiv GM/(\RT c^2)$; the escape speed is $v_{\mathrm{esc}}=\sqrt{2GM/\RT}$.
			All entries satisfy $\kappa < 10^{-3}$, confirming the validity of the Newtonian approximation.
			The upper panel collects the pure Bohm reference ($\sigma_i=0$, $\xi=0$); the lower panel presents two-species configurations with non-zero degeneracy pressure and mediator coupling.
			The ``Fig.'' column shows the marker (shape and colour) used to identify each configuration in Fig.~\ref{fig:gw_predictions}, whose caption defines the colour coding by family.}
		\label{tab:examples}
		\begin{tabular}{@{}cllccccc@{}}
			\toprule
			Fig. & $M$ [$\Msun$] & $m_1$ [$\eV$] & $\RT$ & $\rho_c$ [g\,cm$^{-3}$] & $\bar{\rho}$ [g\,cm$^{-3}$] & $v_{\mathrm{esc}}$ [km\,s$^{-1}$] & $\kappa$ \\
			\midrule
			\multicolumn{8}{c}{\textit{Pure Bohm reference ($\sigma_i=0$, $\xi=0$)}} \\[2pt]
			\multicolumn{8}{c}{\textit{Sub-solar mass}} \\[2pt]
			\mkTriU{mkOrange} & $10^{-8}$ & $5.4\times 10^{-7}$  & $10\,\mathrm{km}$    & $8.6\times 10^{7}$  & $4.8\times 10^{6}$  & 16   & $1.5\times 10^{-9}$ \\[4pt]
			\multicolumn{8}{c}{\textit{Solar mass}} \\[2pt]
			\mkCirc{mkBlue} & $1$       & $2\times 10^{-13}$   & $1.0\,\Rsun$          & $2.2\times 10^{1}$  & $1.2$               & 604  & $2.1\times 10^{-6}$ \\
			\mkSqr{mkBlue} & $1$       & $6\times 10^{-14}$   & $11.6\,\Rsun$         & $1.6\times 10^{-2}$ & $9.0\times 10^{-4}$ & 181  & $1.8\times 10^{-7}$ \\
			\mkDia{mkBlue} & $1$       & $2\times 10^{-14}$   & $105\,\Rsun$          & $2.2\times 10^{-5}$ & $1.2\times 10^{-6}$ & 60   & $2.0\times 10^{-8}$ \\[4pt]
			\multicolumn{8}{c}{\textit{Supra-solar mass}} \\[2pt]
			\mkPent{mkRed} & $5$       & $8\times 10^{-13}$   & $9\,108\,\mathrm{km}$   & $5.7\times 10^{7}$  & $3.1\times 10^{6}$  & $12\,071$ & $8.1\times 10^{-4}$ \\
			\midrule
			\multicolumn{8}{c}{\textit{Two-species with pressure ($\sigma_1=1$, $\sigma_2=10$, $q=10$, $f=0.5$, $\xi=0$)}} \\[2pt]
			\mkDia{mkGreen} & $10^{-4}$ & $4.0\times 10^{-8}$  & $3.2\,\mathrm{km}$    & $2.4\times 10^{16}$ & $1.4\times 10^{12}$ & $2\,880$ & $4.6\times 10^{-5}$ \\
			\mkCirc{mkGreen} & $10^{-2}$ & $4.0\times 10^{-11}$ & $3.2\times 10^{4}\,\mathrm{km}$ & $2.4\times 10^{6}$ & $1.4\times 10^{2}$  & 288  & $4.6\times 10^{-7}$ \\
			\mkPent{mkGreen} & $10^{-1}$ & $4.0\times 10^{-11}$ & $3\,244\,\mathrm{km}$ & $2.3\times 10^{10}$  & $1.4\times 10^{6}$  & $2\,860$ & $4.6\times 10^{-5}$ \\
			\mkStar{mkGreen} & $1$       & $3.2\times 10^{-14}$   & $726\,\Rsun$          & $6.1\times 10^{-5}$ & $3.6\times 10^{-9}$ & 23   & $2.9\times 10^{-9}$ \\
			\mkTriD{mkGreen} & $5$       & $2.6\times 10^{-12}$   & $1.5\times 10^{4}\,\mathrm{km}$ & $1.2\times 10^{10}$  & $6.3\times 10^{5}$ & $9\,406$ & $4.9\times 10^{-4}$ \\[4pt]
			\multicolumn{8}{c}{\textit{Two-species with mediator ($\sigma_1=0.5$, $\sigma_2=5$, $q=10$, $f=0.5$, $\xi=1$)}} \\[2pt]
			\mkPent{mkPurple} & $10^{-5}$ & $1.4\times 10^{-8}$  & $695\,\mathrm{km}$    & $8.7\times 10^{7}$  & $1.4\times 10^{4}$  & 62   & $2.1\times 10^{-8}$ \\
			\mkSqr{mkPurple} & $10^{-3}$ & $6.6\times 10^{-9}$  & $32\,\mathrm{km}$     & $8.9\times 10^{13}$ & $1.4\times 10^{10}$ & $2\,880$ & $4.6\times 10^{-5}$ \\
			\mkDia{mkPurple} & $10^{-2}$ & $7.2\times 10^{-10}$  & $269\,\mathrm{km}$    & $1.5\times 10^{12}$ & $2.4\times 10^{8}$  & $3\,141$ & $5.5\times 10^{-5}$ \\
			\mkHex{mkPurple} & $1$       & $5.4\times 10^{-14}$   & $673\,\Rsun$          & $2.9\times 10^{-5}$ & $4.6\times 10^{-9}$ & 24   & $3.2\times 10^{-9}$ \\
			\mkTriU{mkPurple} & $10^{-8}$ & $2.8\times 10^{-6}$    & $18\,\mathrm{km}$    & $5.0\times 10^{9}$  & $8.1\times 10^{5}$  & 12   & $8.2\times 10^{-10}$ \\
			\bottomrule
		\end{tabular}
	\end{table*}

	\section{Discussion and conclusions}
	\label{sec:conclusions}

	\subsection{Dynamical stability of the ground state}

	The nodeless ground state of the SP system is linearly stable under radial perturbations, as demonstrated both by direct numerical evolution \citep{1991PhRvL..66.1659S,1994PhRvL..72.2516S,2004PhRvD..69l4033G} and by the absence of growing modes in the linearised perturbation spectrum.
	Excited states with one or more nodes are dynamically unstable and decay to the ground state via gravitational cooling on a dynamical timescale \citep{1994PhRvL..72.2516S}.
	This result is reassuring: once formed, the ground-state configurations persist indefinitely in the absence of external perturbations.
	For the fermionic case with $\sigma\neq 0$ the perturbation spectrum is modified by degeneracy pressure and stability must be reassessed; the stability of self-interacting generalisations has a richer structure, depending on the sign and magnitude of the coupling \citep{2011PhRvD..84d3531C}.

	\subsection{Gravitational-wave signatures}
	\label{sec:gw_signatures}

	Were dark-fermion stars to exist in astrophysical environments, their inspiral and merger in binary systems would generate gravitational-wave emission potentially detectable by next-generation observatories.
	For an equal-mass binary of total mass $M_{\mathrm{tot}}$ at contact separation $a = 2\RT$, the gravitational-wave frequency at contact is \citep{Maggiore2007gw}
	\begin{equation}
		f_{\mathrm{GW}} = \frac{1}{\pi}\sqrt{\frac{G M_{\mathrm{tot}}}{(2\RT)^3}}\,,
		\label{eq:fGW_contact}
	\end{equation}
	and the characteristic strain at luminosity distance $d_L$ is
	\begin{equation}
		h_c = \frac{4G}{c^4}\,\frac{\mathcal{M}_c^{5/3}\,(\pi f_{\mathrm{GW}})^{2/3}}{d_L}\,,
		\label{eq:hc_strain}
	\end{equation}
	where $\mathcal{M}_c = M_{\mathrm{tot}}/2^{6/5}$ is the chirp mass for equal-mass components.

\begin{figure}[H]
		\centering
		\includegraphics[width=\linewidth]{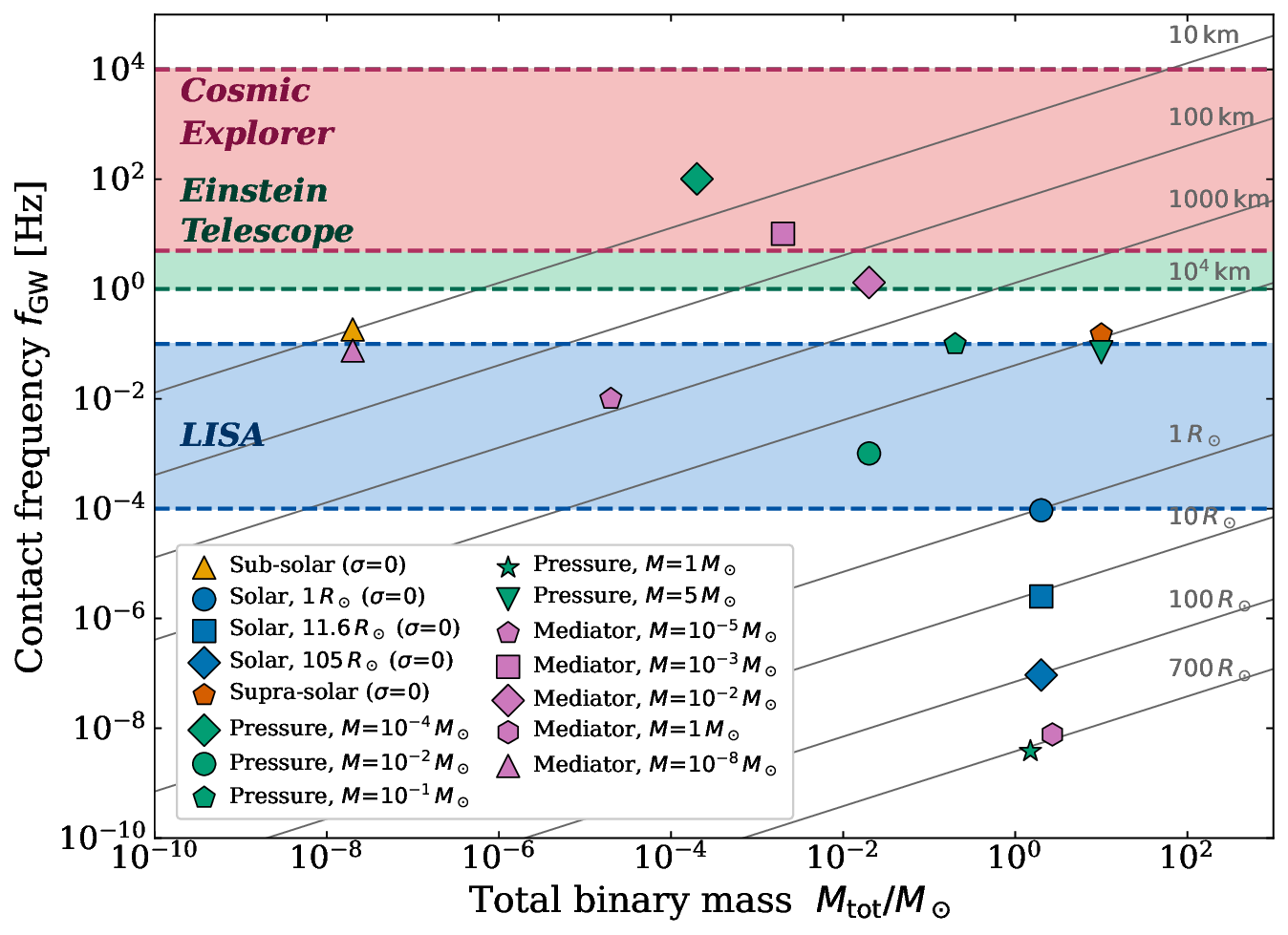}
		\caption{Gravitational-wave contact frequency $f_{\mathrm{GW}}$ (Eq.~\eqref{eq:fGW_contact}) as a function of total binary mass for equal-mass dark-fermion star binaries.
			Diagonal grey lines indicate loci of constant stellar radius (labelled).
			Shaded horizontal bands, bounded by dashed lines, mark the approximate design-sensitivity windows of three planned or proposed detectors: LISA ($10^{-4}$--$10^{-1}\,\mathrm{Hz}$; blue; \citealt{2017arXiv170200786A}), the Einstein Telescope (ET, $1$--$10^{4}\,\mathrm{Hz}$; green; \citealt{2010CQGra..27s4002P}), and Cosmic Explorer (CE, $5$--$10^{4}\,\mathrm{Hz}$; pink; \citealt{2019BAAS...51g..35R}).
			Markers show all fifteen illustrative configurations of Table~\ref{tab:examples}, colour-coded by family: the pure Bohm solutions ($\sigma_i=0$, $\xi=0$) in orange (sub-solar mass), blue (solar mass) and red (supra-solar mass); the two-species models with thermal pressure ($\sigma_1=1$, $\sigma_2=10$) in green; and the two-species models with mediator coupling ($\xi=1$) in purple. The marker shape identifies the individual configuration and reproduces the symbol shown in the ``Fig.'' column of Table~\ref{tab:examples}.
			Compact configurations ($\RT\lesssim 10^{4}\,\mathrm{km}$) fall within or above the LISA band, whilst the most extended solar-mass objects ($\RT\gtrsim 600\,\Rsun$) lie far below it.}
		\label{fig:gw_predictions}
	\end{figure}

	Figure~\ref{fig:gw_predictions} shows the contact gravitational-wave frequency as a function of the total binary mass for several representative compactnesses, together with the design-sensitivity frequency windows of LISA \citep{2017arXiv170200786A}, the Einstein Telescope \citep{2010CQGra..27s4002P}, and Cosmic Explorer \citep{2019BAAS...51g..35R}.
	The contact frequency sets the highest gravitational-wave frequency emitted before merger; any inspiral signal sweeps upward through the detector band until $f_{\mathrm{GW}}$ is reached.
	Compact configurations with $\RT \sim 10\,\mathrm{km}$ and $M \sim 10^{-8}\,\Msun$ produce contact frequencies of order $10^{-1}\,\mathrm{Hz}$, placing them near the upper edge of the LISA band, whilst supra-solar objects ($M\sim 5\,\Msun$, $\RT\sim 10^{4}\,\mathrm{km}$) yield frequencies of order $10^{-1}\,\mathrm{Hz}$, likewise accessible to the LISA band.
	A dedicated signal-analysis study would be required to establish precise detectability thresholds; the present estimates serve to identify the frequency windows in which dark-fermion star mergers ought to be sought.

	The illustrative configurations of Table~\ref{tab:examples} populate distinct regions of the $f_{\mathrm{GW}}$--$M_{\mathrm{tot}}$ plane (Fig.~\ref{fig:gw_predictions}).
	The sub-solar object ($M=10^{-8}\,\Msun$) yields $f_{\mathrm{GW}}\simeq 0.18\,\mathrm{Hz}$, in the decihertz regime that proposed missions such as DECIGO \citep{2011CQGra..28i4011K} and the Big Bang Observer \citep{2006CQGra..23.4887H} are designed to explore.
	The supra-solar case ($M=5\,\Msun$, $\RT\simeq 9\,100\,\mathrm{km}$) produces $f_{\mathrm{GW}}\simeq 0.15\,\mathrm{Hz}$, whilst the most extended solar-mass entry ($\RT\simeq 726\,\Rsun$) yields $f_{\mathrm{GW}}\sim 5\times 10^{-9}\,\mathrm{Hz}$, well below the LISA band.
	In each case $m_1$, through its control of $\RT$, determines whether a given merger falls within the sensitivity window of space-borne or ground-based interferometers.

	\subsection{Microlensing and complementary observational signatures}
	\label{sec:microlensing}

	Gravitational microlensing offers an independent and complementary detection channel, since it requires only that the object possess sufficient mass to deflect background starlight and is entirely agnostic to the existence of a binary companion.
	For a point-mass lens of mass $M$ at luminosity distance $d_L$ from the observer (with the source at $d_S\gg d_L$), the Einstein radius is
	\begin{equation}
		R_E = \sqrt{\frac{4GM}{c^2}\,d_L}\,,
		\label{eq:einstein_radius}
	\end{equation}
	and the corresponding angular Einstein radius $\theta_E=R_E/d_L$.
	The microlensing timescale, defined as the time for the lens to traverse one Einstein radius at transverse velocity $v_\perp$, is
	\begin{equation}
		t_E = \frac{R_E}{v_\perp}\,.
		\label{eq:tE_microlensing}
	\end{equation}

	At a fiducial distance $d_L=8\,\mathrm{kpc}$ and transverse velocity $v_\perp=200\,\mathrm{km\,s^{-1}}$, the sub-solar and supra-solar configurations of Table~\ref{tab:examples} are compact relative to $R_E$ and produce standard point-lens events with timescales $t_E\simeq 21\,\mathrm{h}$ and $87\,\mathrm{days}$, respectively, well within the sensitivity of OGLE \citep{2015AcA....65....1U,2017Natur.548..183M}, MOA \citep{2001MNRAS.327..868B}, MACHO \citep{2000ApJ...542..281A} and EROS-2 \citep{2007A&A...469..387T}.
	The extended solar-mass configurations ($\RT\sim 1$--$10^2\,\Rsun$) exceed $R_E\simeq 6.9\,\Rsun$, so the point-lens approximation breaks down and finite-size corrections must be included \citep{2018PhRvD..97b3518N}; a detailed finite-lens calculation using the density profiles derived here lies beyond the present scope.
	For the supra-solar case, distinguishing the dark-fermion star from an ordinary stellar remnant would require the absence of an electromagnetic counterpart or an anomalous astrometric signal detectable by \textit{Gaia} \citep{2022ApJ...933...83L}.
	Microlensing, the gravitational-wave channel (Sec.~\ref{sec:gw_signatures}), astrometric perturbations and pulsar-timing residuals \citep{2019PhRvD.100b3003D} are complementary in scope, together constraining the abundance, mass function and internal structure of dark-fermion stars.

	\subsection{Observational constraints on the particle mass}
	\label{sec:constraints}

	The superradiance constraints frequently invoked for ultralight dark matter apply exclusively to \emph{bosonic} fields.
	For spin-0 and spin-1 particles the superradiant instability of rotating black holes enables stringent exclusions in the mass window $m\sim 10^{-13}$--$10^{-10}\,\eV$ \citep{2011PhRvD..83d4026A,2020Superradiance}.
	For spin-$\tfrac{1}{2}$ fermions the Pauli exclusion principle prevents exponential amplification, and no superradiant instability develops \citep{1973PhRvL..31.1265U,2023PhRvD.108h4024B}.
	The illustrative masses employed here ($m_1\sim 10^{-14}$--$10^{-7}\,\eV$) are therefore \emph{not} excluded by black-hole superradiance, a point that merits emphasis given the prevalence of bosonic constraints in the literature.

	The relevant fermionic bounds arise from phase-space considerations.
	The classical Tremaine--Gunn argument \citep{1979PhRvL..42..407T} requires $m\gtrsim \mathcal{O}(10^{2})\,\eV$ for a single species constituting the totality of dark matter; this bound is substantially relaxed, to $m\gtrsim \mathcal{O}(10^{-14})\,\eV$, when a large number of quasi-degenerate fermionic species is invoked \citep{2021PhRvD.103e5014D}, or when the dark fermions constitute only a sub-component of the dark sector.
	All $m_1$ values in Table~\ref{tab:examples} satisfy the relaxed bound; however, for the two-species entries the lightest secondary species ($m_2=m_1/q$ with $q=10$) lies at or below the Tremaine--Gunn threshold for the most extended configurations and would require either a richer dark-sector spectrum or a sub-dominant mass fraction to remain consistent.
	Additional constraints from galactic rotation curves \citep{2022PhRvD.105h3015B}, phase-space densities in dwarf spheroidals \citep{2018MNRAS.475.5385D} and laboratory searches with atomic clocks \citep{2025PhRvL.134c1001F} primarily target bosonic couplings; the fermionic hypothesis in this mass range is not excluded by current data, though any astrophysical interpretation must be cross-checked against the latest phase-space and small-scale structure bounds.

	\subsection{Connection to the dark matter problem}

	The configurations studied here contribute to our understanding of the dark matter problem in a specific, and deliberately circumscribed, manner.
	The rigid mass--radius product at $\xi=0$ (Eq.~\eqref{eq:invariants}) admits no adjustable equation of state: once the total mass $M$ is specified, the radius is entirely determined by the single microphysical parameter $m_1$.
	This predictive rigidity, inherited from the scaling symmetry of the SP system and the fixed Kirzhnits coefficient $\lamB=1/9$, is the central virtue of the model.
	Consequently, any detection of a self-gravitating dark-fermion star with independently measured mass and radius would immediately constrain $m_1$, or in the two-species case the combination $(m_1, q, f)$, thereby furnishing a direct bridge between astrophysical observables and the dark-sector Lagrangian.
	In the Thomas--Fermi regime, the Bohm term is confined to a thin shell near the stellar surface, yet it imprints a species-dependent signature, outward quantum pressure for the heavier species, inward surface tension for the lighter, that distinguishes these objects both from conventional compact remnants and from purely bosonic solitons.
	This distinction is by no means trivial: whilst bosonic solitons are governed by a single scalar-field mass, the fermionic equilibria derived here encode the richer structure of a multi-component dark sector through the mass ratio $q$ and the central composition $f$.
	The present results should therefore be understood as establishing a baseline: the irreducible quantum-gradient contribution to the equilibrium of a self-gravitating degenerate dark-fermion fluid, upon which more elaborate models incorporating finite temperature, self-interactions beyond the polytropic approximation, and phase-space coarse-graining may be constructed as the relevant physics demands \citep{2014NatPh..10..496S,2017PhRvD..95d3541H}.

	\subsection{Formation channels}

	Self-gravitating dark-sector fields can relax towards solitonic configurations via gravitational cooling \citep{1994PhRvL..72.2516S}, and it is reasonable to suppose that analogous mechanisms operate for degenerate fermion clumps in sufficiently high-density environments.
	In the bosonic case, dense axion miniclusters have been proposed as formation sites \citep{1993PhRvL..71.3051K}, with more recent work quantifying kinetic-regime condensation and subsequent growth \citep{2018PhRvL.121o1301L,2019PhRvD.100f3528E}; a comprehensive review is given by \citet{2021A&ARv..29....7F}.
	The fermionic analogue remains largely unexplored, and establishing whether gravitational cooling or alternative dissipative mechanisms can assemble the configurations derived here constitutes an open question whose resolution would ultimately determine the astrophysical relevance of dark-fermion stars.

	\subsection{Summary}

	I have shown that a two-component self-gravitating degenerate dark-fermion fluid, described within an orbital-free density-functional framework with the Kirzhnits gradient coefficient $\lamB=1/9$, admits equilibria in which the Bohm potential contributes a species-dependent surface-energy correction analogous to the nuclear liquid-drop model.
	The heavier fermion species establishes an outward quantum-pressure wall, the lighter species provides an inward surface tension, and degeneracy pressure furnishes the dominant bulk confinement: a tripartite balance whose structure is governed by the dimensionless pressure parameter $\sigma_i$ and the mass ratio $q$.
	The predictive rigidity of the resulting mass--radius relation, $M\RT\simeq 9.95\,\lamB\hbar^2/(Gm_1^2)$ in the SP limit, is a noteworthy feature, for it implies that a single microphysical parameter, $m_1$, determines the equilibrium once the total mass is specified, distinguishing these configurations from phenomenological models with adjustable equations of state.

	The observational channels through which these objects might reveal themselves, gravitational-wave contact frequencies in the LISA and Einstein Telescope bands, microlensing signatures accessible to OGLE \citep{2015AcA....65....1U} and MOA \citep{2001MNRAS.327..868B}, and astrometric perturbations, are complementary in scope and sensitivity.
	The mass--radius relations and benchmarked dimensionless invariants presented here furnish a reproducible, first-principles reference for constraining the dark-fermion mass in multi-component dark sectors.
	In this sense, the quantum-gradient structure of a degenerate Fermi fluid leaves an indelible imprint on the equilibrium of self-gravitating dark matter, one that current and forthcoming gravitational-wave observatories are, in principle, well placed to detect.

	\begin{acknowledgments}
		I.~L. thanks the Funda\c{c}\~ao para a Ci\^encia e Tecnologia (FCT), Portugal, for the financial support to the Center for Astrophysics and Gravitation (CENTRA/IST/ULisboa) through grant No.\ UID/PRR/00099/2025 (\doi{10.54499/UID/PRR/00099/2025}) and grant No.\ UID/00099/2025 (\doi{10.54499/UID/00099/2025}).
	\end{acknowledgments}


\begin{thebibliography}{91}%
\makeatletter
\providecommand \@ifxundefined [1]{%
 \@ifx{#1\undefined}
}%
\providecommand \@ifnum [1]{%
 \ifnum #1\expandafter \@firstoftwo
 \else \expandafter \@secondoftwo
 \fi
}%
\providecommand \@ifx [1]{%
 \ifx #1\expandafter \@firstoftwo
 \else \expandafter \@secondoftwo
 \fi
}%
\providecommand \natexlab [1]{#1}%
\providecommand \enquote  [1]{``#1''}%
\providecommand \bibnamefont  [1]{#1}%
\providecommand \bibfnamefont [1]{#1}%
\providecommand \citenamefont [1]{#1}%
\providecommand \href@noop [0]{\@secondoftwo}%
\providecommand \href [0]{\begingroup \@sanitize@url \@href}%
\providecommand \@href[1]{\@@startlink{#1}\@@href}%
\providecommand \@@href[1]{\endgroup#1\@@endlink}%
\providecommand \@sanitize@url [0]{\catcode `\\12\catcode `\$12\catcode
  `\&12\catcode `\#12\catcode `\^12\catcode `\_12\catcode `\%12\relax}%
\providecommand \@@startlink[1]{}%
\providecommand \@@endlink[0]{}%
\providecommand \url  [0]{\begingroup\@sanitize@url \@url }%
\providecommand \@url [1]{\endgroup\@href {#1}{\urlprefix }}%
\providecommand \urlprefix  [0]{URL }%
\providecommand \Eprint [0]{\href }%
\providecommand \doibase [0]{https://doi.org/}%
\providecommand \selectlanguage [0]{\@gobble}%
\providecommand \bibinfo  [0]{\@secondoftwo}%
\providecommand \bibfield  [0]{\@secondoftwo}%
\providecommand \translation [1]{[#1]}%
\providecommand \BibitemOpen [0]{}%
\providecommand \bibitemStop [0]{}%
\providecommand \bibitemNoStop [0]{.\EOS\space}%
\providecommand \EOS [0]{\spacefactor3000\relax}%
\providecommand \BibitemShut  [1]{\csname bibitem#1\endcsname}%
\let\auto@bib@innerbib\@empty
\bibitem [{\citenamefont {{Planck Collaboration}}\ \emph
  {et~al.}(2020)\citenamefont {{Planck Collaboration}}, \citenamefont
  {{Aghanim}}, \citenamefont {{Akrami}}, \citenamefont {{Ashdown}},
  \citenamefont {{Aumont}}, \citenamefont {{Baccigalupi}}, \citenamefont
  {{Ballardini}}, \citenamefont {{Banday}}, \citenamefont {{Barreiro}},
  \citenamefont {{Bartolo}} \emph {et~al.}}]{2020A&A...641A...6P}%
  \BibitemOpen
  \bibfield  {author} {\bibinfo {author} {\bibnamefont {{Planck
  Collaboration}}}, \bibinfo {author} {\bibfnamefont {N.}~\bibnamefont
  {{Aghanim}}}, \bibinfo {author} {\bibfnamefont {Y.}~\bibnamefont {{Akrami}}},
  \bibinfo {author} {\bibfnamefont {M.}~\bibnamefont {{Ashdown}}}, \bibinfo
  {author} {\bibfnamefont {J.}~\bibnamefont {{Aumont}}}, \bibinfo {author}
  {\bibfnamefont {C.}~\bibnamefont {{Baccigalupi}}}, \bibinfo {author}
  {\bibfnamefont {M.}~\bibnamefont {{Ballardini}}}, \bibinfo {author}
  {\bibfnamefont {A.~J.}\ \bibnamefont {{Banday}}}, \bibinfo {author}
  {\bibfnamefont {R.~B.}\ \bibnamefont {{Barreiro}}}, \bibinfo {author}
  {\bibfnamefont {N.}~\bibnamefont {{Bartolo}}}, \emph {et~al.},\ }\bibfield
  {title} {\bibinfo {title} {{Planck 2018 results. VI. Cosmological
  parameters}},\ }\href {https://doi.org/10.1051/0004-6361/201833910}
  {\bibfield  {journal} {\bibinfo  {journal} {\aap}\ }\textbf {\bibinfo
  {volume} {641}},\ \bibinfo {eid} {A6} (\bibinfo {year} {2020})},\ \Eprint
  {https://arxiv.org/abs/1807.06209} {arXiv:1807.06209 [astro-ph.CO]}
  \BibitemShut {NoStop}%
\bibitem [{\citenamefont {{Bullock}}\ and\ \citenamefont
  {{Boylan-Kolchin}}(2017)}]{2017ARA&A..55..343B}%
  \BibitemOpen
  \bibfield  {author} {\bibinfo {author} {\bibfnamefont {J.~S.}\ \bibnamefont
  {{Bullock}}}\ and\ \bibinfo {author} {\bibfnamefont {M.}~\bibnamefont
  {{Boylan-Kolchin}}},\ }\bibfield  {title} {\bibinfo {title} {{Small-Scale
  Challenges to the {$\Lambda$CDM} Paradigm}},\ }\href
  {https://doi.org/10.1146/annurev-astro-091916-055313} {\bibfield  {journal}
  {\bibinfo  {journal} {\araa}\ }\textbf {\bibinfo {volume} {55}},\ \bibinfo
  {pages} {343} (\bibinfo {year} {2017})},\ \Eprint
  {https://arxiv.org/abs/1707.04256} {arXiv:1707.04256 [astro-ph.CO]}
  \BibitemShut {NoStop}%
\bibitem [{\citenamefont {{Navarro}}\ \emph {et~al.}(1996)\citenamefont
  {{Navarro}}, \citenamefont {{Frenk}},\ and\ \citenamefont
  {{White}}}]{1996ApJ...462..563N}%
  \BibitemOpen
  \bibfield  {author} {\bibinfo {author} {\bibfnamefont {J.~F.}\ \bibnamefont
  {{Navarro}}}, \bibinfo {author} {\bibfnamefont {C.~S.}\ \bibnamefont
  {{Frenk}}},\ and\ \bibinfo {author} {\bibfnamefont {S.~D.~M.}\ \bibnamefont
  {{White}}},\ }\bibfield  {title} {\bibinfo {title} {{The Structure of Dark
  Matter Halos}},\ }\href {https://doi.org/10.1086/177173} {\bibfield
  {journal} {\bibinfo  {journal} {\apj}\ }\textbf {\bibinfo {volume} {462}},\
  \bibinfo {pages} {563} (\bibinfo {year} {1996})},\ \Eprint
  {https://arxiv.org/abs/astro-ph/9508025} {arXiv:astro-ph/9508025}
  \BibitemShut {NoStop}%
\bibitem [{\citenamefont {{Navas}}\ \emph {et~al.}(2024)\citenamefont {{Navas}}
  \emph {et~al.}}]{2024PhRvD.110c0001N}%
  \BibitemOpen
  \bibfield  {author} {\bibinfo {author} {\bibfnamefont {S.}~\bibnamefont
  {{Navas}}} \emph {et~al.} (\bibinfo {collaboration} {Particle Data Group}),\
  }\bibfield  {title} {\bibinfo {title} {{Review of Particle Physics}},\ }\href
  {https://doi.org/10.1103/PhysRevD.110.030001} {\bibfield  {journal} {\bibinfo
   {journal} {\prd}\ }\textbf {\bibinfo {volume} {110}},\ \bibinfo {eid}
  {030001} (\bibinfo {year} {2024})}\BibitemShut {NoStop}%
\bibitem [{\citenamefont {{Spergel}}\ and\ \citenamefont
  {{Steinhardt}}(2000)}]{2000PhRvL..84.3760S}%
  \BibitemOpen
  \bibfield  {author} {\bibinfo {author} {\bibfnamefont {D.~N.}\ \bibnamefont
  {{Spergel}}}\ and\ \bibinfo {author} {\bibfnamefont {P.~J.}\ \bibnamefont
  {{Steinhardt}}},\ }\bibfield  {title} {\bibinfo {title} {{Observational
  Evidence for Self-Interacting Cold Dark Matter}},\ }\href
  {https://doi.org/10.1103/PhysRevLett.84.3760} {\bibfield  {journal} {\bibinfo
   {journal} {\prl}\ }\textbf {\bibinfo {volume} {84}},\ \bibinfo {pages}
  {3760} (\bibinfo {year} {2000})},\ \Eprint
  {https://arxiv.org/abs/astro-ph/9909386} {arXiv:astro-ph/9909386}
  \BibitemShut {NoStop}%
\bibitem [{\citenamefont {{Hu}}\ \emph {et~al.}(2000)\citenamefont {{Hu}},
  \citenamefont {{Barkana}},\ and\ \citenamefont
  {{Gruzinov}}}]{2000PhRvL..85.1158H}%
  \BibitemOpen
  \bibfield  {author} {\bibinfo {author} {\bibfnamefont {W.}~\bibnamefont
  {{Hu}}}, \bibinfo {author} {\bibfnamefont {R.}~\bibnamefont {{Barkana}}},\
  and\ \bibinfo {author} {\bibfnamefont {A.}~\bibnamefont {{Gruzinov}}},\
  }\bibfield  {title} {\bibinfo {title} {{Fuzzy Cold Dark Matter: The Wave
  Properties of Ultralight Particles}},\ }\href
  {https://doi.org/10.1103/PhysRevLett.85.1158} {\bibfield  {journal} {\bibinfo
   {journal} {\prl}\ }\textbf {\bibinfo {volume} {85}},\ \bibinfo {pages}
  {1158} (\bibinfo {year} {2000})},\ \Eprint
  {https://arxiv.org/abs/astro-ph/0003365} {arXiv:astro-ph/0003365 [astro-ph]}
  \BibitemShut {NoStop}%
\bibitem [{\citenamefont {{Marsh}}(2016)}]{2016PhR...643....1M}%
  \BibitemOpen
  \bibfield  {author} {\bibinfo {author} {\bibfnamefont {D.~J.~E.}\
  \bibnamefont {{Marsh}}},\ }\bibfield  {title} {\bibinfo {title} {{Axion
  cosmology}},\ }\href {https://doi.org/10.1016/j.physrep.2016.06.005}
  {\bibfield  {journal} {\bibinfo  {journal} {\physrep}\ }\textbf {\bibinfo
  {volume} {643}},\ \bibinfo {pages} {1} (\bibinfo {year} {2016})},\ \Eprint
  {https://arxiv.org/abs/1510.07633} {arXiv:1510.07633 [astro-ph.CO]}
  \BibitemShut {NoStop}%
\bibitem [{\citenamefont {{Ferreira}}(2021)}]{2021A&ARv..29....7F}%
  \BibitemOpen
  \bibfield  {author} {\bibinfo {author} {\bibfnamefont {E.~G.~M.}\
  \bibnamefont {{Ferreira}}},\ }\bibfield  {title} {\bibinfo {title}
  {{Ultra-light dark matter}},\ }\href
  {https://doi.org/10.1007/s00159-021-00135-6} {\bibfield  {journal} {\bibinfo
  {journal} {Astron.~Astrophys.~Rev.}\ }\textbf {\bibinfo {volume} {29}},\
  \bibinfo {eid} {7} (\bibinfo {year} {2021})},\ \Eprint
  {https://arxiv.org/abs/2005.03254} {arXiv:2005.03254 [astro-ph.CO]}
  \BibitemShut {NoStop}%
\bibitem [{\citenamefont {{Hui}}\ \emph {et~al.}(2017)\citenamefont {{Hui}},
  \citenamefont {{Ostriker}}, \citenamefont {{Tremaine}},\ and\ \citenamefont
  {{Witten}}}]{2017PhRvD..95d3541H}%
  \BibitemOpen
  \bibfield  {author} {\bibinfo {author} {\bibfnamefont {L.}~\bibnamefont
  {{Hui}}}, \bibinfo {author} {\bibfnamefont {J.~P.}\ \bibnamefont
  {{Ostriker}}}, \bibinfo {author} {\bibfnamefont {S.}~\bibnamefont
  {{Tremaine}}},\ and\ \bibinfo {author} {\bibfnamefont {E.}~\bibnamefont
  {{Witten}}},\ }\bibfield  {title} {\bibinfo {title} {{Ultralight scalars as
  cosmological dark matter}},\ }\href
  {https://doi.org/10.1103/PhysRevD.95.043541} {\bibfield  {journal} {\bibinfo
  {journal} {\prd}\ }\textbf {\bibinfo {volume} {95}},\ \bibinfo {eid} {043541}
  (\bibinfo {year} {2017})},\ \Eprint {https://arxiv.org/abs/1610.08297}
  {arXiv:1610.08297 [astro-ph.CO]} \BibitemShut {NoStop}%
\bibitem [{\citenamefont {{Schive}}\ \emph
  {et~al.}(2014{\natexlab{a}})\citenamefont {{Schive}}, \citenamefont
  {{Chiueh}},\ and\ \citenamefont {{Broadhurst}}}]{2014NatPh..10..496S}%
  \BibitemOpen
  \bibfield  {author} {\bibinfo {author} {\bibfnamefont {H.-Y.}\ \bibnamefont
  {{Schive}}}, \bibinfo {author} {\bibfnamefont {T.}~\bibnamefont {{Chiueh}}},\
  and\ \bibinfo {author} {\bibfnamefont {T.}~\bibnamefont {{Broadhurst}}},\
  }\bibfield  {title} {\bibinfo {title} {{Cosmic structure as the quantum
  interference of a coherent dark wave}},\ }\href
  {https://doi.org/10.1038/nphys2996} {\bibfield  {journal} {\bibinfo
  {journal} {Nature Physics}\ }\textbf {\bibinfo {volume} {10}},\ \bibinfo
  {pages} {496} (\bibinfo {year} {2014}{\natexlab{a}})},\ \Eprint
  {https://arxiv.org/abs/1406.6586} {arXiv:1406.6586 [astro-ph.GA]}
  \BibitemShut {NoStop}%
\bibitem [{\citenamefont {{Schive}}\ \emph
  {et~al.}(2014{\natexlab{b}})\citenamefont {{Schive}}, \citenamefont
  {{Chiueh}}, \citenamefont {{Broadhurst}},\ and\ \citenamefont
  {{Huang}}}]{2014PhRvL.113z1302S}%
  \BibitemOpen
  \bibfield  {author} {\bibinfo {author} {\bibfnamefont {H.-Y.}\ \bibnamefont
  {{Schive}}}, \bibinfo {author} {\bibfnamefont {T.}~\bibnamefont {{Chiueh}}},
  \bibinfo {author} {\bibfnamefont {T.}~\bibnamefont {{Broadhurst}}},\ and\
  \bibinfo {author} {\bibfnamefont {K.-W.}\ \bibnamefont {{Huang}}},\
  }\bibfield  {title} {\bibinfo {title} {{Contrasting Galaxy Formation from
  Quantum Wave Dark Matter, {$\psi$}DM, with {$\Lambda$}CDM}},\ }\href
  {https://doi.org/10.1103/PhysRevLett.113.261302} {\bibfield  {journal}
  {\bibinfo  {journal} {\prl}\ }\textbf {\bibinfo {volume} {113}},\ \bibinfo
  {eid} {261302} (\bibinfo {year} {2014}{\natexlab{b}})},\ \Eprint
  {https://arxiv.org/abs/1407.7762} {arXiv:1407.7762 [astro-ph.GA]}
  \BibitemShut {NoStop}%
\bibitem [{\citenamefont {{Kaup}}(1968)}]{1968PhRv..172.1331K}%
  \BibitemOpen
  \bibfield  {author} {\bibinfo {author} {\bibfnamefont {D.~J.}\ \bibnamefont
  {{Kaup}}},\ }\bibfield  {title} {\bibinfo {title} {{Klein-Gordon Geon}},\
  }\href {https://doi.org/10.1103/PhysRev.172.1331} {\bibfield  {journal}
  {\bibinfo  {journal} {Phys.~Rev.}\ }\textbf {\bibinfo {volume} {172}},\
  \bibinfo {pages} {1331} (\bibinfo {year} {1968})}\BibitemShut {NoStop}%
\bibitem [{\citenamefont {{Ruffini}}\ and\ \citenamefont
  {{Bonazzola}}(1969)}]{1969PhRv..187.1767R}%
  \BibitemOpen
  \bibfield  {author} {\bibinfo {author} {\bibfnamefont {R.}~\bibnamefont
  {{Ruffini}}}\ and\ \bibinfo {author} {\bibfnamefont {S.}~\bibnamefont
  {{Bonazzola}}},\ }\bibfield  {title} {\bibinfo {title} {{Systems of
  Self-Gravitating Particles in General Relativity and the Concept of an
  Equation of State}},\ }\href {https://doi.org/10.1103/PhysRev.187.1767}
  {\bibfield  {journal} {\bibinfo  {journal} {Phys.~Rev.}\ }\textbf {\bibinfo
  {volume} {187}},\ \bibinfo {pages} {1767} (\bibinfo {year}
  {1969})}\BibitemShut {NoStop}%
\bibitem [{\citenamefont {{Liebling}}\ and\ \citenamefont
  {{Palenzuela}}(2023)}]{2023LRR....26....1L}%
  \BibitemOpen
  \bibfield  {author} {\bibinfo {author} {\bibfnamefont {S.~L.}\ \bibnamefont
  {{Liebling}}}\ and\ \bibinfo {author} {\bibfnamefont {C.}~\bibnamefont
  {{Palenzuela}}},\ }\bibfield  {title} {\bibinfo {title} {{Dynamical Boson
  Stars}},\ }\href {https://doi.org/10.1007/s41114-023-00043-4} {\bibfield
  {journal} {\bibinfo  {journal} {\lrr}\ }\textbf {\bibinfo {volume} {26}},\
  \bibinfo {eid} {1} (\bibinfo {year} {2023})},\ \Eprint
  {https://arxiv.org/abs/1202.5809} {arXiv:1202.5809 [gr-qc]} \BibitemShut
  {NoStop}%
\bibitem [{\citenamefont {{Lieb}}(1977)}]{1977StAM...57...93L}%
  \BibitemOpen
  \bibfield  {author} {\bibinfo {author} {\bibfnamefont {E.~H.}\ \bibnamefont
  {{Lieb}}},\ }\bibfield  {title} {\bibinfo {title} {{Existence and uniqueness
  of the minimizing solution of Choquard's nonlinear equation}},\ }\href
  {https://doi.org/10.1002/sapm197757293} {\bibfield  {journal} {\bibinfo
  {journal} {Stud.~Appl.~Math.}\ }\textbf {\bibinfo {volume} {57}},\ \bibinfo
  {pages} {93} (\bibinfo {year} {1977})}\BibitemShut {NoStop}%
\bibitem [{\citenamefont {{Moroz}}\ \emph {et~al.}(1998)\citenamefont
  {{Moroz}}, \citenamefont {{Penrose}},\ and\ \citenamefont
  {{Tod}}}]{1998CQGra..15.2733M}%
  \BibitemOpen
  \bibfield  {author} {\bibinfo {author} {\bibfnamefont {I.~M.}\ \bibnamefont
  {{Moroz}}}, \bibinfo {author} {\bibfnamefont {R.}~\bibnamefont {{Penrose}}},\
  and\ \bibinfo {author} {\bibfnamefont {P.}~\bibnamefont {{Tod}}},\ }\bibfield
   {title} {\bibinfo {title} {{Spherically-symmetric solutions of the
  Schr{\"o}dinger--Newton equations}},\ }\href
  {https://doi.org/10.1088/0264-9381/15/9/019} {\bibfield  {journal} {\bibinfo
  {journal} {Class.~Quantum Grav.}\ }\textbf {\bibinfo {volume} {15}},\
  \bibinfo {pages} {2733} (\bibinfo {year} {1998})}\BibitemShut {NoStop}%
\bibitem [{\citenamefont {{Chavanis}}(2011)}]{2011PhRvD..84d3531C}%
  \BibitemOpen
  \bibfield  {author} {\bibinfo {author} {\bibfnamefont {P.-H.}\ \bibnamefont
  {{Chavanis}}},\ }\bibfield  {title} {\bibinfo {title} {{Mass-radius relation
  of Newtonian self-gravitating Bose-Einstein condensates with short-range
  interactions. I. Analytical results}},\ }\href
  {https://doi.org/10.1103/PhysRevD.84.043531} {\bibfield  {journal} {\bibinfo
  {journal} {\prd}\ }\textbf {\bibinfo {volume} {84}},\ \bibinfo {pages}
  {043531} (\bibinfo {year} {2011})},\ \Eprint
  {https://arxiv.org/abs/1103.2050} {arXiv:1103.2050} \BibitemShut {NoStop}%
\bibitem [{\citenamefont {{Chavanis}}\ and\ \citenamefont
  {{Delfini}}(2011)}]{2011PhRvD..84d3532C}%
  \BibitemOpen
  \bibfield  {author} {\bibinfo {author} {\bibfnamefont {P.-H.}\ \bibnamefont
  {{Chavanis}}}\ and\ \bibinfo {author} {\bibfnamefont {L.}~\bibnamefont
  {{Delfini}}},\ }\bibfield  {title} {\bibinfo {title} {{Mass-radius relation
  of Newtonian self-gravitating Bose--Einstein condensates with short-range
  interactions. II. Numerical results}},\ }\href
  {https://doi.org/10.1103/PhysRevD.84.043532} {\bibfield  {journal} {\bibinfo
  {journal} {\prd}\ }\textbf {\bibinfo {volume} {84}},\ \bibinfo {eid} {043532}
  (\bibinfo {year} {2011})},\ \Eprint {https://arxiv.org/abs/1103.2698}
  {arXiv:1103.2698 [astro-ph.CO]} \BibitemShut {NoStop}%
\bibitem [{\citenamefont {{Lee}}\ and\ \citenamefont
  {{Pang}}(1987)}]{1987PhRvD..35.3678L}%
  \BibitemOpen
  \bibfield  {author} {\bibinfo {author} {\bibfnamefont {T.~D.}\ \bibnamefont
  {{Lee}}}\ and\ \bibinfo {author} {\bibfnamefont {Y.}~\bibnamefont {{Pang}}},\
  }\bibfield  {title} {\bibinfo {title} {{Fermion Soliton Stars and Black
  Holes}},\ }\href {https://doi.org/10.1103/PhysRevD.35.3678} {\bibfield
  {journal} {\bibinfo  {journal} {\prd}\ }\textbf {\bibinfo {volume} {35}},\
  \bibinfo {pages} {3678} (\bibinfo {year} {1987})}\BibitemShut {NoStop}%
\bibitem [{\citenamefont {{Wise}}\ and\ \citenamefont
  {{Zhang}}(2014)}]{2014PhRvD..90e5030W}%
  \BibitemOpen
  \bibfield  {author} {\bibinfo {author} {\bibfnamefont {M.~B.}\ \bibnamefont
  {{Wise}}}\ and\ \bibinfo {author} {\bibfnamefont {Y.}~\bibnamefont
  {{Zhang}}},\ }\bibfield  {title} {\bibinfo {title} {{Stable Bound States of
  Asymmetric Dark Matter}},\ }\href
  {https://doi.org/10.1103/PhysRevD.90.055030} {\bibfield  {journal} {\bibinfo
  {journal} {\prd}\ }\textbf {\bibinfo {volume} {90}},\ \bibinfo {eid} {055030}
  (\bibinfo {year} {2014})},\ \Eprint {https://arxiv.org/abs/1407.4121}
  {arXiv:1407.4121 [hep-ph]} \BibitemShut {NoStop}%
\bibitem [{\citenamefont {{Del Grosso}}\ \emph {et~al.}(2023)\citenamefont
  {{Del Grosso}}, \citenamefont {{Franciolini}}, \citenamefont {{Pani}},\ and\
  \citenamefont {{Urbano}}}]{2023PhRvD.108d4024D}%
  \BibitemOpen
  \bibfield  {author} {\bibinfo {author} {\bibfnamefont {L.}~\bibnamefont {{Del
  Grosso}}}, \bibinfo {author} {\bibfnamefont {G.}~\bibnamefont
  {{Franciolini}}}, \bibinfo {author} {\bibfnamefont {P.}~\bibnamefont
  {{Pani}}},\ and\ \bibinfo {author} {\bibfnamefont {A.}~\bibnamefont
  {{Urbano}}},\ }\bibfield  {title} {\bibinfo {title} {{Fermion soliton
  stars}},\ }\href {https://doi.org/10.1103/PhysRevD.108.044024} {\bibfield
  {journal} {\bibinfo  {journal} {\prd}\ }\textbf {\bibinfo {volume} {108}},\
  \bibinfo {eid} {044024} (\bibinfo {year} {2023})},\ \Eprint
  {https://arxiv.org/abs/2301.08709} {arXiv:2301.08709 [hep-ph]} \BibitemShut
  {NoStop}%
\bibitem [{\citenamefont {{Coleman}}(1985)}]{1985NuPhB.262..263C}%
  \BibitemOpen
  \bibfield  {author} {\bibinfo {author} {\bibfnamefont {S.}~\bibnamefont
  {{Coleman}}},\ }\bibfield  {title} {\bibinfo {title} {{Q-balls}},\ }\href
  {https://doi.org/10.1016/0550-3213(85)90383-X} {\bibfield  {journal}
  {\bibinfo  {journal} {\nphysb}\ }\textbf {\bibinfo {volume} {262}},\ \bibinfo
  {pages} {263} (\bibinfo {year} {1985})}\BibitemShut {NoStop}%
\bibitem [{\citenamefont {{Tulin}}\ and\ \citenamefont
  {{Yu}}(2018)}]{2018PhR...730....1T}%
  \BibitemOpen
  \bibfield  {author} {\bibinfo {author} {\bibfnamefont {S.}~\bibnamefont
  {{Tulin}}}\ and\ \bibinfo {author} {\bibfnamefont {H.-B.}\ \bibnamefont
  {{Yu}}},\ }\bibfield  {title} {\bibinfo {title} {{Dark Matter
  Self-interactions and Small Scale Structure}},\ }\href
  {https://doi.org/10.1016/j.physrep.2017.11.004} {\bibfield  {journal}
  {\bibinfo  {journal} {\physrep}\ }\textbf {\bibinfo {volume} {730}},\
  \bibinfo {pages} {1} (\bibinfo {year} {2018})},\ \Eprint
  {https://arxiv.org/abs/1705.02358} {arXiv:1705.02358 [hep-ph]} \BibitemShut
  {NoStop}%
\bibitem [{\citenamefont {{Kaplan}}\ \emph {et~al.}(2010)\citenamefont
  {{Kaplan}}, \citenamefont {{Krnjaic}}, \citenamefont {{Rehermann}},\ and\
  \citenamefont {{Wells}}}]{2010JCAP...05..021K}%
  \BibitemOpen
  \bibfield  {author} {\bibinfo {author} {\bibfnamefont {D.~E.}\ \bibnamefont
  {{Kaplan}}}, \bibinfo {author} {\bibfnamefont {G.~Z.}\ \bibnamefont
  {{Krnjaic}}}, \bibinfo {author} {\bibfnamefont {K.~R.}\ \bibnamefont
  {{Rehermann}}},\ and\ \bibinfo {author} {\bibfnamefont {C.~M.}\ \bibnamefont
  {{Wells}}},\ }\bibfield  {title} {\bibinfo {title} {{Atomic dark matter}},\
  }\href {https://doi.org/10.1088/1475-7516/2010/05/021} {\bibfield  {journal}
  {\bibinfo  {journal} {\jcap}\ }\textbf {\bibinfo {volume} {2010}},\ \bibinfo
  {eid} {021} (\bibinfo {year} {2010})},\ \Eprint
  {https://arxiv.org/abs/0909.0753} {arXiv:0909.0753 [hep-ph]} \BibitemShut
  {NoStop}%
\bibitem [{\citenamefont {{Press}}\ and\ \citenamefont
  {{Spergel}}(1985)}]{1985ApJ...296..679P}%
  \BibitemOpen
  \bibfield  {author} {\bibinfo {author} {\bibfnamefont {W.~H.}\ \bibnamefont
  {{Press}}}\ and\ \bibinfo {author} {\bibfnamefont {D.~N.}\ \bibnamefont
  {{Spergel}}},\ }\bibfield  {title} {\bibinfo {title} {Capture by the sun of a
  galactic population of weakly interacting, massive particles},\ }\href
  {https://doi.org/10.1086/163485} {\bibfield  {journal} {\bibinfo  {journal}
  {\apj}\ }\textbf {\bibinfo {volume} {296}},\ \bibinfo {pages} {679} (\bibinfo
  {year} {1985})}\BibitemShut {NoStop}%
\bibitem [{\citenamefont {{Gould}}(1987)}]{1987ApJ...321..560G}%
  \BibitemOpen
  \bibfield  {author} {\bibinfo {author} {\bibfnamefont {A.}~\bibnamefont
  {{Gould}}},\ }\bibfield  {title} {\bibinfo {title} {Weakly interacting
  massive particle distribution in and evaporation from the sun},\ }\href
  {https://doi.org/10.1086/165652} {\bibfield  {journal} {\bibinfo  {journal}
  {\apj}\ }\textbf {\bibinfo {volume} {321}},\ \bibinfo {pages} {560} (\bibinfo
  {year} {1987})}\BibitemShut {NoStop}%
\bibitem [{\citenamefont {{Frandsen}}\ and\ \citenamefont
  {{Sarkar}}(2010)}]{2010PhRvL.105a1301F}%
  \BibitemOpen
  \bibfield  {author} {\bibinfo {author} {\bibfnamefont {M.~T.}\ \bibnamefont
  {{Frandsen}}}\ and\ \bibinfo {author} {\bibfnamefont {S.}~\bibnamefont
  {{Sarkar}}},\ }\bibfield  {title} {\bibinfo {title} {Asymmetric dark matter
  and the sun},\ }\href {https://doi.org/10.1103/PhysRevLett.105.011301}
  {\bibfield  {journal} {\bibinfo  {journal} {\prl}\ }\textbf {\bibinfo
  {volume} {105}},\ \bibinfo {pages} {011301} (\bibinfo {year} {2010})},\
  \Eprint {https://arxiv.org/abs/1003.4505} {arXiv:1003.4505} \BibitemShut
  {NoStop}%
\bibitem [{\citenamefont {{Taoso}}\ \emph {et~al.}(2010)\citenamefont
  {{Taoso}}, \citenamefont {{Iocco}}, \citenamefont {{Meynet}}, \citenamefont
  {{Bertone}},\ and\ \citenamefont {{Eggenberger}}}]{2010PhRvD..82h3509T}%
  \BibitemOpen
  \bibfield  {author} {\bibinfo {author} {\bibfnamefont {M.}~\bibnamefont
  {{Taoso}}}, \bibinfo {author} {\bibfnamefont {F.}~\bibnamefont {{Iocco}}},
  \bibinfo {author} {\bibfnamefont {G.}~\bibnamefont {{Meynet}}}, \bibinfo
  {author} {\bibfnamefont {G.}~\bibnamefont {{Bertone}}},\ and\ \bibinfo
  {author} {\bibfnamefont {P.}~\bibnamefont {{Eggenberger}}},\ }\bibfield
  {title} {\bibinfo {title} {Effect of low mass dark matter particles on the
  sun},\ }\href {https://doi.org/10.1103/PhysRevD.82.083509} {\bibfield
  {journal} {\bibinfo  {journal} {\prd}\ }\textbf {\bibinfo {volume} {82}},\
  \bibinfo {pages} {083509} (\bibinfo {year} {2010})},\ \Eprint
  {https://arxiv.org/abs/1005.5711} {arXiv:1005.5711} \BibitemShut {NoStop}%
\bibitem [{\citenamefont {{Lopes}}\ and\ \citenamefont
  {{Silk}}(2002)}]{2002PhRvL..88o1303L}%
  \BibitemOpen
  \bibfield  {author} {\bibinfo {author} {\bibfnamefont {I.~P.}\ \bibnamefont
  {{Lopes}}}\ and\ \bibinfo {author} {\bibfnamefont {J.}~\bibnamefont
  {{Silk}}},\ }\bibfield  {title} {\bibinfo {title} {{Solar Neutrinos: Probing
  the Quasi-isothermal Solar Core Produced by Supersymmetric Dark Matter
  Particles}},\ }\href {https://doi.org/10.1103/PhysRevLett.88.151303}
  {\bibfield  {journal} {\bibinfo  {journal} {\prl}\ }\textbf {\bibinfo
  {volume} {88}},\ \bibinfo {eid} {151303} (\bibinfo {year} {2002})},\ \Eprint
  {https://arxiv.org/abs/astro-ph/0112390} {arXiv:astro-ph/0112390}
  \BibitemShut {NoStop}%
\bibitem [{\citenamefont {{Lopes}}\ and\ \citenamefont
  {{Silk}}(2011)}]{2011ApJ...733L..51L}%
  \BibitemOpen
  \bibfield  {author} {\bibinfo {author} {\bibfnamefont {I.}~\bibnamefont
  {{Lopes}}}\ and\ \bibinfo {author} {\bibfnamefont {J.}~\bibnamefont
  {{Silk}}},\ }\bibfield  {title} {\bibinfo {title} {{Dark Matter Burning in
  Nuclear Star Clusters}},\ }\href
  {https://doi.org/10.1088/2041-8205/733/2/L51} {\bibfield  {journal} {\bibinfo
   {journal} {\apjl}\ }\textbf {\bibinfo {volume} {733}},\ \bibinfo {eid} {L51}
  (\bibinfo {year} {2011})},\ \Eprint {https://arxiv.org/abs/1104.1118}
  {arXiv:1104.1118} \BibitemShut {NoStop}%
\bibitem [{\citenamefont {{Lopes}}\ and\ \citenamefont
  {{Silk}}(2012)}]{2012ApJ...757..130L}%
  \BibitemOpen
  \bibfield  {author} {\bibinfo {author} {\bibfnamefont {I.}~\bibnamefont
  {{Lopes}}}\ and\ \bibinfo {author} {\bibfnamefont {J.}~\bibnamefont
  {{Silk}}},\ }\bibfield  {title} {\bibinfo {title} {{Solar Constraints on
  Asymmetric Dark Matter}},\ }\href
  {https://doi.org/10.1088/0004-637X/757/2/130} {\bibfield  {journal} {\bibinfo
   {journal} {\apj}\ }\textbf {\bibinfo {volume} {757}},\ \bibinfo {eid} {130}
  (\bibinfo {year} {2012})},\ \Eprint {https://arxiv.org/abs/1209.3631}
  {arXiv:1209.3631} \BibitemShut {NoStop}%
\bibitem [{\citenamefont {{Lopes}}\ and\ \citenamefont
  {{Silk}}(2014)}]{2014ApJ...786...25L}%
  \BibitemOpen
  \bibfield  {author} {\bibinfo {author} {\bibfnamefont {I.}~\bibnamefont
  {{Lopes}}}\ and\ \bibinfo {author} {\bibfnamefont {J.}~\bibnamefont
  {{Silk}}},\ }\bibfield  {title} {\bibinfo {title} {{A Particle Dark Matter
  Footprint on the First Generation of Stars}},\ }\href
  {https://doi.org/10.1088/0004-637X/786/1/25} {\bibfield  {journal} {\bibinfo
  {journal} {\apj}\ }\textbf {\bibinfo {volume} {786}},\ \bibinfo {eid} {25}
  (\bibinfo {year} {2014})},\ \Eprint {https://arxiv.org/abs/1402.0682}
  {arXiv:1402.0682} \BibitemShut {NoStop}%
\bibitem [{\citenamefont {{Lopes}}\ \emph
  {et~al.}(2014{\natexlab{a}})\citenamefont {{Lopes}}, \citenamefont
  {{Kadota}},\ and\ \citenamefont {{Silk}}}]{2014ApJ...780L..15L}%
  \BibitemOpen
  \bibfield  {author} {\bibinfo {author} {\bibfnamefont {I.}~\bibnamefont
  {{Lopes}}}, \bibinfo {author} {\bibfnamefont {K.}~\bibnamefont {{Kadota}}},\
  and\ \bibinfo {author} {\bibfnamefont {J.}~\bibnamefont {{Silk}}},\
  }\bibfield  {title} {\bibinfo {title} {{Constraint on Light Dipole Dark
  Matter from Helioseismology}},\ }\href
  {https://doi.org/10.1088/2041-8205/780/2/L15} {\bibfield  {journal} {\bibinfo
   {journal} {\apjl}\ }\textbf {\bibinfo {volume} {780}},\ \bibinfo {eid} {L15}
  (\bibinfo {year} {2014}{\natexlab{a}})},\ \Eprint
  {https://arxiv.org/abs/1310.0673} {arXiv:1310.0673} \BibitemShut {NoStop}%
\bibitem [{\citenamefont {{Lopes}}\ \emph
  {et~al.}(2014{\natexlab{b}})\citenamefont {{Lopes}}, \citenamefont
  {{Panci}},\ and\ \citenamefont {{Silk}}}]{2014ApJ...795..162L}%
  \BibitemOpen
  \bibfield  {author} {\bibinfo {author} {\bibfnamefont {I.}~\bibnamefont
  {{Lopes}}}, \bibinfo {author} {\bibfnamefont {P.}~\bibnamefont {{Panci}}},\
  and\ \bibinfo {author} {\bibfnamefont {J.}~\bibnamefont {{Silk}}},\
  }\bibfield  {title} {\bibinfo {title} {{Helioseismology with Long-range Dark
  Matter-Baryon Interactions}},\ }\href
  {https://doi.org/10.1088/0004-637X/795/2/162} {\bibfield  {journal} {\bibinfo
   {journal} {\apj}\ }\textbf {\bibinfo {volume} {795}},\ \bibinfo {eid} {162}
  (\bibinfo {year} {2014}{\natexlab{b}})},\ \Eprint
  {https://arxiv.org/abs/1402.0682} {arXiv:1402.0682} \BibitemShut {NoStop}%
\bibitem [{\citenamefont {{Lopes}}\ \emph {et~al.}(2019)\citenamefont
  {{Lopes}}, \citenamefont {{Lopes}},\ and\ \citenamefont
  {{Silk}}}]{2019ApJ...880L..25L}%
  \BibitemOpen
  \bibfield  {author} {\bibinfo {author} {\bibfnamefont {J.}~\bibnamefont
  {{Lopes}}}, \bibinfo {author} {\bibfnamefont {I.}~\bibnamefont {{Lopes}}},\
  and\ \bibinfo {author} {\bibfnamefont {J.}~\bibnamefont {{Silk}}},\
  }\bibfield  {title} {\bibinfo {title} {{Asteroseismology of Red Clump Stars
  as a Probe of the Dark Matter Content of the Galaxy Central Region}},\ }\href
  {https://doi.org/10.3847/2041-8213/ab2fdd} {\bibfield  {journal} {\bibinfo
  {journal} {\apjl}\ }\textbf {\bibinfo {volume} {880}},\ \bibinfo {eid} {L25}
  (\bibinfo {year} {2019})},\ \Eprint {https://arxiv.org/abs/1907.10089}
  {arXiv:1907.10089} \BibitemShut {NoStop}%
\bibitem [{\citenamefont {{Casanellas}}\ and\ \citenamefont
  {{Lopes}}(2011)}]{2011MNRAS.410..535C}%
  \BibitemOpen
  \bibfield  {author} {\bibinfo {author} {\bibfnamefont {J.}~\bibnamefont
  {{Casanellas}}}\ and\ \bibinfo {author} {\bibfnamefont {I.}~\bibnamefont
  {{Lopes}}},\ }\bibfield  {title} {\bibinfo {title} {Towards the use of
  asteroseismology to investigate the nature of dark matter},\ }\href
  {https://doi.org/10.1111/j.1365-2966.2010.17463.x} {\bibfield  {journal}
  {\bibinfo  {journal} {\mnras}\ }\textbf {\bibinfo {volume} {410}},\ \bibinfo
  {pages} {535} (\bibinfo {year} {2011})},\ \Eprint
  {https://arxiv.org/abs/1008.0646} {arXiv:1008.0646} \BibitemShut {NoStop}%
\bibitem [{\citenamefont {{Casanellas}}\ and\ \citenamefont
  {{Lopes}}(2013)}]{2013ApJ...765L..21C}%
  \BibitemOpen
  \bibfield  {author} {\bibinfo {author} {\bibfnamefont {J.}~\bibnamefont
  {{Casanellas}}}\ and\ \bibinfo {author} {\bibfnamefont {I.}~\bibnamefont
  {{Lopes}}},\ }\bibfield  {title} {\bibinfo {title} {First asteroseismic
  limits on the nature of dark matter},\ }\href
  {https://doi.org/10.1088/2041-8205/765/1/L21} {\bibfield  {journal} {\bibinfo
   {journal} {\apjl}\ }\textbf {\bibinfo {volume} {765}},\ \bibinfo {pages}
  {L21} (\bibinfo {year} {2013})},\ \Eprint {https://arxiv.org/abs/1212.2985}
  {arXiv:1212.2985} \BibitemShut {NoStop}%
\bibitem [{\citenamefont {{Rato}}\ \emph {et~al.}(2021)\citenamefont {{Rato}},
  \citenamefont {{Lopes}},\ and\ \citenamefont
  {{Lopes}}}]{2021MNRAS.507.3434R}%
  \BibitemOpen
  \bibfield  {author} {\bibinfo {author} {\bibfnamefont {J.}~\bibnamefont
  {{Rato}}}, \bibinfo {author} {\bibfnamefont {J.}~\bibnamefont {{Lopes}}},\
  and\ \bibinfo {author} {\bibfnamefont {I.}~\bibnamefont {{Lopes}}},\
  }\bibfield  {title} {\bibinfo {title} {{On asymmetric dark matter constraints
  from the asteroseismology of a subgiant star}},\ }\href
  {https://doi.org/10.1093/mnras/stab2372} {\bibfield  {journal} {\bibinfo
  {journal} {\mnras}\ }\textbf {\bibinfo {volume} {507}},\ \bibinfo {pages}
  {3434} (\bibinfo {year} {2021})},\ \Eprint {https://arxiv.org/abs/2108.11478}
  {arXiv:2108.11478} \BibitemShut {NoStop}%
\bibitem [{\citenamefont {{de Lavallaz}}\ and\ \citenamefont
  {{Fairbairn}}(2010)}]{2010PhRvD..81l3521D}%
  \BibitemOpen
  \bibfield  {author} {\bibinfo {author} {\bibfnamefont {A.}~\bibnamefont {{de
  Lavallaz}}}\ and\ \bibinfo {author} {\bibfnamefont {M.}~\bibnamefont
  {{Fairbairn}}},\ }\bibfield  {title} {\bibinfo {title} {Neutron stars as dark
  matter probes},\ }\href {https://doi.org/10.1103/PhysRevD.81.123521}
  {\bibfield  {journal} {\bibinfo  {journal} {\prd}\ }\textbf {\bibinfo
  {volume} {81}},\ \bibinfo {pages} {123521} (\bibinfo {year} {2010})},\
  \Eprint {https://arxiv.org/abs/0912.4264} {arXiv:0912.4264} \BibitemShut
  {NoStop}%
\bibitem [{\citenamefont {{Kouvaris}}\ and\ \citenamefont
  {{Tinyakov}}(2011)}]{2011PhRvD..83h3512K}%
  \BibitemOpen
  \bibfield  {author} {\bibinfo {author} {\bibfnamefont {C.}~\bibnamefont
  {{Kouvaris}}}\ and\ \bibinfo {author} {\bibfnamefont {P.}~\bibnamefont
  {{Tinyakov}}},\ }\bibfield  {title} {\bibinfo {title} {Constraining
  asymmetric dark matter through observations of compact stars},\ }\href
  {https://doi.org/10.1103/PhysRevD.83.083512} {\bibfield  {journal} {\bibinfo
  {journal} {\prd}\ }\textbf {\bibinfo {volume} {83}},\ \bibinfo {pages}
  {083512} (\bibinfo {year} {2011})},\ \Eprint
  {https://arxiv.org/abs/1012.2039} {arXiv:1012.2039} \BibitemShut {NoStop}%
\bibitem [{\citenamefont {{Bramante}}\ and\ \citenamefont
  {{Raj}}(2024)}]{2024PhR..1052....1B}%
  \BibitemOpen
  \bibfield  {author} {\bibinfo {author} {\bibfnamefont {J.}~\bibnamefont
  {{Bramante}}}\ and\ \bibinfo {author} {\bibfnamefont {N.}~\bibnamefont
  {{Raj}}},\ }\bibfield  {title} {\bibinfo {title} {Dark matter in compact
  stars},\ }\href {https://doi.org/10.1016/j.physrep.2023.12.001} {\bibfield
  {journal} {\bibinfo  {journal} {Phys. Rep.}\ }\textbf {\bibinfo {volume}
  {1052}},\ \bibinfo {pages} {1} (\bibinfo {year} {2024})},\ \Eprint
  {https://arxiv.org/abs/2307.14435} {arXiv:2307.14435} \BibitemShut {NoStop}%
\bibitem [{\citenamefont {{Panotopoulos}}\ \emph {et~al.}(2024)\citenamefont
  {{Panotopoulos}}, \citenamefont {{Rinc{\'o}n}},\ and\ \citenamefont
  {{Lopes}}}]{2024PhLB..85638901P}%
  \BibitemOpen
  \bibfield  {author} {\bibinfo {author} {\bibfnamefont {G.}~\bibnamefont
  {{Panotopoulos}}}, \bibinfo {author} {\bibfnamefont {{\'A}.}~\bibnamefont
  {{Rinc{\'o}n}}},\ and\ \bibinfo {author} {\bibfnamefont {I.}~\bibnamefont
  {{Lopes}}},\ }\bibfield  {title} {\bibinfo {title} {{Anisotropic dark energy
  stars within vanishing complexity factor formalism: Hydrostatic equilibrium,
  radial oscillations, and observational implications}},\ }\href
  {https://doi.org/10.1016/j.physletb.2024.138901} {\bibfield  {journal}
  {\bibinfo  {journal} {Physics Letters B}\ }\textbf {\bibinfo {volume}
  {856}},\ \bibinfo {eid} {138901} (\bibinfo {year} {2024})},\ \Eprint
  {https://arxiv.org/abs/2407.17335} {arXiv:2407.17335 [gr-qc]} \BibitemShut
  {NoStop}%
\bibitem [{\citenamefont {{Buras-Stubbs}}\ and\ \citenamefont
  {{Lopes}}(2026)}]{2026PhRvD.113d3049B}%
  \BibitemOpen
  \bibfield  {author} {\bibinfo {author} {\bibfnamefont {Z.}~\bibnamefont
  {{Buras-Stubbs}}}\ and\ \bibinfo {author} {\bibfnamefont {I.}~\bibnamefont
  {{Lopes}}},\ }\bibfield  {title} {\bibinfo {title} {{Rotational behavior of
  exotic compact objects}},\ }\href {https://doi.org/10.1103/fffq-myw5}
  {\bibfield  {journal} {\bibinfo  {journal} {\prd}\ }\textbf {\bibinfo
  {volume} {113}},\ \bibinfo {eid} {043049} (\bibinfo {year} {2026})},\ \Eprint
  {https://arxiv.org/abs/2602.06660} {arXiv:2602.06660 [astro-ph.HE]}
  \BibitemShut {NoStop}%
\bibitem [{\citenamefont {{Madelung}}(1927)}]{1927ZPhy...40..322M}%
  \BibitemOpen
  \bibfield  {author} {\bibinfo {author} {\bibfnamefont {E.}~\bibnamefont
  {{Madelung}}},\ }\bibfield  {title} {\bibinfo {title} {{Quantentheorie in
  hydrodynamischer Form}},\ }\href {https://doi.org/10.1007/BF01400372}
  {\bibfield  {journal} {\bibinfo  {journal} {\zphy}\ }\textbf {\bibinfo
  {volume} {40}},\ \bibinfo {pages} {322} (\bibinfo {year} {1927})}\BibitemShut
  {NoStop}%
\bibitem [{\citenamefont {{Bohm}}(1952)}]{1952PhRv...85..166B}%
  \BibitemOpen
  \bibfield  {author} {\bibinfo {author} {\bibfnamefont {D.}~\bibnamefont
  {{Bohm}}},\ }\bibfield  {title} {\bibinfo {title} {{A Suggested
  Interpretation of the Quantum Theory in Terms of ``Hidden'' Variables. I}},\
  }\href {https://doi.org/10.1103/PhysRev.85.166} {\bibfield  {journal}
  {\bibinfo  {journal} {Phys.~Rev.}\ }\textbf {\bibinfo {volume} {85}},\
  \bibinfo {pages} {166} (\bibinfo {year} {1952})}\BibitemShut {NoStop}%
\bibitem [{\citenamefont {{von Weizs{\"a}cker}}(1935)}]{1935ZPhy...96..431W}%
  \BibitemOpen
  \bibfield  {author} {\bibinfo {author} {\bibfnamefont {C.~F.}\ \bibnamefont
  {{von Weizs{\"a}cker}}},\ }\bibfield  {title} {\bibinfo {title} {{Zur Theorie
  der Kernmassen}},\ }\href {https://doi.org/10.1007/BF01337700} {\bibfield
  {journal} {\bibinfo  {journal} {\zphy}\ }\textbf {\bibinfo {volume} {96}},\
  \bibinfo {pages} {431} (\bibinfo {year} {1935})}\BibitemShut {NoStop}%
\bibitem [{\citenamefont {{Potekhin}}\ \emph {et~al.}(2025)\citenamefont
  {{Potekhin}}, \citenamefont {{Chugunov}}, \citenamefont {{Shchechilin}},\
  and\ \citenamefont {{Chamel}}}]{2025PhyU...68..691P}%
  \BibitemOpen
  \bibfield  {author} {\bibinfo {author} {\bibfnamefont {A.~Y.}\ \bibnamefont
  {{Potekhin}}}, \bibinfo {author} {\bibfnamefont {A.~I.}\ \bibnamefont
  {{Chugunov}}}, \bibinfo {author} {\bibfnamefont {N.~N.}\ \bibnamefont
  {{Shchechilin}}},\ and\ \bibinfo {author} {\bibfnamefont {N.}~\bibnamefont
  {{Chamel}}},\ }\bibfield  {title} {\bibinfo {title} {{On variational trial
  functions in extended Thomas--Fermi method}},\ }\href
  {https://doi.org/10.3367/UFNe.2024.11.039804} {\bibfield  {journal} {\bibinfo
   {journal} {Physics Uspekhi}\ }\textbf {\bibinfo {volume} {68}},\ \bibinfo
  {pages} {691} (\bibinfo {year} {2025})},\ \Eprint
  {https://arxiv.org/abs/2411.11021} {arXiv:2411.11021 [nucl-th]} \BibitemShut
  {NoStop}%
\bibitem [{\citenamefont {{Manfredi}}\ and\ \citenamefont
  {{Haas}}(2001)}]{2001PhRvB..64g5316M}%
  \BibitemOpen
  \bibfield  {author} {\bibinfo {author} {\bibfnamefont {G.}~\bibnamefont
  {{Manfredi}}}\ and\ \bibinfo {author} {\bibfnamefont {F.}~\bibnamefont
  {{Haas}}},\ }\bibfield  {title} {\bibinfo {title} {{Self-consistent fluid
  model for a quantum electron gas}},\ }\href
  {https://doi.org/10.1103/PhysRevB.64.075316} {\bibfield  {journal} {\bibinfo
  {journal} {\prb}\ }\textbf {\bibinfo {volume} {64}},\ \bibinfo {eid} {075316}
  (\bibinfo {year} {2001})},\ \Eprint {https://arxiv.org/abs/cond-mat/0203394}
  {arXiv:cond-mat/0203394} \BibitemShut {NoStop}%
\bibitem [{\citenamefont {{Michta}}\ \emph {et~al.}(2015)\citenamefont
  {{Michta}}, \citenamefont {{Graziani}},\ and\ \citenamefont
  {{Bonitz}}}]{2015CoPP...55..437M}%
  \BibitemOpen
  \bibfield  {author} {\bibinfo {author} {\bibfnamefont {D.}~\bibnamefont
  {{Michta}}}, \bibinfo {author} {\bibfnamefont {F.}~\bibnamefont
  {{Graziani}}},\ and\ \bibinfo {author} {\bibfnamefont {M.}~\bibnamefont
  {{Bonitz}}},\ }\bibfield  {title} {\bibinfo {title} {{Quantum Hydrodynamics
  for Plasmas: A Thomas--Fermi Theory Perspective}},\ }\href
  {https://doi.org/10.1002/ctpp.201500024} {\bibfield  {journal} {\bibinfo
  {journal} {\copp}\ }\textbf {\bibinfo {volume} {55}},\ \bibinfo {pages} {437}
  (\bibinfo {year} {2015})},\ \Eprint {https://arxiv.org/abs/1504.04973}
  {arXiv:1504.04973} \BibitemShut {NoStop}%
\bibitem [{\citenamefont {{Prsa}}\ \emph {et~al.}(2016)\citenamefont {{Prsa}},
  \citenamefont {{Harmanec}}, \citenamefont {{Torres}}, \citenamefont
  {{Mamajek}}, \citenamefont {{Asplund}}, \citenamefont {{Capitaine}},
  \citenamefont {{Christensen-Dalsgaard}}, \citenamefont {{Depagne}},
  \citenamefont {{Haberreiter}}, \citenamefont {{Hekker}} \emph
  {et~al.}}]{2016AJ....152...41P}%
  \BibitemOpen
  \bibfield  {author} {\bibinfo {author} {\bibfnamefont {A.}~\bibnamefont
  {{Prsa}}}, \bibinfo {author} {\bibfnamefont {P.}~\bibnamefont {{Harmanec}}},
  \bibinfo {author} {\bibfnamefont {G.}~\bibnamefont {{Torres}}}, \bibinfo
  {author} {\bibfnamefont {E.}~\bibnamefont {{Mamajek}}}, \bibinfo {author}
  {\bibfnamefont {M.}~\bibnamefont {{Asplund}}}, \bibinfo {author}
  {\bibfnamefont {N.}~\bibnamefont {{Capitaine}}}, \bibinfo {author}
  {\bibfnamefont {J.}~\bibnamefont {{Christensen-Dalsgaard}}}, \bibinfo
  {author} {\bibfnamefont {E.}~\bibnamefont {{Depagne}}}, \bibinfo {author}
  {\bibfnamefont {M.}~\bibnamefont {{Haberreiter}}}, \bibinfo {author}
  {\bibfnamefont {S.}~\bibnamefont {{Hekker}}}, \emph {et~al.},\ }\bibfield
  {title} {\bibinfo {title} {{Nominal Values for Selected Solar and Planetary
  Quantities: IAU 2015 Resolution B3}},\ }\href
  {https://doi.org/10.3847/0004-6256/152/2/41} {\bibfield  {journal} {\bibinfo
  {journal} {AJ}\ }\textbf {\bibinfo {volume} {152}},\ \bibinfo {eid} {41}
  (\bibinfo {year} {2016})},\ \Eprint {https://arxiv.org/abs/1605.09788}
  {arXiv:1605.09788 [astro-ph.SR]} \BibitemShut {NoStop}%
\bibitem [{\citenamefont {{Serot}}\ and\ \citenamefont
  {{Walecka}}(1997)}]{1997IJMPE...6..515S}%
  \BibitemOpen
  \bibfield  {author} {\bibinfo {author} {\bibfnamefont {B.~D.}\ \bibnamefont
  {{Serot}}}\ and\ \bibinfo {author} {\bibfnamefont {J.~D.}\ \bibnamefont
  {{Walecka}}},\ }\bibfield  {title} {\bibinfo {title} {{Recent Progress in
  Quantum Hadrodynamics}},\ }\href {https://doi.org/10.1142/S0218301397000299}
  {\bibfield  {journal} {\bibinfo  {journal} {Int.~J.~Mod.~Phys.~E}\ }\textbf
  {\bibinfo {volume} {6}},\ \bibinfo {pages} {515} (\bibinfo {year} {1997})},\
  \Eprint {https://arxiv.org/abs/nucl-th/9701058} {arXiv:nucl-th/9701058}
  \BibitemShut {NoStop}%
\bibitem [{\citenamefont {{Weinberg}}(1972)}]{Weinberg1972gc}%
  \BibitemOpen
  \bibfield  {author} {\bibinfo {author} {\bibfnamefont {S.}~\bibnamefont
  {{Weinberg}}},\ }\href
  {https://ui.adsabs.harvard.edu/abs/1972gcpa.book.....W} {\emph {\bibinfo
  {title} {{Gravitation and Cosmology: Principles and Applications of the
  General Theory of Relativity}}}}\ (\bibinfo  {publisher} {John Wiley \&
  Sons},\ \bibinfo {address} {New York},\ \bibinfo {year} {1972})\BibitemShut
  {NoStop}%
\bibitem [{\citenamefont {{Will}}(2018)}]{Will2018teg}%
  \BibitemOpen
  \bibfield  {author} {\bibinfo {author} {\bibfnamefont {C.~M.}\ \bibnamefont
  {{Will}}},\ }\href {https://doi.org/10.1017/9781316338612} {\emph {\bibinfo
  {title} {{Theory and Experiment in Gravitational Physics}}}},\ \bibinfo
  {edition} {2nd}\ ed.\ (\bibinfo  {publisher} {Cambridge University Press},\
  \bibinfo {address} {Cambridge},\ \bibinfo {year} {2018})\BibitemShut
  {NoStop}%
\bibitem [{\citenamefont {{Moldabekov}}\ \emph
  {et~al.}(2018{\natexlab{a}})\citenamefont {{Moldabekov}}, \citenamefont
  {{Bonitz}},\ and\ \citenamefont {{Ramazanov}}}]{2018PhPl...25c2115M}%
  \BibitemOpen
  \bibfield  {author} {\bibinfo {author} {\bibfnamefont {Z.~A.}\ \bibnamefont
  {{Moldabekov}}}, \bibinfo {author} {\bibfnamefont {M.}~\bibnamefont
  {{Bonitz}}},\ and\ \bibinfo {author} {\bibfnamefont {T.~S.}\ \bibnamefont
  {{Ramazanov}}},\ }\bibfield  {title} {\bibinfo {title} {{Theoretical
  foundations of quantum hydrodynamics for plasmas}},\ }\href
  {https://doi.org/10.1063/1.5003910} {\bibfield  {journal} {\bibinfo
  {journal} {\pop}\ }\textbf {\bibinfo {volume} {25}},\ \bibinfo {pages}
  {031903} (\bibinfo {year} {2018}{\natexlab{a}})}\BibitemShut {NoStop}%
\bibitem [{\citenamefont {{Chandrasekhar}}(1939)}]{1939isss.book.....C}%
  \BibitemOpen
  \bibfield  {author} {\bibinfo {author} {\bibfnamefont {S.}~\bibnamefont
  {{Chandrasekhar}}},\ }\href
  {https://ui.adsabs.harvard.edu/abs/1939isss.book.....C} {\emph {\bibinfo
  {title} {{An Introduction to the Study of Stellar Structure}}}}\ (\bibinfo
  {publisher} {University of Chicago Press},\ \bibinfo {address} {Chicago},\
  \bibinfo {year} {1939})\BibitemShut {NoStop}%
\bibitem [{\citenamefont {{Shapiro}}\ and\ \citenamefont
  {{Teukolsky}}(1983)}]{1983bhwd.book.....S}%
  \BibitemOpen
  \bibfield  {author} {\bibinfo {author} {\bibfnamefont {S.~L.}\ \bibnamefont
  {{Shapiro}}}\ and\ \bibinfo {author} {\bibfnamefont {S.~A.}\ \bibnamefont
  {{Teukolsky}}},\ }\href {https://doi.org/10.1002/9783527617661} {\emph
  {\bibinfo {title} {{Black Holes, White Dwarfs, and Neutron Stars: The Physics
  of Compact Objects}}}}\ (\bibinfo  {publisher} {Wiley},\ \bibinfo {address}
  {New York},\ \bibinfo {year} {1983})\BibitemShut {NoStop}%
\bibitem [{\citenamefont {{Guzm{\'a}n}}\ and\ \citenamefont
  {{Ure{\~n}a-L{\'o}pez}}(2004)}]{2004PhRvD..69l4033G}%
  \BibitemOpen
  \bibfield  {author} {\bibinfo {author} {\bibfnamefont {F.~S.}\ \bibnamefont
  {{Guzm{\'a}n}}}\ and\ \bibinfo {author} {\bibfnamefont {L.~A.}\ \bibnamefont
  {{Ure{\~n}a-L{\'o}pez}}},\ }\bibfield  {title} {\bibinfo {title} {{Evolution
  of the Schr{\"o}dinger-Newton system for a self-gravitating scalar field}},\
  }\href {https://doi.org/10.1103/PhysRevD.69.124033} {\bibfield  {journal}
  {\bibinfo  {journal} {\prd}\ }\textbf {\bibinfo {volume} {69}},\ \bibinfo
  {eid} {124033} (\bibinfo {year} {2004})},\ \Eprint
  {https://arxiv.org/abs/gr-qc/0404014} {arXiv:gr-qc/0404014 [gr-qc]}
  \BibitemShut {NoStop}%
\bibitem [{\citenamefont {{Colpi}}\ \emph {et~al.}(1986)\citenamefont
  {{Colpi}}, \citenamefont {{Shapiro}},\ and\ \citenamefont
  {{Wasserman}}}]{1986PhRvL..57.2485C}%
  \BibitemOpen
  \bibfield  {author} {\bibinfo {author} {\bibfnamefont {M.}~\bibnamefont
  {{Colpi}}}, \bibinfo {author} {\bibfnamefont {S.~L.}\ \bibnamefont
  {{Shapiro}}},\ and\ \bibinfo {author} {\bibfnamefont {I.}~\bibnamefont
  {{Wasserman}}},\ }\bibfield  {title} {\bibinfo {title} {{Boson Stars:
  Gravitational Equilibria of Self-Interacting Scalar Fields}},\ }\href
  {https://doi.org/10.1103/PhysRevLett.57.2485} {\bibfield  {journal} {\bibinfo
   {journal} {\prl}\ }\textbf {\bibinfo {volume} {57}},\ \bibinfo {pages}
  {2485} (\bibinfo {year} {1986})}\BibitemShut {NoStop}%
\bibitem [{\citenamefont {{Moldabekov}}\ \emph {et~al.}(2015)\citenamefont
  {{Moldabekov}}, \citenamefont {{Schoof}}, \citenamefont {{Ludwig}},
  \citenamefont {{Bonitz}},\ and\ \citenamefont
  {{Ramazanov}}}]{2015PhPl...22d4501M}%
  \BibitemOpen
  \bibfield  {author} {\bibinfo {author} {\bibfnamefont {Z.~A.}\ \bibnamefont
  {{Moldabekov}}}, \bibinfo {author} {\bibfnamefont {T.}~\bibnamefont
  {{Schoof}}}, \bibinfo {author} {\bibfnamefont {P.}~\bibnamefont {{Ludwig}}},
  \bibinfo {author} {\bibfnamefont {M.}~\bibnamefont {{Bonitz}}},\ and\
  \bibinfo {author} {\bibfnamefont {T.~S.}\ \bibnamefont {{Ramazanov}}},\
  }\bibfield  {title} {\bibinfo {title} {{Statically screened ion potential and
  Bohm potential in a quantum plasma}},\ }\href
  {https://doi.org/10.1063/1.4916571} {\bibfield  {journal} {\bibinfo
  {journal} {\pop}\ }\textbf {\bibinfo {volume} {22}},\ \bibinfo {pages}
  {044501} (\bibinfo {year} {2015})},\ \Eprint
  {https://arxiv.org/abs/1508.01120} {arXiv:1508.01120} \BibitemShut {NoStop}%
\bibitem [{\citenamefont {{Moldabekov}}\ \emph
  {et~al.}(2018{\natexlab{b}})\citenamefont {{Moldabekov}}, \citenamefont
  {{Bonitz}},\ and\ \citenamefont {{Ramazanov}}}]{2018CoPP...58..290M}%
  \BibitemOpen
  \bibfield  {author} {\bibinfo {author} {\bibfnamefont {Z.~A.}\ \bibnamefont
  {{Moldabekov}}}, \bibinfo {author} {\bibfnamefont {M.}~\bibnamefont
  {{Bonitz}}},\ and\ \bibinfo {author} {\bibfnamefont {T.~S.}\ \bibnamefont
  {{Ramazanov}}},\ }\bibfield  {title} {\bibinfo {title} {{Gradient correction
  and Bohm potential for two- and one-dimensional electron gases at a finite
  temperature}},\ }\href {https://doi.org/10.1002/ctpp.201700113} {\bibfield
  {journal} {\bibinfo  {journal} {\copp}\ }\textbf {\bibinfo {volume} {58}},\
  \bibinfo {pages} {290} (\bibinfo {year} {2018}{\natexlab{b}})},\ \Eprint
  {https://arxiv.org/abs/1709.05310} {arXiv:1709.05310} \BibitemShut {NoStop}%
\bibitem [{\citenamefont {{Wyatt}}(2005)}]{Wyatt:2005uc}%
  \BibitemOpen
  \bibfield  {author} {\bibinfo {author} {\bibfnamefont {R.~E.}\ \bibnamefont
  {{Wyatt}}},\ }\href {https://doi.org/10.1007/0-387-28145-2} {\emph {\bibinfo
  {title} {{Quantum Dynamics with Trajectories. Introduction to quantum
  hydrodynamics}}}}\ (\bibinfo  {publisher} {Springer},\ \bibinfo {year}
  {2005})\BibitemShut {NoStop}%
\bibitem [{\citenamefont {{Gardner}}\ and\ \citenamefont
  {{Ringhofer}}(1996)}]{1996PhRvE..53..157G}%
  \BibitemOpen
  \bibfield  {author} {\bibinfo {author} {\bibfnamefont {C.~L.}\ \bibnamefont
  {{Gardner}}}\ and\ \bibinfo {author} {\bibfnamefont {C.}~\bibnamefont
  {{Ringhofer}}},\ }\bibfield  {title} {\bibinfo {title} {{Smooth quantum
  potential for the hydrodynamic model}},\ }\href
  {https://doi.org/10.1103/PhysRevE.53.157} {\bibfield  {journal} {\bibinfo
  {journal} {\pre}\ }\textbf {\bibinfo {volume} {53}},\ \bibinfo {pages} {157}
  (\bibinfo {year} {1996})}\BibitemShut {NoStop}%
\bibitem [{\citenamefont {{Chavanis}}(2020)}]{2020EPJP..135..290C}%
  \BibitemOpen
  \bibfield  {author} {\bibinfo {author} {\bibfnamefont {P.-H.}\ \bibnamefont
  {{Chavanis}}},\ }\bibfield  {title} {\bibinfo {title} {{Statistical mechanics
  of self-gravitating systems in general relativity: I. The quantum Fermi
  gas}},\ }\href {https://doi.org/10.1140/epjp/s13360-020-00268-0} {\bibfield
  {journal} {\bibinfo  {journal} {Eur. Phys. J. Plus}\ }\textbf {\bibinfo
  {volume} {135}},\ \bibinfo {pages} {290} (\bibinfo {year} {2020})},\ \Eprint
  {https://arxiv.org/abs/1908.10806} {arXiv:1908.10806} \BibitemShut {NoStop}%
\bibitem [{\citenamefont {{Seidel}}\ and\ \citenamefont
  {{Suen}}(1991)}]{1991PhRvL..66.1659S}%
  \BibitemOpen
  \bibfield  {author} {\bibinfo {author} {\bibfnamefont {E.}~\bibnamefont
  {{Seidel}}}\ and\ \bibinfo {author} {\bibfnamefont {W.-M.}\ \bibnamefont
  {{Suen}}},\ }\bibfield  {title} {\bibinfo {title} {{Oscillating Soliton
  Stars}},\ }\href {https://doi.org/10.1103/PhysRevLett.66.1659} {\bibfield
  {journal} {\bibinfo  {journal} {\prl}\ }\textbf {\bibinfo {volume} {66}},\
  \bibinfo {pages} {1659} (\bibinfo {year} {1991})}\BibitemShut {NoStop}%
\bibitem [{\citenamefont {{Seidel}}\ and\ \citenamefont
  {{Suen}}(1994)}]{1994PhRvL..72.2516S}%
  \BibitemOpen
  \bibfield  {author} {\bibinfo {author} {\bibfnamefont {E.}~\bibnamefont
  {{Seidel}}}\ and\ \bibinfo {author} {\bibfnamefont {W.-M.}\ \bibnamefont
  {{Suen}}},\ }\bibfield  {title} {\bibinfo {title} {{Formation of Solitonic
  Stars through Gravitational Cooling}},\ }\href
  {https://doi.org/10.1103/PhysRevLett.72.2516} {\bibfield  {journal} {\bibinfo
   {journal} {\prl}\ }\textbf {\bibinfo {volume} {72}},\ \bibinfo {pages}
  {2516} (\bibinfo {year} {1994})},\ \Eprint
  {https://arxiv.org/abs/gr-qc/9309015} {arXiv:gr-qc/9309015 [gr-qc]}
  \BibitemShut {NoStop}%
\bibitem [{\citenamefont {{Maggiore}}(2007)}]{Maggiore2007gw}%
  \BibitemOpen
  \bibfield  {author} {\bibinfo {author} {\bibfnamefont {M.}~\bibnamefont
  {{Maggiore}}},\ }\href
  {https://doi.org/10.1093/acprof:oso/9780198570745.001.0001} {\emph {\bibinfo
  {title} {{Gravitational Waves: Volume 1: Theory and Experiments}}}}\
  (\bibinfo  {publisher} {Oxford University Press},\ \bibinfo {address}
  {Oxford},\ \bibinfo {year} {2007})\BibitemShut {NoStop}%
\bibitem [{\citenamefont {{Amaro-Seoane}}\ \emph {et~al.}(2017)\citenamefont
  {{Amaro-Seoane}}, \citenamefont {{Audley}}, \citenamefont {{Babak}} \emph
  {et~al.}}]{2017arXiv170200786A}%
  \BibitemOpen
  \bibfield  {author} {\bibinfo {author} {\bibfnamefont {P.}~\bibnamefont
  {{Amaro-Seoane}}}, \bibinfo {author} {\bibfnamefont {H.}~\bibnamefont
  {{Audley}}}, \bibinfo {author} {\bibfnamefont {S.}~\bibnamefont {{Babak}}},
  \emph {et~al.},\ }\bibfield  {title} {\bibinfo {title} {{Laser Interferometer
  Space Antenna}},\ }\href {https://doi.org/10.48550/arXiv.1702.00786}
  {\bibfield  {journal} {\bibinfo  {journal} {arXiv e-prints}\ ,\ \bibinfo
  {eid} {arXiv:1702.00786}} (\bibinfo {year} {2017})},\ \Eprint
  {https://arxiv.org/abs/1702.00786} {arXiv:1702.00786 [astro-ph.IM]}
  \BibitemShut {NoStop}%
\bibitem [{\citenamefont {{Punturo}}\ \emph {et~al.}(2010)\citenamefont
  {{Punturo}}, \citenamefont {{Abernathy}}, \citenamefont {{Acernese}} \emph
  {et~al.}}]{2010CQGra..27s4002P}%
  \BibitemOpen
  \bibfield  {author} {\bibinfo {author} {\bibfnamefont {M.}~\bibnamefont
  {{Punturo}}}, \bibinfo {author} {\bibfnamefont {M.}~\bibnamefont
  {{Abernathy}}}, \bibinfo {author} {\bibfnamefont {F.}~\bibnamefont
  {{Acernese}}}, \emph {et~al.},\ }\bibfield  {title} {\bibinfo {title} {{The
  Einstein Telescope: a third-generation gravitational wave observatory}},\
  }\href {https://doi.org/10.1088/0264-9381/27/19/194002} {\bibfield  {journal}
  {\bibinfo  {journal} {Class.~Quantum Grav.}\ }\textbf {\bibinfo {volume}
  {27}},\ \bibinfo {eid} {194002} (\bibinfo {year} {2010})}\BibitemShut
  {NoStop}%
\bibitem [{\citenamefont {{Reitze}}\ \emph {et~al.}(2019)\citenamefont
  {{Reitze}}, \citenamefont {{Adhikari}}, \citenamefont {{Ballmer}} \emph
  {et~al.}}]{2019BAAS...51g..35R}%
  \BibitemOpen
  \bibfield  {author} {\bibinfo {author} {\bibfnamefont {D.}~\bibnamefont
  {{Reitze}}}, \bibinfo {author} {\bibfnamefont {R.~X.}\ \bibnamefont
  {{Adhikari}}}, \bibinfo {author} {\bibfnamefont {S.}~\bibnamefont
  {{Ballmer}}}, \emph {et~al.},\ }\bibfield  {title} {\bibinfo {title} {{Cosmic
  Explorer: The U.S. Contribution to Gravitational-Wave Astronomy beyond
  LIGO}},\ }\href {https://ui.adsabs.harvard.edu/abs/2019BAAS...51g..35R}
  {\bibfield  {journal} {\bibinfo  {journal} {Bull.~Am.~Astron.~Soc.}\ }\textbf
  {\bibinfo {volume} {51}},\ \bibinfo {pages} {35} (\bibinfo {year} {2019})},\
  \Eprint {https://arxiv.org/abs/1907.04833} {arXiv:1907.04833 [astro-ph.IM]}
  \BibitemShut {NoStop}%
\bibitem [{\citenamefont {{Kawamura}}\ \emph {et~al.}(2011)\citenamefont
  {{Kawamura}}, \citenamefont {{Ando}}, \citenamefont {{Seto}} \emph
  {et~al.}}]{2011CQGra..28i4011K}%
  \BibitemOpen
  \bibfield  {author} {\bibinfo {author} {\bibfnamefont {S.}~\bibnamefont
  {{Kawamura}}}, \bibinfo {author} {\bibfnamefont {M.}~\bibnamefont {{Ando}}},
  \bibinfo {author} {\bibfnamefont {N.}~\bibnamefont {{Seto}}}, \emph
  {et~al.},\ }\bibfield  {title} {\bibinfo {title} {{The Japanese space
  gravitational wave antenna: DECIGO}},\ }\href
  {https://doi.org/10.1088/0264-9381/28/9/094011} {\bibfield  {journal}
  {\bibinfo  {journal} {Class.~Quantum Grav.}\ }\textbf {\bibinfo {volume}
  {28}},\ \bibinfo {eid} {094011} (\bibinfo {year} {2011})}\BibitemShut
  {NoStop}%
\bibitem [{\citenamefont {{Harry}}\ \emph {et~al.}(2006)\citenamefont
  {{Harry}}, \citenamefont {{Fritschel}}, \citenamefont {{Shaddock}},
  \citenamefont {{Folkner}},\ and\ \citenamefont
  {{Phinney}}}]{2006CQGra..23.4887H}%
  \BibitemOpen
  \bibfield  {author} {\bibinfo {author} {\bibfnamefont {G.~M.}\ \bibnamefont
  {{Harry}}}, \bibinfo {author} {\bibfnamefont {P.}~\bibnamefont
  {{Fritschel}}}, \bibinfo {author} {\bibfnamefont {D.~A.}\ \bibnamefont
  {{Shaddock}}}, \bibinfo {author} {\bibfnamefont {W.}~\bibnamefont
  {{Folkner}}},\ and\ \bibinfo {author} {\bibfnamefont {E.~S.}\ \bibnamefont
  {{Phinney}}},\ }\bibfield  {title} {\bibinfo {title} {{Laser interferometry
  for the Big Bang Observer}},\ }\href
  {https://doi.org/10.1088/0264-9381/23/15/008} {\bibfield  {journal} {\bibinfo
   {journal} {Class.~Quantum Grav.}\ }\textbf {\bibinfo {volume} {23}},\
  \bibinfo {pages} {4887} (\bibinfo {year} {2006})}\BibitemShut {NoStop}%
\bibitem [{\citenamefont {{Udalski}}\ \emph {et~al.}(2015)\citenamefont
  {{Udalski}}, \citenamefont {{Szyma{\'n}ski}},\ and\ \citenamefont
  {{Szyma{\'n}ski}}}]{2015AcA....65....1U}%
  \BibitemOpen
  \bibfield  {author} {\bibinfo {author} {\bibfnamefont {A.}~\bibnamefont
  {{Udalski}}}, \bibinfo {author} {\bibfnamefont {M.~K.}\ \bibnamefont
  {{Szyma{\'n}ski}}},\ and\ \bibinfo {author} {\bibfnamefont {G.}~\bibnamefont
  {{Szyma{\'n}ski}}},\ }\bibfield  {title} {\bibinfo {title} {{OGLE-IV: Fourth
  Phase of the Optical Gravitational Lensing Experiment}},\ }\href@noop {}
  {\bibfield  {journal} {\bibinfo  {journal} {Acta Astron.}\ }\textbf {\bibinfo
  {volume} {65}},\ \bibinfo {pages} {1} (\bibinfo {year} {2015})},\ \Eprint
  {https://arxiv.org/abs/1504.05966} {arXiv:1504.05966} \BibitemShut {NoStop}%
\bibitem [{\citenamefont {{Mr{\'o}z}}\ \emph {et~al.}(2017)\citenamefont
  {{Mr{\'o}z}}, \citenamefont {{Udalski}}, \citenamefont {{Skowron}},
  \citenamefont {{Poleski}}, \citenamefont {{Koz{\l}owski}}, \citenamefont
  {{Szyma{\'n}ski}}, \citenamefont {{Soszy{\'n}ski}}, \citenamefont
  {{Wyrzykowski}}, \citenamefont {{Pietrukowicz}}, \citenamefont {{Ulaczyk}}
  \emph {et~al.}}]{2017Natur.548..183M}%
  \BibitemOpen
  \bibfield  {author} {\bibinfo {author} {\bibfnamefont {P.}~\bibnamefont
  {{Mr{\'o}z}}}, \bibinfo {author} {\bibfnamefont {A.}~\bibnamefont
  {{Udalski}}}, \bibinfo {author} {\bibfnamefont {J.}~\bibnamefont
  {{Skowron}}}, \bibinfo {author} {\bibfnamefont {R.}~\bibnamefont
  {{Poleski}}}, \bibinfo {author} {\bibfnamefont {S.}~\bibnamefont
  {{Koz{\l}owski}}}, \bibinfo {author} {\bibfnamefont {M.~K.}\ \bibnamefont
  {{Szyma{\'n}ski}}}, \bibinfo {author} {\bibfnamefont {I.}~\bibnamefont
  {{Soszy{\'n}ski}}}, \bibinfo {author} {\bibfnamefont {{\L}.}~\bibnamefont
  {{Wyrzykowski}}}, \bibinfo {author} {\bibfnamefont {P.}~\bibnamefont
  {{Pietrukowicz}}}, \bibinfo {author} {\bibfnamefont {K.}~\bibnamefont
  {{Ulaczyk}}}, \emph {et~al.},\ }\bibfield  {title} {\bibinfo {title} {{No
  large population of unbound or wide-orbit Jupiter-mass planets}},\ }\href
  {https://doi.org/10.1038/nature23276} {\bibfield  {journal} {\bibinfo
  {journal} {Nature}\ }\textbf {\bibinfo {volume} {548}},\ \bibinfo {pages}
  {183} (\bibinfo {year} {2017})},\ \Eprint {https://arxiv.org/abs/1707.07634}
  {arXiv:1707.07634} \BibitemShut {NoStop}%
\bibitem [{\citenamefont {{Bond}}\ \emph {et~al.}(2001)\citenamefont {{Bond}},
  \citenamefont {{Abe}}, \citenamefont {{Dodd}} \emph
  {et~al.}}]{2001MNRAS.327..868B}%
  \BibitemOpen
  \bibfield  {author} {\bibinfo {author} {\bibfnamefont {I.~A.}\ \bibnamefont
  {{Bond}}}, \bibinfo {author} {\bibfnamefont {F.}~\bibnamefont {{Abe}}},
  \bibinfo {author} {\bibfnamefont {R.~J.}\ \bibnamefont {{Dodd}}}, \emph
  {et~al.},\ }\bibfield  {title} {\bibinfo {title} {{Real-time difference
  imaging analysis of MOA Galactic bulge observations during 2000}},\ }\href
  {https://doi.org/10.1046/j.1365-8711.2001.04776.x} {\bibfield  {journal}
  {\bibinfo  {journal} {\mnras}\ }\textbf {\bibinfo {volume} {327}},\ \bibinfo
  {pages} {868} (\bibinfo {year} {2001})},\ \Eprint
  {https://arxiv.org/abs/astro-ph/0104028} {arXiv:astro-ph/0104028}
  \BibitemShut {NoStop}%
\bibitem [{\citenamefont {{Alcock}}\ \emph {et~al.}(2000)\citenamefont
  {{Alcock}}, \citenamefont {{Allsman}}, \citenamefont {{Alves}}, \citenamefont
  {{Axelrod}}, \citenamefont {{Becker}}, \citenamefont {{Bennett}},
  \citenamefont {{Cook}}, \citenamefont {{Dalal}}, \citenamefont {{Drake}},
  \citenamefont {{Freeman}} \emph {et~al.}}]{2000ApJ...542..281A}%
  \BibitemOpen
  \bibfield  {author} {\bibinfo {author} {\bibfnamefont {C.}~\bibnamefont
  {{Alcock}}}, \bibinfo {author} {\bibfnamefont {R.~A.}\ \bibnamefont
  {{Allsman}}}, \bibinfo {author} {\bibfnamefont {D.~R.}\ \bibnamefont
  {{Alves}}}, \bibinfo {author} {\bibfnamefont {T.~S.}\ \bibnamefont
  {{Axelrod}}}, \bibinfo {author} {\bibfnamefont {A.~C.}\ \bibnamefont
  {{Becker}}}, \bibinfo {author} {\bibfnamefont {D.~P.}\ \bibnamefont
  {{Bennett}}}, \bibinfo {author} {\bibfnamefont {K.~H.}\ \bibnamefont
  {{Cook}}}, \bibinfo {author} {\bibfnamefont {N.}~\bibnamefont {{Dalal}}},
  \bibinfo {author} {\bibfnamefont {A.~J.}\ \bibnamefont {{Drake}}}, \bibinfo
  {author} {\bibfnamefont {K.~C.}\ \bibnamefont {{Freeman}}}, \emph {et~al.},\
  }\bibfield  {title} {\bibinfo {title} {{The MACHO Project: Microlensing
  Results from 5.7 Years of Large Magellanic Cloud Observations}},\ }\href
  {https://doi.org/10.1086/309512} {\bibfield  {journal} {\bibinfo  {journal}
  {\apj}\ }\textbf {\bibinfo {volume} {542}},\ \bibinfo {pages} {281} (\bibinfo
  {year} {2000})},\ \Eprint {https://arxiv.org/abs/astro-ph/0001272}
  {arXiv:astro-ph/0001272} \BibitemShut {NoStop}%
\bibitem [{\citenamefont {{Tisserand}}\ \emph {et~al.}(2007)\citenamefont
  {{Tisserand}}, \citenamefont {{Le Guillou}}, \citenamefont {{Afonso}},
  \citenamefont {{Albert}}, \citenamefont {{Andersen}}, \citenamefont
  {{Ansari}}, \citenamefont {{Aubourg}}, \citenamefont {{Bareyre}},
  \citenamefont {{Beaulieu}}, \citenamefont {{Charlot}} \emph
  {et~al.}}]{2007A&A...469..387T}%
  \BibitemOpen
  \bibfield  {author} {\bibinfo {author} {\bibfnamefont {P.}~\bibnamefont
  {{Tisserand}}}, \bibinfo {author} {\bibfnamefont {L.}~\bibnamefont {{Le
  Guillou}}}, \bibinfo {author} {\bibfnamefont {C.}~\bibnamefont {{Afonso}}},
  \bibinfo {author} {\bibfnamefont {J.~N.}\ \bibnamefont {{Albert}}}, \bibinfo
  {author} {\bibfnamefont {J.}~\bibnamefont {{Andersen}}}, \bibinfo {author}
  {\bibfnamefont {R.}~\bibnamefont {{Ansari}}}, \bibinfo {author}
  {\bibfnamefont {{\'E}.}~\bibnamefont {{Aubourg}}}, \bibinfo {author}
  {\bibfnamefont {P.}~\bibnamefont {{Bareyre}}}, \bibinfo {author}
  {\bibfnamefont {J.~P.}\ \bibnamefont {{Beaulieu}}}, \bibinfo {author}
  {\bibfnamefont {X.}~\bibnamefont {{Charlot}}}, \emph {et~al.},\ }\bibfield
  {title} {\bibinfo {title} {{Limits on the Macho content of the Galactic Halo
  from the EROS-2 Survey of the Magellanic Clouds}},\ }\href
  {https://doi.org/10.1051/0004-6361:20066017} {\bibfield  {journal} {\bibinfo
  {journal} {\aap}\ }\textbf {\bibinfo {volume} {469}},\ \bibinfo {pages} {387}
  (\bibinfo {year} {2007})},\ \Eprint {https://arxiv.org/abs/astro-ph/0607207}
  {arXiv:astro-ph/0607207} \BibitemShut {NoStop}%
\bibitem [{\citenamefont {{Niikura}}\ \emph {et~al.}(2018)\citenamefont
  {{Niikura}}, \citenamefont {{Takada}}, \citenamefont {{Yasuda}},
  \citenamefont {{Lupton}}, \citenamefont {{Sumi}}, \citenamefont {{More}},
  \citenamefont {{Kurita}}, \citenamefont {{Mukae}},\ and\ \citenamefont
  {{Oguri}}}]{2018PhRvD..97b3518N}%
  \BibitemOpen
  \bibfield  {author} {\bibinfo {author} {\bibfnamefont {H.}~\bibnamefont
  {{Niikura}}}, \bibinfo {author} {\bibfnamefont {M.}~\bibnamefont {{Takada}}},
  \bibinfo {author} {\bibfnamefont {N.}~\bibnamefont {{Yasuda}}}, \bibinfo
  {author} {\bibfnamefont {R.~H.}\ \bibnamefont {{Lupton}}}, \bibinfo {author}
  {\bibfnamefont {T.}~\bibnamefont {{Sumi}}}, \bibinfo {author} {\bibfnamefont
  {S.}~\bibnamefont {{More}}}, \bibinfo {author} {\bibfnamefont
  {T.}~\bibnamefont {{Kurita}}}, \bibinfo {author} {\bibfnamefont
  {S.}~\bibnamefont {{Mukae}}},\ and\ \bibinfo {author} {\bibfnamefont
  {M.}~\bibnamefont {{Oguri}}},\ }\bibfield  {title} {\bibinfo {title}
  {{Microlensing constraints on primordial black holes with Subaru/HSC
  Andromeda observations}},\ }\href
  {https://doi.org/10.1103/PhysRevD.97.023518} {\bibfield  {journal} {\bibinfo
  {journal} {\prd}\ }\textbf {\bibinfo {volume} {97}},\ \bibinfo {pages}
  {023518} (\bibinfo {year} {2018})},\ \Eprint
  {https://arxiv.org/abs/1701.02151} {arXiv:1701.02151} \BibitemShut {NoStop}%
\bibitem [{\citenamefont {{Lam}}\ \emph {et~al.}(2022)\citenamefont {{Lam}},
  \citenamefont {{Lu}}, \citenamefont {{Udalski}}, \citenamefont {{Bond}},
  \citenamefont {{Bennett}}, \citenamefont {{Skowron}}, \citenamefont
  {{Mr{\'o}z}},\ and\ \citenamefont {{Poleski}}}]{2022ApJ...933...83L}%
  \BibitemOpen
  \bibfield  {author} {\bibinfo {author} {\bibfnamefont {C.~Y.}\ \bibnamefont
  {{Lam}}}, \bibinfo {author} {\bibfnamefont {J.~R.}\ \bibnamefont {{Lu}}},
  \bibinfo {author} {\bibfnamefont {A.}~\bibnamefont {{Udalski}}}, \bibinfo
  {author} {\bibfnamefont {I.}~\bibnamefont {{Bond}}}, \bibinfo {author}
  {\bibfnamefont {D.~P.}\ \bibnamefont {{Bennett}}}, \bibinfo {author}
  {\bibfnamefont {J.}~\bibnamefont {{Skowron}}}, \bibinfo {author}
  {\bibfnamefont {P.}~\bibnamefont {{Mr{\'o}z}}},\ and\ \bibinfo {author}
  {\bibfnamefont {R.}~\bibnamefont {{Poleski}}},\ }\bibfield  {title} {\bibinfo
  {title} {{An Isolated Mass-gap Black Hole or Neutron Star Detected with
  Astrometric Microlensing}},\ }\href
  {https://doi.org/10.3847/2041-8213/ac7442} {\bibfield  {journal} {\bibinfo
  {journal} {\apjl}\ }\textbf {\bibinfo {volume} {933}},\ \bibinfo {pages}
  {L23} (\bibinfo {year} {2022})},\ \Eprint {https://arxiv.org/abs/2202.01903}
  {arXiv:2202.01903} \BibitemShut {NoStop}%
\bibitem [{\citenamefont {{Dror}}\ \emph {et~al.}(2019)\citenamefont {{Dror}},
  \citenamefont {{Ramani}}, \citenamefont {{Trickle}},\ and\ \citenamefont
  {{Zurek}}}]{2019PhRvD.100b3003D}%
  \BibitemOpen
  \bibfield  {author} {\bibinfo {author} {\bibfnamefont {J.~A.}\ \bibnamefont
  {{Dror}}}, \bibinfo {author} {\bibfnamefont {H.}~\bibnamefont {{Ramani}}},
  \bibinfo {author} {\bibfnamefont {T.}~\bibnamefont {{Trickle}}},\ and\
  \bibinfo {author} {\bibfnamefont {K.~M.}\ \bibnamefont {{Zurek}}},\
  }\bibfield  {title} {\bibinfo {title} {{Pulsar timing probes of primordial
  black holes and subhalos}},\ }\href
  {https://doi.org/10.1103/PhysRevD.100.023003} {\bibfield  {journal} {\bibinfo
   {journal} {\prd}\ }\textbf {\bibinfo {volume} {100}},\ \bibinfo {pages}
  {023003} (\bibinfo {year} {2019})},\ \Eprint
  {https://arxiv.org/abs/1901.04490} {arXiv:1901.04490} \BibitemShut {NoStop}%
\bibitem [{\citenamefont {{Arvanitaki}}\ and\ \citenamefont
  {{Dubovsky}}(2011)}]{2011PhRvD..83d4026A}%
  \BibitemOpen
  \bibfield  {author} {\bibinfo {author} {\bibfnamefont {A.}~\bibnamefont
  {{Arvanitaki}}}\ and\ \bibinfo {author} {\bibfnamefont {S.}~\bibnamefont
  {{Dubovsky}}},\ }\bibfield  {title} {\bibinfo {title} {{Exploring the string
  axiverse with precision black hole physics}},\ }\href
  {https://doi.org/10.1103/PhysRevD.83.044026} {\bibfield  {journal} {\bibinfo
  {journal} {\prd}\ }\textbf {\bibinfo {volume} {83}},\ \bibinfo {eid} {044026}
  (\bibinfo {year} {2011})},\ \Eprint {https://arxiv.org/abs/1004.3558}
  {arXiv:1004.3558 [hep-th]} \BibitemShut {NoStop}%
\bibitem [{\citenamefont {{Brito}}\ \emph {et~al.}(2020)\citenamefont
  {{Brito}}, \citenamefont {{Cardoso}},\ and\ \citenamefont
  {{Pani}}}]{2020Superradiance}%
  \BibitemOpen
  \bibfield  {author} {\bibinfo {author} {\bibfnamefont {R.}~\bibnamefont
  {{Brito}}}, \bibinfo {author} {\bibfnamefont {V.}~\bibnamefont {{Cardoso}}},\
  and\ \bibinfo {author} {\bibfnamefont {P.}~\bibnamefont {{Pani}}},\ }\href
  {https://doi.org/10.1007/978-3-030-46622-0} {\emph {\bibinfo {title}
  {{Superradiance: New Frontiers in Black Hole Physics}}}},\ \bibinfo {edition}
  {2nd}\ ed.,\ \bibinfo {series} {Lect.~Notes Phys.}, Vol.\ \bibinfo {volume}
  {971}\ (\bibinfo  {publisher} {Springer},\ \bibinfo {year} {2020})\ \Eprint
  {https://arxiv.org/abs/1501.06570} {arXiv:1501.06570 [gr-qc]} \BibitemShut
  {NoStop}%
\bibitem [{\citenamefont {{Unruh}}(1973)}]{1973PhRvL..31.1265U}%
  \BibitemOpen
  \bibfield  {author} {\bibinfo {author} {\bibfnamefont {W.~G.}\ \bibnamefont
  {{Unruh}}},\ }\bibfield  {title} {\bibinfo {title} {{Second Quantization in
  the Kerr Metric}},\ }\href {https://doi.org/10.1103/PhysRevLett.31.1265}
  {\bibfield  {journal} {\bibinfo  {journal} {\prl}\ }\textbf {\bibinfo
  {volume} {31}},\ \bibinfo {pages} {1265} (\bibinfo {year}
  {1973})}\BibitemShut {NoStop}%
\bibitem [{\citenamefont {{Dai}}\ and\ \citenamefont
  {{Stojkovic}}(2023)}]{2023PhRvD.108h4024B}%
  \BibitemOpen
  \bibfield  {author} {\bibinfo {author} {\bibfnamefont {D.-C.}~\bibnamefont
  {{Dai}}}\ and\ \bibinfo {author} {\bibfnamefont {D.}~\bibnamefont {{Stojkovic}}},\
  }\bibfield  {title} {\bibinfo {title} {{Shedding new light on the absence of
  fermionic superradiance and maximal infalling rate of fermions into a black
  hole}},\ }\href {https://doi.org/10.1103/PhysRevD.108.084024} {\bibfield
  {journal} {\bibinfo  {journal} {\prd}\ }\textbf {\bibinfo {volume} {108}},\
  \bibinfo {pages} {084024} (\bibinfo {year} {2023})},\ \Eprint
  {https://arxiv.org/abs/2309.13511} {arXiv:2309.13511} \BibitemShut {NoStop}%
\bibitem [{\citenamefont {{Tremaine}}\ and\ \citenamefont
  {{Gunn}}(1979)}]{1979PhRvL..42..407T}%
  \BibitemOpen
  \bibfield  {author} {\bibinfo {author} {\bibfnamefont {S.}~\bibnamefont
  {{Tremaine}}}\ and\ \bibinfo {author} {\bibfnamefont {J.~E.}\ \bibnamefont
  {{Gunn}}},\ }\bibfield  {title} {\bibinfo {title} {{Dynamical Role of Light
  Neutral Leptons in Cosmology}},\ }\href
  {https://doi.org/10.1103/PhysRevLett.42.407} {\bibfield  {journal} {\bibinfo
  {journal} {\prl}\ }\textbf {\bibinfo {volume} {42}},\ \bibinfo {pages} {407}
  (\bibinfo {year} {1979})}\BibitemShut {NoStop}%
\bibitem [{\citenamefont {{Davoudiasl}}\ \emph {et~al.}(2021)\citenamefont
  {{Davoudiasl}}, \citenamefont {{Denton}},\ and\ \citenamefont
  {{McGady}}}]{2021PhRvD.103e5014D}%
  \BibitemOpen
  \bibfield  {author} {\bibinfo {author} {\bibfnamefont {H.}~\bibnamefont
  {{Davoudiasl}}}, \bibinfo {author} {\bibfnamefont {P.~B.}\ \bibnamefont
  {{Denton}}},\ and\ \bibinfo {author} {\bibfnamefont {D.~A.}\ \bibnamefont
  {{McGady}}},\ }\bibfield  {title} {\bibinfo {title} {{Ultralight fermionic
  dark matter}},\ }\href {https://doi.org/10.1103/PhysRevD.103.055014}
  {\bibfield  {journal} {\bibinfo  {journal} {\prd}\ }\textbf {\bibinfo
  {volume} {103}},\ \bibinfo {eid} {055014} (\bibinfo {year} {2021})},\ \Eprint
  {https://arxiv.org/abs/2008.06505} {arXiv:2008.06505 [hep-ph]} \BibitemShut
  {NoStop}%
\bibitem [{\citenamefont {{Bar}}\ \emph {et~al.}(2022)\citenamefont {{Bar}},
  \citenamefont {{Blum}},\ and\ \citenamefont {{Sun}}}]{2022PhRvD.105h3015B}%
  \BibitemOpen
  \bibfield  {author} {\bibinfo {author} {\bibfnamefont {N.}~\bibnamefont
  {{Bar}}}, \bibinfo {author} {\bibfnamefont {K.}~\bibnamefont {{Blum}}},\ and\
  \bibinfo {author} {\bibfnamefont {C.}~\bibnamefont {{Sun}}},\ }\bibfield
  {title} {\bibinfo {title} {{Galactic rotation curves versus ultralight dark
  matter: A systematic comparison with SPARC data}},\ }\href
  {https://doi.org/10.1103/PhysRevD.105.083015} {\bibfield  {journal} {\bibinfo
   {journal} {\prd}\ }\textbf {\bibinfo {volume} {105}},\ \bibinfo {eid}
  {083015} (\bibinfo {year} {2022})},\ \Eprint
  {https://arxiv.org/abs/2111.03070} {arXiv:2111.03070 [hep-ph]} \BibitemShut
  {NoStop}%
\bibitem [{\citenamefont {{Di Paolo}}\ \emph {et~al.}(2018)\citenamefont {{Di
  Paolo}}, \citenamefont {{Nesti}},\ and\ \citenamefont
  {{Villante}}}]{2018MNRAS.475.5385D}%
  \BibitemOpen
  \bibfield  {author} {\bibinfo {author} {\bibfnamefont {C.}~\bibnamefont {{Di
  Paolo}}}, \bibinfo {author} {\bibfnamefont {F.}~\bibnamefont {{Nesti}}},\
  and\ \bibinfo {author} {\bibfnamefont {F.~L.}\ \bibnamefont {{Villante}}},\
  }\bibfield  {title} {\bibinfo {title} {{Phase space mass bound for fermionic
  dark matter from dwarf spheroidal galaxies}},\ }\href
  {https://doi.org/10.1093/mnras/sty091} {\bibfield  {journal} {\bibinfo
  {journal} {\mnras}\ }\textbf {\bibinfo {volume} {475}},\ \bibinfo {pages}
  {5385} (\bibinfo {year} {2018})},\ \Eprint {https://arxiv.org/abs/1704.06644}
  {arXiv:1704.06644} \BibitemShut {NoStop}%
\bibitem [{\citenamefont {{Filzinger}}\ \emph {et~al.}(2025)\citenamefont
  {{Filzinger}}, \citenamefont {{Lange}}, \citenamefont {{Huntemann}},
  \citenamefont {{Sanner}},\ and\ \citenamefont
  {{Peik}}}]{2025PhRvL.134c1001F}%
  \BibitemOpen
  \bibfield  {author} {\bibinfo {author} {\bibfnamefont {M.}~\bibnamefont
  {{Filzinger}}}, \bibinfo {author} {\bibfnamefont {R.}~\bibnamefont
  {{Lange}}}, \bibinfo {author} {\bibfnamefont {N.}~\bibnamefont
  {{Huntemann}}}, \bibinfo {author} {\bibfnamefont {C.}~\bibnamefont
  {{Sanner}}},\ and\ \bibinfo {author} {\bibfnamefont {E.}~\bibnamefont
  {{Peik}}},\ }\bibfield  {title} {\bibinfo {title} {{Improved Limits on the
  Coupling of Ultralight Bosonic Dark Matter to Photons from Optical-Clock
  Comparisons}},\ }\href {https://doi.org/10.1103/PhysRevLett.134.031001}
  {\bibfield  {journal} {\bibinfo  {journal} {\prl}\ }\textbf {\bibinfo
  {volume} {134}},\ \bibinfo {eid} {031001} (\bibinfo {year} {2025})},\ \Eprint
  {https://arxiv.org/abs/2312.13723} {arXiv:2312.13723 [physics.atom-ph]}
  \BibitemShut {NoStop}%
\bibitem [{\citenamefont {{Kolb}}\ and\ \citenamefont
  {{Tkachev}}(1993)}]{1993PhRvL..71.3051K}%
  \BibitemOpen
  \bibfield  {author} {\bibinfo {author} {\bibfnamefont {E.~W.}\ \bibnamefont
  {{Kolb}}}\ and\ \bibinfo {author} {\bibfnamefont {I.~I.}\ \bibnamefont
  {{Tkachev}}},\ }\bibfield  {title} {\bibinfo {title} {{Axion Miniclusters and
  Bose Stars}},\ }\href {https://doi.org/10.1103/PhysRevLett.71.3051}
  {\bibfield  {journal} {\bibinfo  {journal} {\prl}\ }\textbf {\bibinfo
  {volume} {71}},\ \bibinfo {pages} {3051} (\bibinfo {year} {1993})},\ \Eprint
  {https://arxiv.org/abs/hep-ph/9303313} {arXiv:hep-ph/9303313 [hep-ph]}
  \BibitemShut {NoStop}%
\bibitem [{\citenamefont {{Levkov}}\ \emph {et~al.}(2018)\citenamefont
  {{Levkov}}, \citenamefont {{Panin}},\ and\ \citenamefont
  {{Tkachev}}}]{2018PhRvL.121o1301L}%
  \BibitemOpen
  \bibfield  {author} {\bibinfo {author} {\bibfnamefont {D.~G.}\ \bibnamefont
  {{Levkov}}}, \bibinfo {author} {\bibfnamefont {A.~G.}\ \bibnamefont
  {{Panin}}},\ and\ \bibinfo {author} {\bibfnamefont {I.~I.}\ \bibnamefont
  {{Tkachev}}},\ }\bibfield  {title} {\bibinfo {title} {{Gravitational
  Bose-Einstein Condensation in the Kinetic Regime}},\ }\href
  {https://doi.org/10.1103/PhysRevLett.121.151301} {\bibfield  {journal}
  {\bibinfo  {journal} {\prl}\ }\textbf {\bibinfo {volume} {121}},\ \bibinfo
  {eid} {151301} (\bibinfo {year} {2018})},\ \Eprint
  {https://arxiv.org/abs/1804.05857} {arXiv:1804.05857 [astro-ph.CO]}
  \BibitemShut {NoStop}%
\bibitem [{\citenamefont {{Eggemeier}}\ and\ \citenamefont
  {{Niemeyer}}(2019)}]{2019PhRvD.100f3528E}%
  \BibitemOpen
  \bibfield  {author} {\bibinfo {author} {\bibfnamefont {B.}~\bibnamefont
  {{Eggemeier}}}\ and\ \bibinfo {author} {\bibfnamefont {J.~C.}\ \bibnamefont
  {{Niemeyer}}},\ }\bibfield  {title} {\bibinfo {title} {{Formation and mass
  growth of dark matter halos in fuzzy dark matter simulations}},\ }\href
  {https://doi.org/10.1103/PhysRevD.100.063528} {\bibfield  {journal} {\bibinfo
   {journal} {\prd}\ }\textbf {\bibinfo {volume} {100}},\ \bibinfo {eid}
  {063528} (\bibinfo {year} {2019})},\ \Eprint
  {https://arxiv.org/abs/1906.01348} {arXiv:1906.01348 [astro-ph.CO]}
  \BibitemShut {NoStop}%
\end{thebibliography}
%

\end{document}